\documentclass{emulateapj}
\usepackage{multirow}
\usepackage{epsfig}
 \usepackage{url}
\usepackage{bmpsize}
\usepackage{amsmath}
\usepackage[toc,page]{appendix}
\usepackage[caption=false]{subfig}

\begin{document}

\title{The Scatter in the Hot Gas Content of Early-Type Galaxies}

\author{Yuanyuan Su\altaffilmark{1$\dagger$},
Jimmy A.\ Irwin\altaffilmark{2}, Raymond E.\ White III\altaffilmark{2}, and Michael C.\ Cooper\altaffilmark{1}}
\affil{$^1$Department of Physics and Astronomy, University of California, Irvine, 4129 Frederick Reines Hall, Irvine, CA 92697, USA}

\affil{$^2$Department of Physics and Astronomy, University of 
Alabama, Box 870324, Tuscaloosa, AL 35487, USA}
\altaffiltext{$\dagger$}{Email: yuanyuas@uci.edu}

\begin{abstract}
Optically-similar early-type galaxies are observed to have a large and poorly understood range in the amount of hot, X-ray-emitting gas they contain.To investigate the origin of this diversity, we studied the hot gas properties of all 42 early-type galaxies in the multiwavelength ATLAS$^{\rm 3D}$ survey that have sufficiently deep {\sl Chandra} X-ray observations. We related their hot gas properties to a number of internal and external physical quantities. To characterize the amount of hot gas relative to the stellar light, we use the ratio of the gaseous X-ray luminosity to the stellar $K$-band luminosity, $L_{X_{\rm gas}}/L_K$; we also use the deviations  of $L_{X_{\rm gas}}$ from the best-fit $L_{X_{\rm gas}}$--$L_K$ relation (denoted $\Delta L_{X_{\rm gas}}$). We quantitatively confirm previous suggestions that various effects conspire to produce the large scatter in the observed $L_X/L_K$ relation. In particular, we find that the deviations $\Delta L_{X_{\rm gas}}$ are most strongly positively correlated with the (low rates of) star formation and the hot gas temperatures in the sample galaxies. This suggests that mild stellar feedback may energize the gas without pushing it out of the host galaxies. We also find that galaxies in high galaxy density environments tend to be massive slow-rotators, while galaxies in low galaxy density environments tend to be low mass, fast-rotators. Moreover, cold gas in clusters and fields may have different origins. The star formation rate increases with cold gas mass for field galaxies but it appears to be uncorrelated with cold gas for cluster galaxies.
\end{abstract}


\section{\bf Introduction}
\smallskip

Present day early-type galaxies (elliptical and lenticular galaxies; ETGs) tend to contain substantial 
atmospheres of hot gas.  The thermal and chemical properties of this gas reservoir  
preserve much of the  history of the galaxy.  
Such hot gas can also regulate star formation and impact the evolution of the host galaxy.
In particular, there have been a growing number of 
studies that show ETGs may contain sizable cold gas halos (Oosterloo et al.\ 2010; Young et al.\ 2011; Serra et al.\ 2012).
Lagos et al.\ (2014) found through numerical simulations that 90\% of the cold gas in ETGs may be supplied by the radiative cooling of their hot atmospheres.  
The O${\rm VII}$ line has been detected in the ISM of several ETGs, revealing the existence of weak cooling flows (Pinto et al.\ 2014). 
Given that hot gas is heavily involved in the formation and evolution of ETGs,
it is crucial to understand what mechanisms are capable of modifying the hot gas content.

The primary source of the hot ISM is
gas lost from aging stars, while at least some fraction comes from minor mergers and accretion of the intergalactic medium (Mathews \& Brighenti 2003). Such gas is heated through supernova explosions and the thermalization of stellar motions. It radiates mainly in X-rays via thermal bremsstrahlung and line emission. 
Galaxies with comparable stellar masses should have similar amounts of gas deposited into the ISM. Thus, we expect to observe a tight relation between the X-ray luminosity ($L_{X}$) and the optical luminosity ($L_{\rm opt}$).  In contrast, a puzzlingly large variation in X-ray luminosity among ETGs of similar optical luminosities has been reported in observations. 
The discussion of this discrepancy dates back to the time of the {\sl Einstein Observatory} (e.g.\ Canizares et al.\ 1987; Fabbiano et al.\ 1992). 
As one of the early investigations using the modest {\sl ROSAT} measurements,  
O'Sullivan et al.\ (2001) found a scatter in $L_{X}/L_{\rm opt}$ of up to two orders of magnitude for 430 nearby ETGs. Later, the {\sl Chandra} X-ray Observatory, with its high sensitivity, large spectral range, and in particular its superb $0.5^{\prime\prime}$ angular resolution, found that ETGs contain detectable X-ray emitting point sources such as low mass X-ray binaries (LMXBs),
in addition to hot gas (e.g.\ Sarazin 2000). 
Recent studies show that additional stellar components such as cataclysmic variables and active binaries (CV/ABs) contribute to the diffuse X-ray emission as well
(e.g.\ Revnivtsev et al.\ 2008). 
Galaxies with low $L_{X}/L_{\rm opt}$
generally have a larger fraction of their flux provided by stellar point sources than do X-ray luminous gas-rich galaxies. After excluding the contamination of LMXBs and other stellar emission, current studies with {\sl Chandra} observations reveal an even larger scatter in the gaseous content relative to stars, as measured by the $L_{X_{\rm gas}}/L_{\rm opt}$ relation for ETGs (e.g.\ Boroson et al.\ 2011). We might naively expect that the $L_{X_{\rm gas}}/L_{\rm opt}$ relation may be related to galaxy ages, since older galaxies have more time to accumulate gaseous atmospheres; however the scatter in $L_{X_{\rm gas}}/L_{\rm opt}$ greatly exceeds the range in ages for these ETGs.

Energy feedback 
may modify the hot gas content of ETGs, since
gas can be driven from  galaxies by supernovae (SNe) heating and active galactic nuclear (AGN) activity.  
David et al.\ (2006) found that heating by Type Ia supernova (SNIa) is energetically sufficient to generate galactic winds in these galaxies even if the present SNIa rate is overestimated. AGN heating may also instigate gaseous outflows. X-ray cavities filled with radio emission have been frequently detected in the ISM of elliptical galaxies (e.g.\ Finoguenov \& Jones 2002, Forman et al.\ 2005), which indicates the effects of AGN activity.
If the feedback is relatively mild, gas  may just be redistributed out to larger radii (and eventually fall back) and the feedback energy could heat up more gas into an X-ray emitting phase (Chevalier \& Clegg 1985). 
These internal feedback processes are particularly important for isolated galaxies, since environmental effects such as ram pressure stripping 
and pressure confinement by intracluster gas would be insignificant.

Deeply connected to the feedback scenario is the dependence of a galaxy's
gaseous content upon the depth of the galaxy's gravitational potential:
it is easier for more massive galaxies to retain their gas against loss mechanisms such as ram pressure stripping and internally driven outflows. 
Mathews et al.\ (2006) studied a number of massive ETGs and found that their hydrostatic masses seem to correlate with $L_X/L_{\rm opt}$; they concluded that the role of feedback becomes more important in less massive galaxies. 
However, the  total masses of ETGs can be very difficult to measure accurately. In some cases, hot gas temperatures and stellar velocity dispersions can be regarded as proxies for total masses.
The depths of gravitational potentials can also be influenced by the rotation and flattening of galaxies.
It has been noticed since the {\sl Einstein Observatory} epoch that flattened ETGs have smaller $L_X/L_{\rm opt}$ compared to their rounder counterparts (Eskridge et al.\ 1995). 
Sarzi et al.\ (2013) demonstrated that slowly rotating galaxies have larger X-ray halos, although 
it has always been difficult to disentangle rotation and flatness, since fast rotational systems would also be flatter due to centrifugal forces. 
The ATLAS$^{\rm 3D}$ project proposes a kinematic classification of ETGs into fast- and
slow-rotators to distinguish ``true ellipticals" from ``true lenticulars;" this kinematic classification is less subject to
the projection effects which confounded prior morphological studies
(Emsellem et al.\ 2011).

Environmental influences are another widely discussed 
possible cause of the scatter in $L_{X_{\rm gas}}/L_{\rm opt}$ 
(White \& Sarazin 1991; Mathews \& Brighenti 1998; Brown \& Bregman 2000). Most galaxies, in particular early-type, reside in groups and clusters. The hot gas content of such galaxies can be affected by their environments through various processes, including interactions between galaxies (harassment), 
interactions between galaxies and intracluster gas (ram pressure stripping and pressure confinement), 
and interactions between galaxies and the gravitational potential of the cluster (tidal stripping).
Perhaps the most commonly discussed case is ram pressure stripping, the process through which the hydrodynamic drag of intracluster gas pulls 
the ISM out of  galaxies moving through it (Gunn \& Gott 1972).  
Simulations suggest that nearly all cluster galaxies experience 
ram pressure stripping throughout their lifetime (Bruggen \& De Lucia 2008). 
Indeed, nothing is more convincing than witnessing a galaxy losing its hot gas. 
There are several X-ray observations of on-going ram pressure stripping in ETGs. These observations are characterized by hot gas tails displaced from, and trailing behind, the host galaxy's stellar distributions  
(M86 -- Randall et al.\ 2008; NGC~4472 -- Irwin \& Sarazin 1996; NGC~1404 -- Machacek et al.\ 2005; NGC~1400 -- Su et al.\ 2014). 
Most such studies focus only on individual galaxies. 
The next logical step is to undertake a statistical study of a sample of ETGs experiencing varying amounts of ram pressure.

Our primary goal is to understand what causes the large scatter in the $L_{X_{\rm gas}}/L_{\rm opt}$ relation, using a statistically large sample of ETGs.
In this paper we investigate the hot gas content of all 42 ETGs in the ATLAS$^{\rm 3D}$ survey that have 
sufficiently deep {\sl Chandra} X-ray observations. 
We look for correlations between $L_{X_{\rm gas}}/L_K$ and various internal and external
factors which may affect the hot gas content of ETGs, in order to infer their relative importance.
In addition, 15 of the 42 galaxies in this sample reside in the Virgo Cluster, which allows us to statistically study the effects of ram pressure stripping in a cluster environment.

We assume $H_0 = 70$ km~s$^{-1}$ Mpc$^{-1}$, $\Omega_{\Lambda}=0.7$, and $\Omega_M=0.3$. 
Throughout this paper uncertainties are given at the $1\sigma$ confidence level unless otherwise stated.  
We report our sample selection process and introduce their properties in \S2;
observations and data reduction are described in \S3; we report our results in \S4; 
we discuss the implications of our results in \S5; 
we summarize our main conclusions in \S6.
In the Appendix, we examine statistical and systematic uncertainties and provide additional figures. 
\bigskip

\section{\bf Sample Selections} 
\smallskip

 \begin{figure} 
 \vspace{-3mm}
\epsscale{1.2}
\plotone{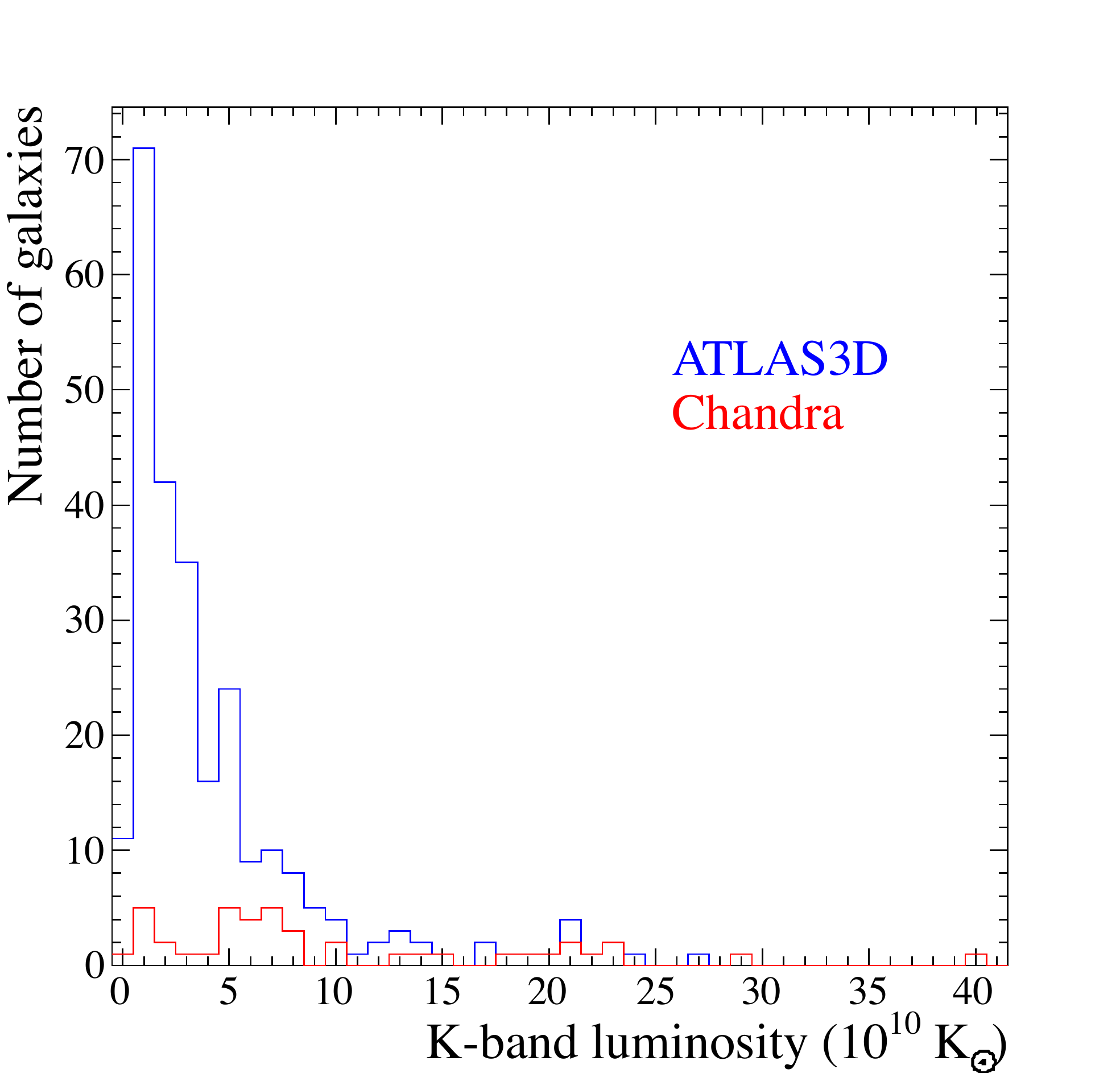}
\figcaption{\label{fig:sam} K-band luminosity distribution of all 260 galaxies in ATLAS$^{\rm 3D}$ and 42 galaxies in our sample. K-band luminosities shown in this figure are taken from Cappellari et al. (2011).}
\end{figure}

We want to study a statistically significant sample of ETGs that is as diverse as possible and with published values of multiple relevant physical characteristics of the galaxies. Our main goal is to relate the hot gas properties of these galaxies to those observables which can reflect their formation and evolution from different aspects. 
ATLAS$^{\rm 3D}$ 
is a volume-limited (1.16$\times10^5$ Mpc$^3$), multi-wavelength survey of a complete sample ($L_K>6\times10^9~L_{\odot}$) of ETGs  (Cappellari et al.\ 2013). 
To our knowledge, it provides the richest and most complete information on the global stellar kinematics, dynamics, and different phases of the interstellar medium (but not X-ray emitting hot gas) of a large sample of ETGs. Even more importantly, the values of these observables were derived through a uniform analysis. 
We aim to study all galaxies in ATLAS$^{\rm 3D}$ that have {\sl Chandra} observations
that are  deep enough to constrain their hot gas properties.

\begin{table*}
  \caption{Chandra Observational Log}
  \label{tab:distance}
  \begin{center}
    \leavevmode
    \begin{tabular}{lcccc} \hline \hline   
\colhead{Name}&\colhead{Instrument}&\colhead{Obs-ID}&\colhead{Exposure (ks)}&\\ \hline
IC1024 & ACIS-S & 14901&19.8\\
NGC~821 & ACIS-S&4006, 4408, 5691,5692, 6310, 6313, 6314&13.9,13.6,39.6, 27.6, 31.8,49, 39.6\\
NGC~1023&ACIS-S&8197,8198,8464,8465&46.8,47.3,42.4,44.2\\
NGC~1266&ACIS-S&11578&29.5\\
NGC~2768&ACIS-S&9528&61.8\\
NGC~2778&ACIS-S&11777&29.2\\
NGC~3377&ACIS-S&2934&38.8\\
NGC~3379&ACIS-S&1587, 7073, 7074, 7075, 7076&29.9, 81.6, 67.9, 79.4, 66.5\\
NGC~3384&ACIS-S&13829&92.5\\
NGC~3599&ACIS-S&9566&19.7\\
NGC~3607&ACIS-I&2038&27\\
NGC~3608&ACIS-I&2038&27\\
NGC~3665&ACIS-S&3222&17.9\\
NGC~4203&ACIS-S&10535&41.6\\
NGC~4261&ACIS-S&9569&99\\
NGC~4278&ACIS-S&4741, 7077, 7079, 7080, 7081, 7808&35,111, 106,56,112, 52\\
NGC~4342 & ACIS-S & 4687, 12955& 35.4, 68.1\\
NGC~4365 & ACIS-S& 2015, 5921, 5922, 5923, 5924, 7224& 39, 38, 39, 35, 25, 10\\
NGC~4374 & ACIS-S & 5908, 6131, 803 & 45.3, 38.9, 27.3 \\
NGC~4382 & ACIS-S & 2016 & 39.7\\
NGC~4406 & ACIS-S & 318, 963 & 11, 14  \\
NGC~4458&ACIS-S&14905&29.4\\
NGC~4459 & ACIS-S & 2927, 11784&  9.8, 29.8 \\ 
NGC~4472 & ACIS-S & 321, 11247& 31.5, 38.6\\
NGC~4473 & ACIS-S & 4688 & 25.5 \\
NGC~4477 & ACIS-S & 9527 & 35.0  \\
NGC~4494&ACIS-S&2079&21.7\\
NGC~4526 & ACIS-S & 3925 & 39.6 \\
NGC~4552 & ACIS-S & 2072 & 51.3 \\
NGC~4564 & ACIS-S & 4008 & 15.7 \\
NGC~4596 & ACIS-S & 11785 & 31.0 \\
NGC~4621 & ACIS-S & 2068 & 21.3 \\
NGC~4636 & ACIS-S & 323 & 43.5 \\
NGC~4649 & ACIS-S & 785, 8182, 8507& 15.8, 36.9, 17.1 \\
NGC~4697 & ACIS-S& 784, 4727, 4728, 4729, 4730 &38, 40, 34, 22, 35\\
NGC~4710 & ACIS-S & 9512 & 28.8 \\
NGC~5422&ACIS-S&9511, 9772&17.9, 18.4\\
NGC~5576&ACIS-S&11781&29.7\\
NGC~5813&ACIS-S&12952&143\\
NGC~5846&ACIS-S&788, 7062, 8448, 8449&21,23, 8, 19 \\
&ACIS-I&7923&87\\
NGC~5866&ACIS-S&2879&30.6\\
NGC~7457&ACIS-S&11786&28.7\\ \hline
    \end{tabular}
  \end{center}
\end{table*}

We require that each galaxy has been observed with {\sl Chandra} for at least 15 ksec.\footnote{We did not include NGC~4486A since this galaxy is too close to M87. It is extremely difficult to disentangle its X-ray emission from that of M87. We did not include M87 because it is the  central galaxy of the Virgo cluster, so it is difficult to separate
its galactic gas from intracluster gas.} 
Our screening criteria yielded 42 ETGs, as listed in Table~1.
We take their distances and half-light radii ($r_e$) from ATLAS$^{\rm 3D}$(Cappellari et al.\ 2013). All galaxies in our sample reside at distances between 10 and 31\,Mpc, with their $r_e$ ranging from  0.5 to 8.6 kpc.  Galaxies in our sample are diverse in luminosity and environment and are representative of ETGs. 
From the literature we obtain stellar ages, stellar velocity dispersions ($\sigma$), atomic gas masses ($M_{\rm HI}$), molecular gas masses ($M_{\rm H_2}$), total masses (within $1\,r_e$) ($M_{\rm tot}$), a rotational parameter ($\lambda$),
and galaxy ellipticities ($\epsilon$), with most values from ATLAS$^{\rm 3D}$ and measured with consistent analyses. 
We take the star formation rate (SFR) of almost all galaxies in our sample from Amblard et al.\ (2014). These values were derived from their spectral energy distributions (SEDs), fitting from the UV to millimeter wavelengths. If not listed in Amblard et al.\ (2014), we take  SFR values from Davis et al.\ (2014), which are derived from 
{\sl WISE} 22$\mu$m data.
We list  these observables in Tables~2 and 3.

It should be noted that {\sl Chandra} archived galaxies do not form a complete sample. 
In Figure~\ref{fig:sam}, we compare the K-band luminosity distribution of 42 galaxies in our sample and 260 galaxies in the ATLAS$^{\rm 3D}$ survey. Very faint galaxies ($L_K\lesssim 10^{10}\,$L$_{K\odot}$) have been underrepresented in our sample. 
To our knowledge, this is however the largest sample available that is suitable for associating hot gas properties of ETGs with internal and external factors that may be related to galaxy evolution.  

\begin{table*}
\caption{Properties of early-type galaxies in our sample from ATLAS3D}
  \begin{center}
    \leavevmode
\begin{tabular}{lccccccccc}\hline \hline 
{Name}&{Distance$^{\rm a}$}&{$r_{\rm e}$$^{\rm a}$}&{$M_{\rm tot}$$^{\rm b}$}&{Rotation $\lambda$$^{\rm c}$}&{Ellipticity $\epsilon$$^{\rm c}$}&{$\lambda/\sqrt{\epsilon}$}&{$M_{\rm HI}$$^{\rm d}$}&{$M_{\rm H_2}$$^{\rm e}$}&{$\sigma$$^{\rm b}$}\\
&{(Mpc)}&{(kpc, arcmin)}&{($10^{10}$M$_{\odot}$)} &{} &{}&{}&($10^{8}$M$_{\odot}$)& {($10^{8}$M$_{\odot}$)}&{(km\,s$^{-1}$)}\\ \hline
IC1024&24.2&1.293, 0.187&1.469&0.691&0.587&0.902&--&3.981&77.98\\
NGC~821 & 23.4&4.507, 0.664& 12.455&0.273&0.392&0.436&	$<$0.081	&$<$0.331& 179.5\\
NGC~1023 &11.1&2.575, 0.798&6.592&0.391&	0.363&0.649	&19.498	&0.062&166.7\\
NGC~1266&29.9&2.891, 0.340&2.576&0.638	&0.193&1.452&	--	&19.498&79.07\\
NGC~2768 &21.8&6.499, 1.052&34.198 &0.253	&0.472&0.368	&0.646	&0.437&198.2 \\
NGC~2778&22.3&1.696, 0.264&3.141&0.572	&0.224&1.209	&$<$0.115&$<$0.302&132.1\\
NGC~3377&10.9&1.870, 0.591&2.938&0.522&	0.503&0.736	&$<$0.033&$<$0.091&128.2\\
NGC~3379&10.3&1.975, 0.664&8.222&0.157	&0.104&0.487	&$<$0.031	&$<$0.052&185.8\\
NGC~3384&11.3&1.773, 0.539&3.664&0.407	&0.065&1.596	&0.178	&0.129&130.0\\
NGC~3599&19.8&2.236, 0.391&0.989&0.282	&0.08&0.997	&$<$0.107	&0.234&63.68\\
NGC~3607 &22.2 &4.163, 0.648&21.979 &0.209	&0.185&0.486	&$<$0.083	&2.630  &206.5 \\
NGC~3608 & 22.3 &3.158, 0.492&9.036  & 0.043	&0.19&0.099&	0.145&$<$0.380 &169.0\\
NGC~3665&33.1&4.876, 0.515&35.975&0.41	&0.216&0.882	&$<$0.269	&8.128&216.3\\
NGC~4203&14.7&2.095, 0.492&4.018&0.305	&0.154&0.777	&14.125	&0.245&129.1\\
NGC~4261 &30.8 &5.539, 0.634 &52.723 & 0.085&	0.222&0.18&0.5$^{\rm f}$&	$<$0.479 &265.5\\
NGC~4278 & 15.6&2.350, 0.527&11.912& 0.178	&0.103&0.555	&6.310	&$<$0.282&212.8\\
NGC~4342 &16.5 &0.518, 0.110&3.319&0.528&	0.442&0.794&	--	&$<$0.174&242.1\\
NGC~4365 & 23.3& 5.831, 0.875&33.497&0.088	&0.254&0.175	&$<$0.4$^{\rm f}$&$<$0.417 &221.3\\
NGC~4374 &18.5 & 4.653, 0.875&38.459&0.024&	0.147&0.063&	$<$0.182&$<$0.170&258.2\\
NGC~4382 & 17.9 &5.656, 1.101 &28.054&0.163	&0.202&0.363	&$<$0.093	&$<$0.229&179.1\\
NGC~4406 & 16.8  &7.512, 1.555&39.811&0.052	&0.211&0.113	&1	&$<$0.251&190.5\\
NGC~4458&16.4&1.838, 0.391&1.076&0.072	&0.121&0.207	&$<$0.081	&$<$0.204&88.51\\
NGC~4459 & 16.1 &2.817, 0.605 &8.299&0.438	&0.148&1.139	&$<$0.081	&1.738& 158.1\\ 
NGC~4472 & 17.1 & 7.882, 1.592&59.566&0.077	&0.172&0.186	&0.5$^{\rm f}$&$<$0.178&250.0\\
NGC~4473 &15.3 & 2.000, 0.449 &8.472&0.229	&0.421&0.353	&$<$0.072	&$<$0.117&186.6\\
NGC~4477 &16.5 &3.050, 0.648&8.770 &0.446&	0.135&1.214&$<$0.089&	0.347&148.9\\
NGC~4494&16.6&3.943, 0.816&9.840&0.212	&0.173&0.51	&$<$0.069	&$<$0.178&150.0\\
NGC~4526 &16.4 &3.551, 0.744&17.498&0.453	&0.361&0.754	&$<$0.19$^{\rm g}$	&3.890&208.9\\
NGC~4552 &15.8&2.517, 0.565& 15.922&0.049	&0.047&0.226	&$<$0.074&	0.191&224.4\\
NGC~4564 & 15.8&1.517, 0.340&3.828&0.619	&0.56&0.827&	$<$0.081&0.178&154.5\\
NGC~4596 &16.5 &3.050, 0.648 &8.204&0.639	&0.254&1.268	&$<$0.135	&0.204& 125.6\\
NGC~4621 & 14.9 & 3.080, 0.711& 12.882&0.291	&0.365&0.482	&$<$0.072&$<$0.135&197.7\\
NGC~4636 &14.3  &6.167, 1.485&24.889&0.036	&0.094&0.117&	8.1$^{\rm f}$	&$<$0.074&181.6\\
NGC~4649 &17.3 & 5.453, 1.101&52.360&0.127	&0.156&0.322	&$<$0.151	&$<$0.275&267.9\\
NGC~4697 &  11.4 &3.379, 1.028&11.695& 0.322&	0.447&0.482&	$<$0.6$^{\rm f}$	&$<$0.072&169.4\\
NGC~4710 &16.5 &2.368, 0.503 &5.768&0.652	&0.699&0.78	&0.069	&5.248&104.7\\
NGC~5422&30.8&3.115, 0.356&8.892&0.6	&0.604&0.772	&0.741	&$<$0.603&157.4\\
NGC~5576&24.8&2.609, 0.365&7.568&0.102	&0.306&0.184	&$<$0.4$^{\rm f}$&	$<$0.398&155.2\\
NGC~5813&31.3&8.614, 0.959 &39.08&0.071	&0.17&0.172	&$<$1.0$^{\rm f}$&	$<$0.490&210.9\\
NGC~5846 &24.2&6.783, 0.981 &36.559&0.032&	0.062&0.129&3.5	&$<$0.603&223.4\\
NGC~5866 & 14.9& 2.621, 0.605 &10.023& 0.319	&0.566&0.424	&0.091	&2.951&157.0\\
NGC~7457&12.9&2.251, 0.605 &1.652&0.519&	0.47&0.757&	$<$0.041	&0.091&74.64\\ \hline
    \end{tabular}
  \end{center}
\tablecomments{(a).\ Cappellari et al.\ (2011); (b).\ Total mass measured within 1\,$r_{\rm e}$ (Cappellari et al.\ 2013); (c).\ Emsellem et al.\ (2011);
(d).\ Serra et al.\ (2012);
(e).\ Young et al.\ (2011);
(f).\ Serra \& Oosterloo (2010);
(g).\ Lucero \& Young (2013).
}
\end{table*}

In order to probe how different factors affect the ISM of different galaxies,
we also subdivide our sample in three ways, based on the E/S0 dichotomy ($\lambda/\sqrt{\epsilon}$), galaxy mass, and environment. 
In the E/S0 subdivision, we use the stellar dynamical criterion found by the ATLAS$^{\rm 3D}$ project
to distinguish ``true elliptical" galaxies from ``true lenticular" galaxies in our sample.
The ATLAS$^{\rm 3D}$ project found that a combination of a rotational parameter ($\lambda$) and 
the galaxy ellipticity ($\epsilon$) could be used to distinguish slow rotators 
(``true ellipticals," which have $\lambda/\sqrt{\epsilon} < 0.31$)
from fast rotators (``true lenticulars," which have $\lambda/\sqrt{\epsilon} > 0.31$).
Using this criterion for our sample, we will distinguish 
12 ``true elliptical" galaxies from 30 ``true lenticular" galaxies.
We also divide our sample into high mass and low mass systems: 
we regard the  20 galaxies in our sample with $M_{\rm tot}\ge10^{11}$ M$_{\odot}$ as high mass galaxies, while the remaining 22 galaxies are regarded as low mass. 
Finally, we divide our sample environmentally:
galaxies with 15 or fewer SDSS galaxies nearby (Column 4 in Table~3; see \S3.5) are classified as 
field galaxies, while galaxies with more neighbors are classified as belonging to groups and clusters. 
Four galaxies in our sample do not have good SDSS coverage, so we determine their
environments differently: according to the literature, we consider NGC~1023 as a galaxy in a group and we regard NGC~821, NGC~1266, and NGC~7457 as being in relatively isolated environments (Chernin et al.\ 2010; Arnold et al.\ 2014; Malchaney \& Jeltema 2010; Alatalo et al.\ 2014). Consequently, 21 galaxies in our sample are classified to be in high galaxy density environments and the other 21 galaxies are classified to be in low galaxy density environments.

\begin{table*}
\caption{Other properties of early-type galaxies in our sample and their spectral fitting results}
  \begin{center}
    \leavevmode
 \begin{tabular}{lccccccccc} \hline \hline  
{Name}&{Morphology$^{\star}$}&{Age}&{SDSS}&{Virgo$^{\ddagger}$}&{SFR$^{\rm o}$}&{${L_{X_{\rm gas}}}^{\dagger}$}&{${L_K}^{\dagger}$}&{${T_X}^{\dagger}$}&{${M_{X_{\rm gas}}}^{\dagger}$}\\
{}&{} & {(Gyr)}&{} & {}&{(M$_{\odot}/$yr)}&{($10^{40}$ erg\,s$^{-1}$)}&{(10$^{10}$\,L$_{K\odot}$)}&{(keV)}&{($10^8$M$_{\odot}$)}  \\ \hline 
IC1024&S0&--&5&0&$0.527^{+0.218}_{-0.154}$$^{\rm p}$&$0.095^{+0.024}_{-0.032}$&1.116&--&0.112$^{+0.013}_{-0.021}$\\
NGC~821 &E6 & 7.5$^{\rm a}$& --&0&$<0.013^{\rm q}$ & $0.0039^{+0.078}_{-0.0039}$&8.930& 0.15$^{+0.85}_{-0.05}$$^{\ast}$&0.198$^{+0.271}_{-0.198}$\\
NGC~1023 & SB0&4.7$^{\rm g}$ & --& 0&$0.015^{+0.011}_{-0.006}$&$0.659^{+0.725}_{-0.348}$&6.550 &$0.15^{+0.01}_{-0.02}$&1.003$^{+0.450}_{-0.314}$\\
NGC~1266&SB0&1.8$^{\rm b}$&--&0&$1.318^{+0.920}_{-0.542}$&$0.288^{+0.071}_{-0.049}$&3.002&$0.87^{+0.09}_{-0.09}$&0.692$^{+0.080}_{-0.062}$\\
NGC~2768 & E6& 11.2$^{\rm m}$& 5& 0&$0.033^{+0.024}_{-0.014}$& $1.249^{+0.246}_{-0.183}$&15.600&$0.37^{+0.04}_{-0.04}$&4.907$^{+0.462}_{-0.374}$ \\
NGC~2778&E&5.4$^{\rm a}$&4&0&$0.008^{+0.009}_{-0.004}$&$0.007^{+0.010}_{-0.007}$&2.022&--&0.045$^{+0.026}_{-0.045}$\\
NGC~3377&E5-6&3.7$^{\rm a}$&12& 0&$0.004^{+0.003}_{-0.003}$&$0.0032^{+0.0064}_{-0.0032}$&2.274&$0.22^{+0.12}_{-0.07}$$^{\ast}$&0.041$^{+0.030}_{-0.041}$\\
NGC~3379&E1&8.6$^{\rm a}$&12& 0 &$0.026^{+0.014}_{-0.009}$& $0.010^{+0.003}_{-0.0025}$&5.777&$0.25^{+0.03}_{-0.02}$$^{\ast}$&0.066$^{+0.009}_{-0.009}$\\
NGC~3384&SB0&3.2$^{\rm a}$&12&0&$0.008^{+0.004}_{-0.003}$&$0.007^{+0.004}_{-0.004}$&4.038&$0.25^{+0.30}_{-0.10}$&0.071$^{+0.018}_{-0.021}$\\
NGC~3599&SA0&--&13&0&$0.073^{+0.021}_{-0.016}$$^{\rm p}$&$0.015^{+0.016}_{-0.007}$&1.575&--&0.095$^{+0.042}_{-0.026}$\\
NGC~3607 & SA0&3.6$^{\rm c}$ &13&0&$0.042^{+0.032}_{-0.018}$ &$0.746^{+0.106}_{-0.084}$ &12.380&$	0.45^{+0.05}_{-0.05}$&2.078$^{+0.143}_{-0.121}$ \\
NGC~3608 & E2&6.9$^{\rm a}$ &13 &0&$0.015^{+0.007}_{-0.005}$ & $0.358^{+0.075}_{-0.060}$&4.793&$0.39^{+0.07}_{-0.05}$&0.995$^{+0.100}_{-0.087}$\\
NGC~3665&SA0&9.9$^{\rm k}$&1&0&$0.109^{+0.045}_{-0.032}$$^{\rm p}$&$1.919^{+0.385}_{-0.287}$&18.183&$0.45^{+0.09}_{-0.07}$ &4.347$^{+0.416}_{-0.338}$\\
NGC~4203&SAB0&--&2&0&$0.047^{+0.035}_{-0.020}$&$0.039^{+0.014}_{-0.014}$&3.914&$0.83^{+0.10}_{-0.16}$&0.156$^{+0.025}_{-0.030}$\\
NGC~4261 & E2-3&15.5$^{\rm a}$ &51 &0&$0.117^{+0.064}_{-0.042}$ & $4.261^{+0.038}_{-0.038}$&21&$0.81^{+0.01}_{-0.01}$&8.186$^{+0.036}_{-0.036}$\\
NGC~4278 & E1-2 & 10.7$^{\rm c}$&17 & 0&$0.019^{+0.011}_{-0.007}$&$0.088^{+0.007}_{-0.012}$&6.3&$0.65^{+0.06}_{-0.06}$&0.329$^{+0.013}_{-0.023}$\\
NGC~4342 & S0 & -- &32 &1&--&$0.055^{+0.007}_{-0.006}$&1.25&$0.62^{+0.01}_{-0.01}$&0.067$^{+0.004}_{-0.004}$\\
NGC~4365 & E3& 5.9$^{\rm d}$& 43 &0&$0.050^{+0.026}_{-0.017}$&$0.544^{+0.041}_{-0.040}$&16&$0.59^{+0.02}_{-0.02}$&2.447$^{+0.090}_{-0.091}$\\
NGC~4374 & E1 & 12.2$^{\rm a}$ & 46 &1 &0.058$^{+0.026}_{-0.018}$ &$5.423^{+0.544}_{-0.531}$&19.76&$0.76^{+0.004}_{-0.004}$&6.044$^{+0.296}_{-0.303}$\\
NGC~4382 & SA0 & 1.6$^{\rm c}$ &15  &1 &0.002$^{+0.010}_{-0.002}$ &$1.378^{+0.251}_{-0.259}$&20.82&$0.38^{+0.02}_{-0.02}$&5.351$^{+0.467}_{-0.530}$\\
NGC~4406 & E3 & 11$^{\rm e}$  &28 &1&-- &$9.988^{+1.281}_{-1.281}$&26.82&$0.81^{+0.01}_{-0.01}$&14.81$^{+0.921}_{-0.982}$\\
NGC~4458&E0-1&16$^{\rm c}$&53&0&$0.0007^{+0.0005}_{-0.0003}$&$0.004^{+0.010}_{-0.004}$&1.121&--&0.047$^{+0.041}_{-0.047}$\\
NGC~4459 & SA0 &  7.1$^{\rm f}$ &54 &1&$0.071^{+0.070}_{-0.035}$ &$0.181^{+0.018}_{-0.017}$&6.53&$0.57^{+0.05}_{-0.06}$&0.477$^{+0.023}_{-0.022}$\\ 
NGC~4472 & E2 & 7.9$^{\rm a}$ & 51 &1&$0.085^{+0.050}_{-0.031}$&$16.096^{+0.836}_{-0.836}$	&35.51&$	1.05^{+0.002}_{-0.002}$&23.07$^{+0.592}_{-0.607}$\\
NGC~4473 & E5 & 12.2$^{\rm m}$ & 51 & 1&$0.074^{+0.038}_{-0.025}$ &$0.055^{+0.021}_{-0.017}$&5.57&$0.75^{+0.19}_{-0.20}$&0.136$^{+0.023}_{-0.024}$\\
NGC~4477 & SB0 & 9.6$^{\rm g}$ &55  & 1 &$0.030^{+0.021}_{-0.012}$&$0.740^{+	0.027}_{-0.027}$&6.13&$0.61^{+0.02}_{-0.19}$&1.077$^{+0.020}_{-0.020}$ \\
NGC~4494&E1-2&12$^{\rm i}$&0&0&$7.413^{+7.041}_{-3.611}$&$0.014^{+0.012}_{-0.014}$&17.246&$0.75^{+0.43}_{-0.37}$&0.213$^{+0.077}_{-0.213}$\\
NGC~4526 & SAB0 & 1.7$^{\rm h}$ &18 &1&$0.028^{+0.034}_{-0.015}$&$0.506^{+0.109}_{-0.109}$&13.31&$0.29^{+0.02}_{	-0.02}$&2.283$^{+0.234}_{-0.261}$\\
NGC~4552&E0-1&9.6$^{\rm c}$&40&1&$0.041^{+0.048}_{-0.022}$&$2.207^{+0.110}_{-0.811}$&7.83&$0.6^{+0.01}_{-0.01}$&1.796$^{+0.044}_{-0.368}$\\
NGC~4564&E&5.9$^{\rm c}$&50&1&$0.016^{+0.010}_{-0.006}$&$0.005^{+0.008}_{-0.005}$&2.54&--&0.048$^{+0.028}_{-0.048}$\\
NGC~4596 & SB0 & 11$^{\rm n}$ & 36 &1&$0.007^{+0.005}_{-0.003}$$^{\rm p}$ &$0.098^{+0.026}_{-0.020}$&5.67&$0.55^{+0.09}_{-0.11}$&0.399$^{+0.050}_{-0.042}$\\
NGC~4621 & E1-2 & 10.5$^{\rm j}$ & 36 &1&$0.019^{+0.032}_{-0.012}$ &$0.053^{+0.103}_{-0.026}$&5.54&$	0.18^{+0.05}_{-0.07}$&0.362$^{+0.259}_{-0.105}$\\
NGC~4636 & E0-1 & 10.3  &21 & 0&$0.024^{+0.015}_{-0.009}$&$20.028^{+0.506}_{-0.506}$&10.84&$0.76^{+0.002}_{-0.002}$&15.12$^{+0.190}_{-0.192}$\\
NGC~4649 & E5 & 11.7$^{\rm a}$ & 38 & 1&0.129$^{+0.049}_{-0.035}$&$11.090^{+0.993}_{-0.993}$&29.68&$	0.91^{+0.003}_{-0.003}$&9.292$^{+0.407}_{-0.456}$\\
NGC~4697 & E6&  8.9$^{\rm a}$ &0&0&$0.072^{+0.042}_{-0.027}$&$0.232^{+0.044}_{-0.036}$&8.2&$0.32^{+0.04}_{-0.03}$&0.820$^{+0.074}_{-0.066}$\\
NGC~4710 & SA0 & -- &8 &1&$0.326^{+0.084}_{-0.070}$$^{\rm p}$ &$0.087^{+0.013}_{-0.015}$&4.45&$0.74^{+0.08}_{-0.09}$&0.249$^{+0.018}_{-0.023}$\\
NGC~5422&S0&5$^{\rm l}$&6&0&--&$0.017^{+0.014}_{-0.017}$&5.349&$0.15^{+	0.80}_{-0.15}$&0.458$^{+0.163}_{-0.458}$\\
NGC~5576&E3&10.2$^{\rm j}$&6&0&$0.013^{+0.008}_{-0.005}$&$0.036^{+0.042}_{-0.020}$&7.672&$0.49^{+0.44}_{-0.24}$&0.198$^{+0.093}_{-0.066}$\\
NGC~5813&E1-2&18.3$^{\rm a}$&23&0&$0.052^{+0.031}_{-0.019}$&$68.870^{+1.958}_{-1.953}$&22.443&$0.7^{+0.002}_{-0.002}$&46.68$^{+0.659}_{-0.667}$\\
NGC~5846 &E0-1 & 13.5$^{\rm a}$&18& 0&$0.087^{+0.045}_{-0.030}$&$28.260^{+1.170}_{-1.163}$&23.8&$0.71^{+0.01}_{-0.01}$&20.37$^{+0.417}_{-0.424}$\\
NGC~5866 & SA0&  1.8$^{\rm d}$& 4&0&$0.044^{+0.061}_{-0.025}$ &$0.473^{+0.307}_{-0.197}$&8.1&$0.23^{+0.06}_{-0.03}$&0.818$^{+0.232}_{-0.193}$\\
NGC~7457&SA0&2$^{\rm g}$&--&0&$0.002^{+0.003}_{-0.001}$&$0.003^{+0.008}_{-0.003}$&1.727&--&0.047$^{+0.039}_{-0.047}$  \\ \hline 
    \end{tabular}
  \end{center}
\tablecomments{
${\star}$: morphology type taken from NED, not related to the parameter $\lambda/\sqrt{\epsilon}$ that we used to distinguish Es and S0s in this work.
${\ddagger}$: 1 represents member galaxies of the Virgo Cluster; otherwise 0.
$\dagger$: derived within 2\,$r_{\rm e}$.
(a).\ Trager et al.\ (2000); (b).\ Crocker et al.\ (2012); (c).\ Terlevich \& Duncan (2002); (d).\ Howell (2005); (e).\ Zhang et al.\ (2008); (f).\ Lees et al.\ (1991); (g).\ McDermid et al.\ (2006); (h).\ Gallagher et al.\ (2008); (i).\ Humphrey et al.\ (2006); (j).\ Idiart et al.\ (2007); (k).\ Sanchez-Blazquez et al.\ (2006); (l).\ Sil'Chenko (2006); (m).\ Kuntschner et al.\ (2010); (n).\ Sil'Chenko \& Chilingarian (2011); (o) Amblard et al.\ (2014); (p). Davis et al.\ (2014); (q). Shapiro et al.\ (2010). ($^{\ast}$) Boroson et al.\ (2011).
}
\end{table*}

\bigskip
    
%2
\section{\bf Observations and data reduction} 
\smallskip

%2.1

{\sl Chandra} is the best instrument to study the hot gas properties of 
ETGs, which have k$T\lesssim1$ keV. Its superb spatial resolution can resolve out LMXBs and central active galactic nuclei. The emission of other stellar components are also best calibrated with {\sl Chandra}.  
We used {\sl CIAO}~4.5 and {\sl CALDB}~4.5.8 to reduce ACIS-S and ACIS-I data. All data were reprocessed from level 1 events, so that the latest, consistent calibrations were used.  
Only events with grades 0, 2, 3, 4, and 6 are included. We also removed bad pixels, bad columns, and node boundaries.
We filtered background flares with the light curve filtering script {\tt lc\_clean}.  The effective exposure times are shown in Table~1. 
Point sources were detected in a 0.3--7.0 keV image with {\tt wavdetect}, supplied with a 0.5 keV exposure map. The detection threshold was set to 10$^{-6}$ and the scales of {\tt wavdetect} 
ranged from 1 to 8, in steps increasing by a factor of
$\sqrt{2}$. Point sources, including galactic nuclei, were removed.

\subsection{Regions and Background}
We adopt the optical effective radii $r_e$ for the sample galaxies as listed by ATLAS$^{\rm 3D}$, which were taken from the Third Reference Catalogue of Bright Galaxies (RC3, de Vaucouleurs et al.\ 1991).
We chose an X-ray extraction aperture of $2\,r_e$ to determine 
the X-ray luminosity ($L_{X_{\rm gas}}$) and  mass ($M_{X_{\rm gas}}$) of the hot gas.
Local background, extracted from a region away from the source region on the same CCD chip, was used in the X-ray spectral analysis of most galaxies. 
The area of the local background was chosen to be at least twice the area of the source region to ensure a sufficient S/N ratio for background subtraction. In some cases the ISM emission fills the entire S3 chip. 
We adopt other methods to estimate the background emission for galaxies with effective radii larger than 0.9$^{\prime}$.
For galaxies also observed with the S1 chip, we used ``stowed background" data\footnote{particle background inside the detector observed with ACIS stowed. \url{http://cxc.harvard.edu/contrib/maxim/stowed}} for the spectral fitting. 
We fit a spectrum extracted from the S1 chip with a stowed background of the same region on the S1 chip to determine the surface brightness of cosmic X-ray and Galactic emission background, since the S1 chip is more offset and less contaminated by source emission. We then fit the spectrum extracted from the S3 chip with a corresponding stowed background by adding scaled X-ray background components obtained with the S1 chip to the fitting.  
If the S1 observation was not available, we used ACIS ``blank-sky" data\footnote{\url{http://cxc.harvard.edu/contrib/maxim/acisbg/}} for a background estimate. Since the amplitude of the instrumental background components vary with time, we scale the background count rate using
the 10--12 keV count rate ratio of the source and background data.

\subsection{Spectral analysis}

We grouped spectra to have at least one count per energy bin and adopted C-statistics for all our spectral analyses. Redistribution matrix files 
(RMFs) and ancillary response files 
(ARFs) were generated for each region using the {\tt specextract} tool.
Energy bands were restricted to 0.7--7.5 keV, 
where the responses are best calibrated. 
We performed spectral analysis with {\sl Xspec} 12.7.0. 
The model we adopted to fit the diffuse emission in each galaxy is
${\tt phabs}*({\tt apec}+{\tt powerlaw}+{\tt mekal}+{\tt powerlaw})$.   
The absorbing column density $N_{H}$, associated with the photoelectric absorption model component {\tt phabs}, was fixed at the Galactic value 
in the line of sight to each galaxy (Dickey \& Lockman 1990). 
The {\tt apec} component represents thermal emission from the hot gas.
The first {\tt powerlaw} component, with an index fixed at 1.6, represents the contribution from unresolved LMXBs (Irwin et al.\ 2003). 
In addition to hot gas and unresolved LMXBs, faint stellar X-ray sources such as cataclysmic variables (CVs) and coronally active binaries (ABs) also contribute to the X-ray flux.
Revnivtsev et al.\ (2007, 2008, 2009) calibrated the X-ray emission from such old stellar populations in several extremely gas-poor galaxies.
Revnivtsev et al.\ (2008) found
a nearly universal relation for the unresolved X-ray emissivity per $L_{K}$ in old stellar populations:
$L_X/L_K= 5.9\times10^{27}$\,erg\,s$^{-1}$ ${L_{K\odot}}^{-1}$,
where $L_X$ is in the 0.5--2.0 keV band.
We used this relation to derive X-ray estimates for
these stellar sources from the galaxies' $K$-band luminosities $L_K$ within $2\,r_e$.
The {\tt mekal}+{\tt powerlaw} components represent these CV/ABs sources, where the {\tt mekal} thermal emission temperature is fixed at 0.5 keV (and its abundance
is fixed at solar), and the {\tt powerlaw} index is fixed at 1.9 (Revnivtsev et al.\ 2008). 
The ratio of the fluxes of the {\tt mekal}  and  {\tt powerlaw} components was set to 2.03 (Revnivtsev et al.\ 2008). 
In our spectral analysis, we fixed the CV/ABs components at the estimated flux based on the $L_{K}$ of each galaxy. 
The redshift of each galaxy was taken from NED and fixed. We let the gaseous {\tt apec} model temperature, metallicity, and normalization free to vary. For those galaxies which did not have well-constrained metallicities, we fixed them at 0.7\,Z$_{\odot}$, the average of the hot gas metallicities of ETGs 
($kT\lesssim1$\,keV) studied by Su \& Irwin (2013).

 \subsection{Determination of $L_{X_{\rm gas}}$ and $M_{X_{\rm gas}}$}
The X-ray luminosities $L_{X_{\rm gas}}$ estimated in this paper are
hot gas luminosities in the $0.1-2.0$ keV energy range. X-ray emission from CV/ABs and unresolved LMXBs are explicitly excluded, having been removed spectrally as outlined above. We list the best-fit $L_{X_{\rm gas}}$ and gas temperature $T_X$ for each galaxy in Table~3.
Assuming a spherical distribution for the hot gas, we obtained the volume of the extraction region for each galaxy. 
We derived the average hot gas density from
the best fit normalization of the {\tt apec} thermal emission model, which 
is defined as
$$ 
{\rm norm} = \frac{10^{-14}}{4\pi[D_A(1+z)]^2}\int{n_en_H}{dV}, 
$$
where 
$D_A$ is the angular distance to the galaxy, and $V$ is the volume of the 
 source region.
With the hot gas density and volume, we obtained the hot gas mass $M_{X_{\rm gas}}$ for each galaxy in the sample, as listed in Table~3.

\subsection{2MASS photometry}
We used $K$-band luminosities to characterize the stellar masses of the sample galaxies,
since the $K$-band accurately represents
the largely old stellar populations of ETGs. 
These galaxies' $K$-band luminosities were derived from {\sl Two Micron All Sky Survey} ({\sl 2MASS}) (Skrutskie et al.\ 2006) archived images. The regions used for $K$-band photometry  were the same as those used in the X-ray analyses. 
Bright nuclear and foreground sources (detected by eye) were excluded and refilled with a local surface brightness component using the {\tt dmfilth} tool in {\sl CIAO}~4.5.   
Source counts were obtained by subtracting the local background component. 
We converted source counts to the corresponding magnitude and corrected for Galactic extinction. The $K$-band solar luminosity was assumed to be $L_{K\odot}=5.67\times10^{31}$ ergs s$^{-1}$ (Mannucci et al.\ 2005). The sample galaxies' $K$-band luminosities $L_{K}$   are listed in Table~2.

\subsection{Galaxy environment}

The Sloan Digital Sky Survey (SDSS) (York et al.\ 2000) was used to assess the galaxy density around
each galaxy in the sample. Using the main spectroscopic sample from Data Release 7 (DR7, Abazajian et al.\ 2009), we count neighboring galaxies within a projected radius of 500 kpc and within a redshift difference of $\lvert z \rvert<$ 1500 km\,s$^{-1}$. 
Spectroscopic redshifts 
were used, since they are much more accurate than photometric redshifts. 
To compensate for the differences in the detection limit of galaxies at different distances, we included all galaxies brighter than $m_r=17.8$ for the most distant galaxies and used a correspondingly brighter magnitude threshold for more nearby galaxies. The number of galaxies in the neighborhood of each sample galaxy is listed in Table~3. 
Galaxies with the most numerous neighbors tend to reside in the Virgo Cluster. 
Four out of 42 galaxies in our sample are not well covered by the SDSS and are described in \S2.  

 \begin{figure} 
\epsscale{1.2}
\plotone{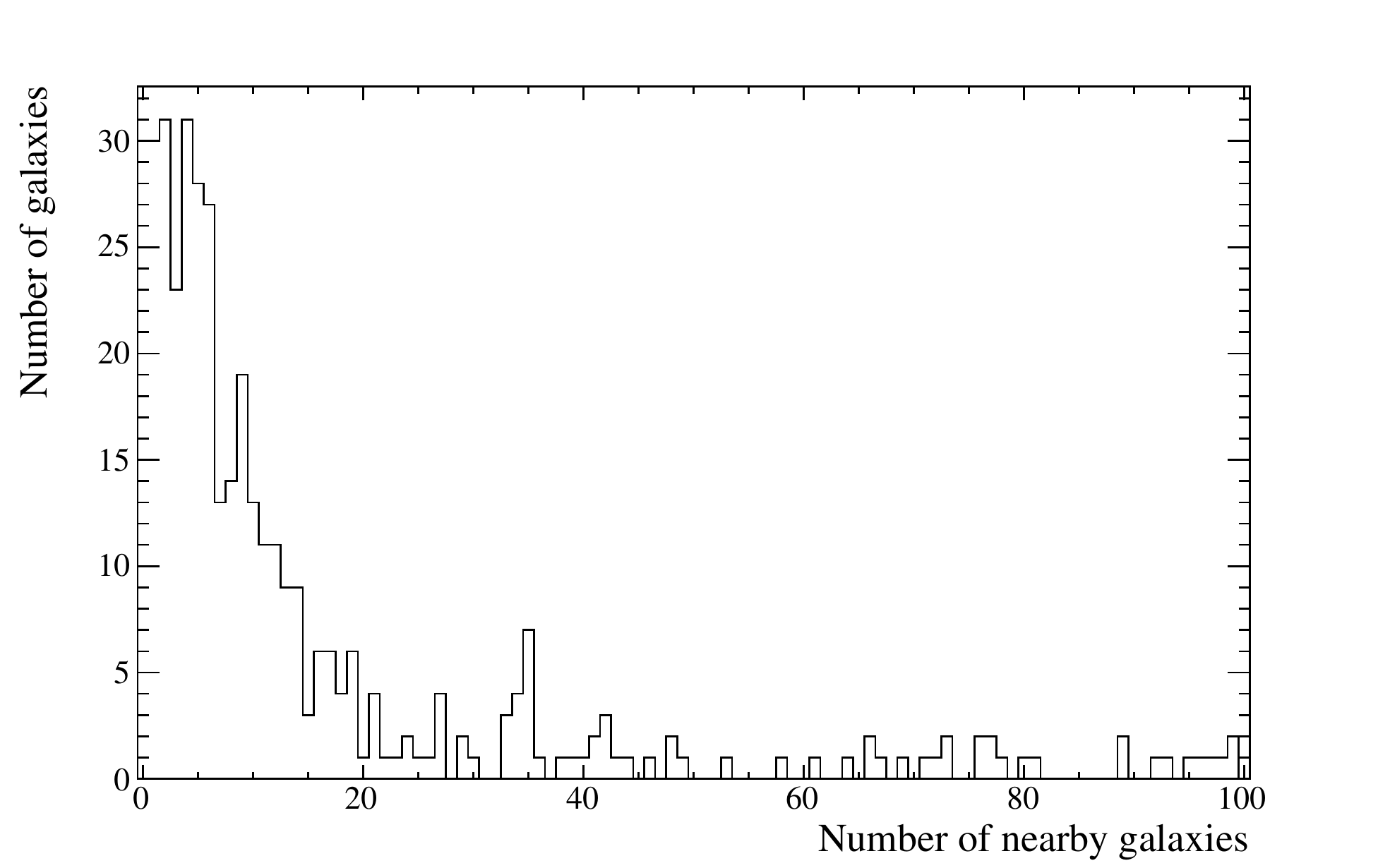}
\figcaption{\label{fig:hist} Histogram of the number of neighbor galaxies within 500 kpc of each galaxy based on RC3 for 430 galaxies taken from O'Sullivan et al.\ (2001). 
We take 15 nearby neighbor galaxies as a natural cut between galaxies in dense environments and galaxies in relatively isolated environment.}
\end{figure}

\begin{figure} 
\epsscale{1.2}
\plotone{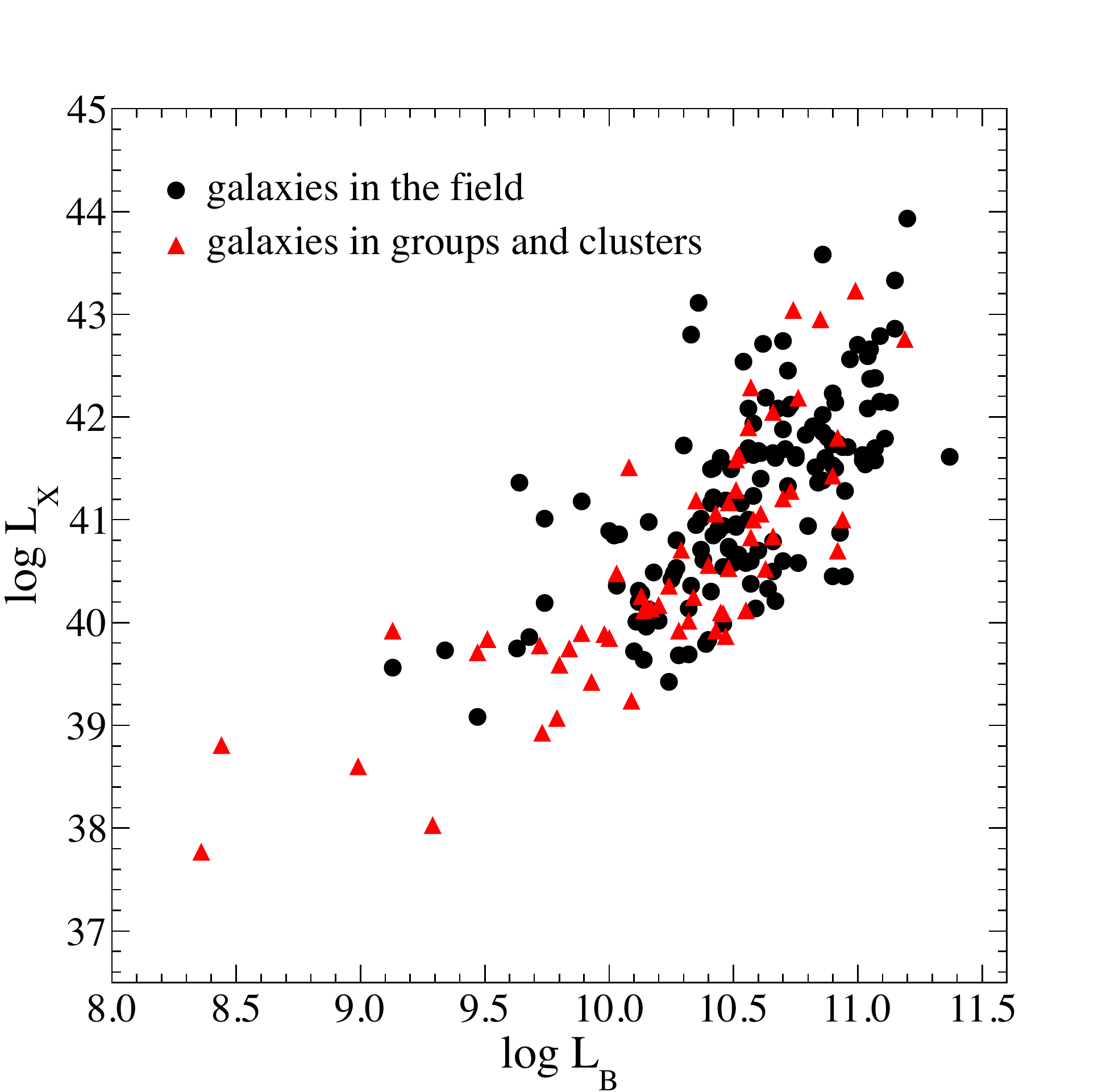}
\figcaption{\label{fig:osullivan} The total X-ray luminosity $L_X$ (including LMXBs and stellar sources) as a function of $L_B$ for 209 ETGs in isolated and dense environments taken from the {\sl ROSAT} survey of O'Sullivan et al.\ (2001). [{\sl see the electronic edition of the journal for a color version of this figure.}]}
\end{figure}

We supplemented this environmental study of 42 galaxies possessing sufficiently deep, 
high resolution {\sl Chandra} X-ray data
by investigating a much larger sample of galaxies
observed with the (lower resolution) {\sl ROSAT} X-ray telescope. 
The O'Sullivan et al.\ (2001) {\sl ROSAT} galaxy survey includes 430 ETGs,
209 of which were detected.
The histogram in Figure~\ref{fig:hist} shows the distribution of the number of neighboring galaxies
around   each of the 430 ETGs in the O'Sullivan et al.\ (2001) sample.
The galaxy sample was taken from NED\footnote{http://ned.ipac.caltech.edu/} and neighbors
were defined to be those galaxies within a 500 kpc projected radius of each {\sl ROSAT} galaxy.
Despite the statistical unevenness of both the NED and {\sl ROSAT} samples,
there seems to be a natural division between high-density and low-density environments 
at the level of $\sim15$ neighbors.
Of the 209 ETGs detected in the O'Sullivan et al.\ (2001) {\sl ROSAT} survey, 
146 are in dense environments, while 63 are in low-density environments.
Figure~\ref{fig:osullivan} plots the X-ray and optical (blue) luminosities
of the 209 {\sl ROSAT} detections; the galaxies are labelled by environment,
with 15 neighbors dividing high-density from low-density environments.
The high- and low-density samples are almost completely overlapping.

\section{\bf Results}

\subsection{The scatter in the $L_{X_{\rm gas}}$--$L_K$ relation}

\begin{figure*}
   \begin{center}
     \leavevmode 
         \epsfxsize=15cm\epsfbox{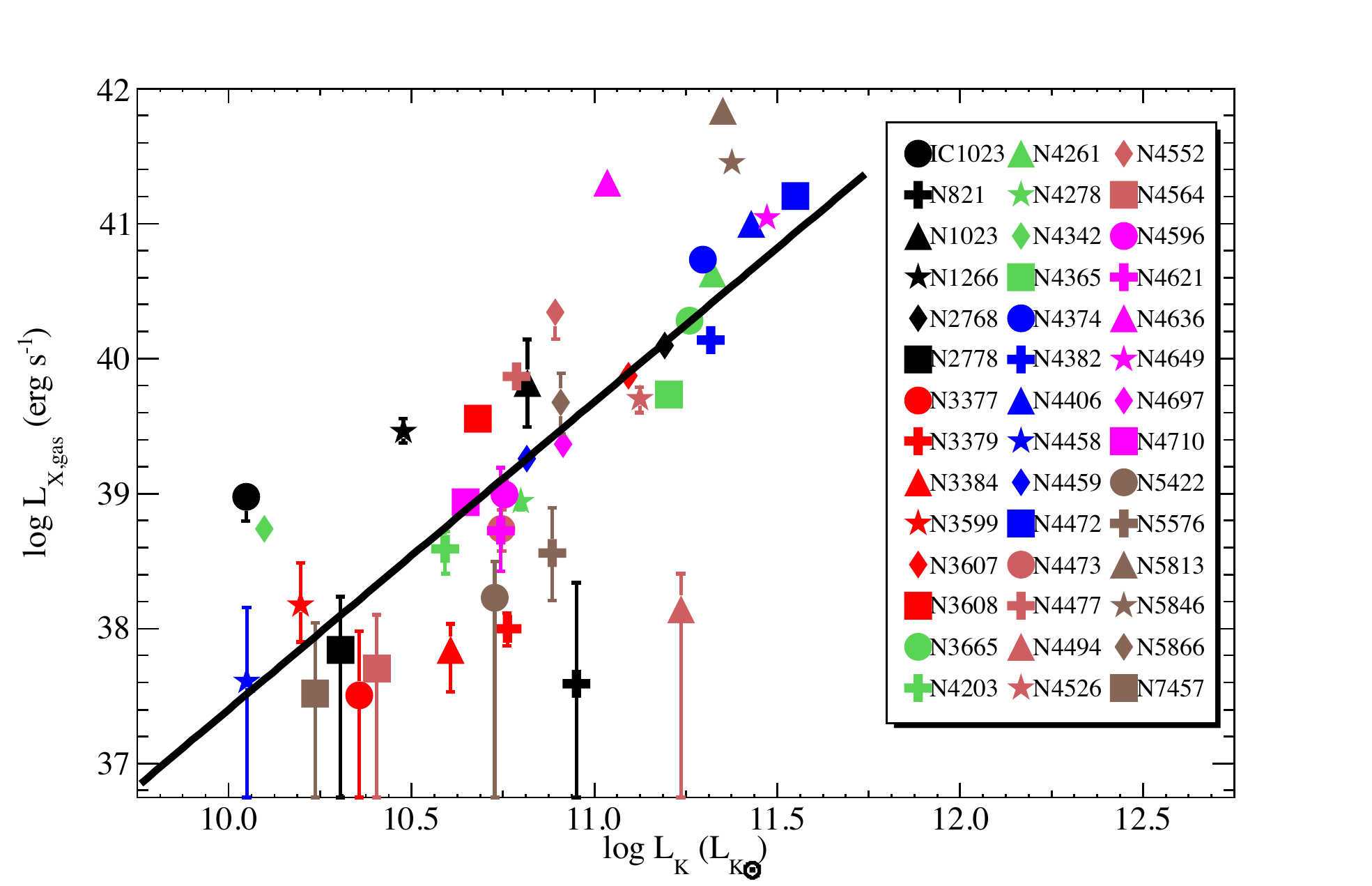}
\caption{Gaseous X-ray luminosities as a function of $K$-band luminosities for all galaxies in this work. 
The black solid line is the best-fit $L_{X_{\rm gas}}$--$L_K$ relation 
( ${\rm log}(L_{X_{\rm gas}})=2.3{\rm log}(L_K)+14.6$). }
\label{fig:lxk} 
\end{center}
\end{figure*}

The gaseous X-ray luminosities $L_{X_{\rm gas}}$ of our 
sample galaxies are plotted 
against the galaxies' stellar $K$-band luminosities $L_{K}$ in Figure~\ref{fig:lxk};
$L_{X_{\rm gas}}$ ranges from $\sim3\times10^{37}$ to $7\times10^{41}$ ergs s$^{-1}$, 
while $L_{K}$ ranges from $\sim10^{10}$ to $3\times10^{11}~L_{K\odot}$ (see Table~3). 
The scatter in the $L_{X_{\rm gas}}$--$L_K$ correlation ranges up to a factor of 1000, demonstrating the diversity of galaxies in our sample. 
We fit the $L_{X_{\rm gas}}$--$L_K$ relation to a single power law,
${\rm log}(L_{X_{\rm gas}})=A~{\rm log}(L_K)+B$, and obtained best-fit values for the slope of $A=2.3\pm 0.3$ and the intercept of $B=14.6\pm 3.3$, as indicated by the  solid black line in Figure~\ref{fig:lxk}. 
The slope we obtained is slightly shallower than (but within the errors of) that obtained 
for a {\sl Chandra} sample of 30 ETGs
by Boroson et al.\ (2011), who found a best-fit slope of $2.6\pm0.4$.
Our sample contains relatively more  faint galaxies, which may have
a shallower $L_{X_{\rm gas}}$--$L_K$ slope than the sample as a whole.
Moreover, our study uses the 0.1--2 keV energy band, 
rather than the 0.5--2 keV band used by 
Boroson et al.\ (2011). 
This would flatten our slope relative to the 0.5--2 keV slope,
since X-ray fainter galaxies tend to have lower temperatures, 
thus have a higher fraction of their total gaseous emission below 0.5 keV 
than do more X-ray luminous galaxies.
To test this, we re-performed our analysis using the 0.5--2.0 keV band and obtained a slope of $2.4\pm0.3$, in better agreement with the Boroson et al.\ (2011) result.
Another major difference between our work and that of Boroson et al.\ (2011) is that we use metric
radii (two optical effective radii) for the extraction regions in both the X-ray and $K$-band analyses, while Boroson et al.\ (2011) use background-limited photometric radii (radii where the diffuse emission reaches the background level) in their X-ray analysis and their $L_K$ values were taken from NED. This could also cause differences in our results; nevertheless, our approach is more self-consistent. 
In the Appendix, we further examine the impact of variations in the best-fit $L_{X_{\rm gas}}$--$L_K$ relation (including using a best-fit broken power law) upon our results.

\begin{figure} 
\epsscale{1.3}
\plotone{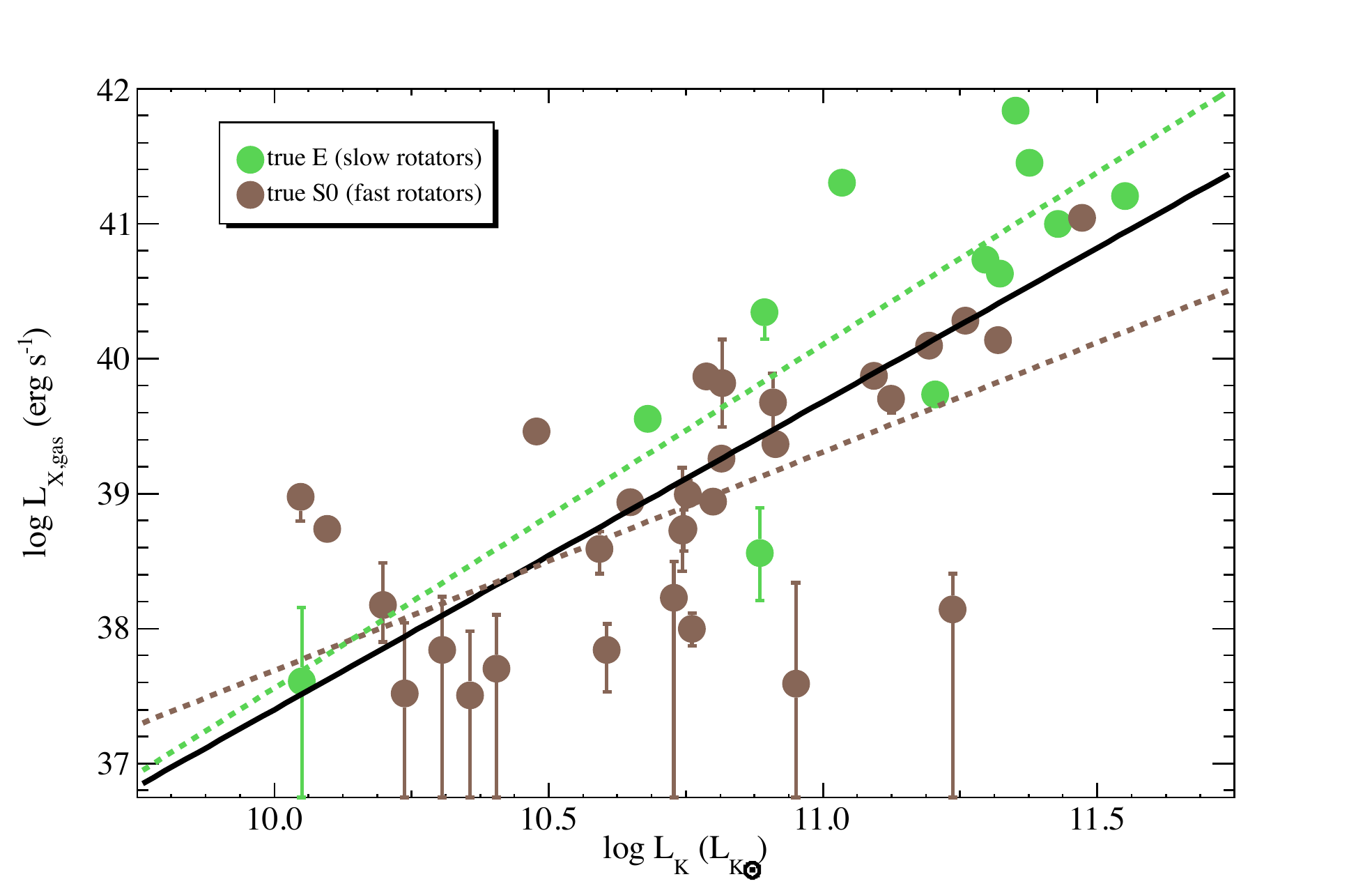}
\plotone{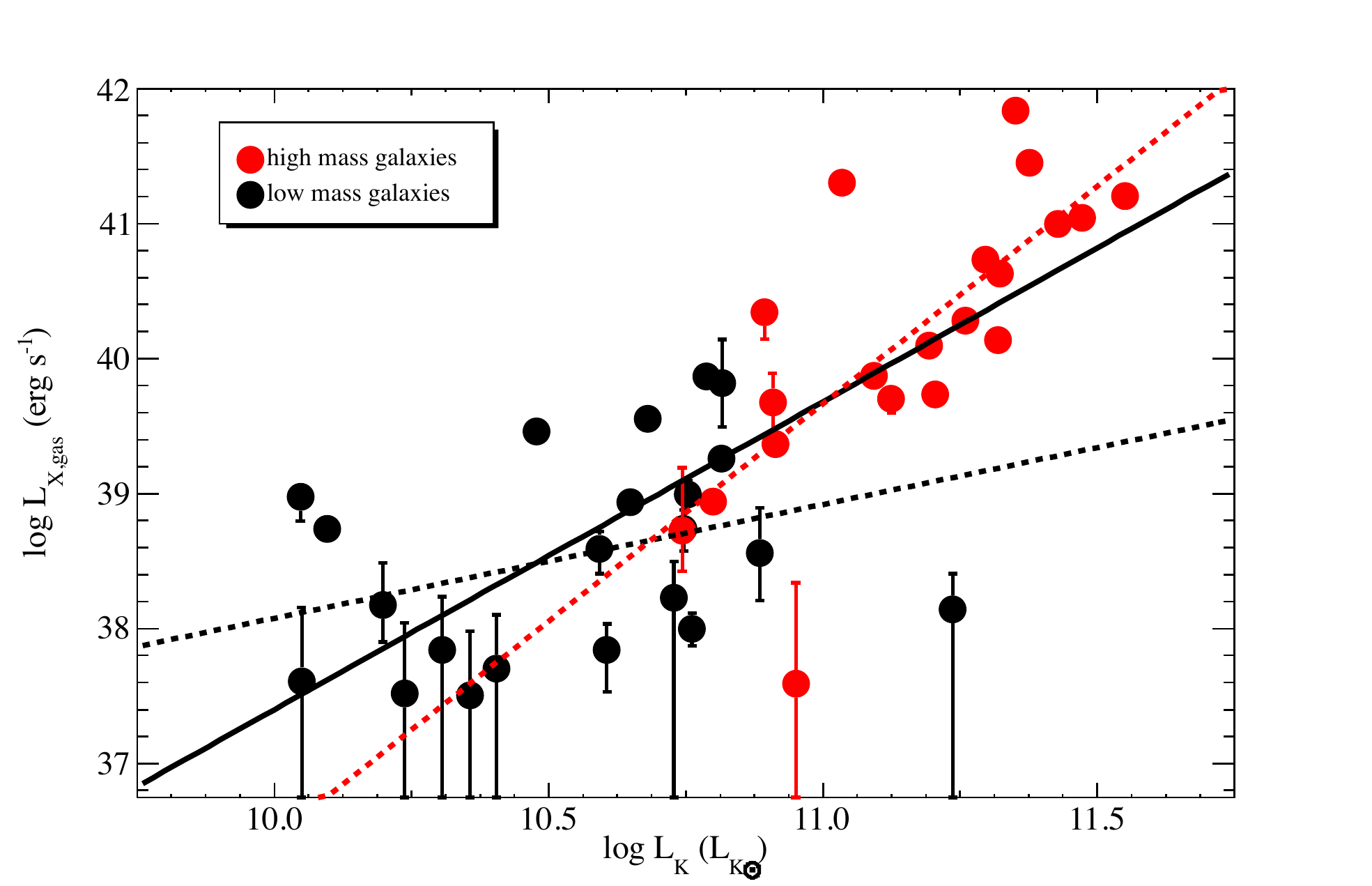}
\plotone{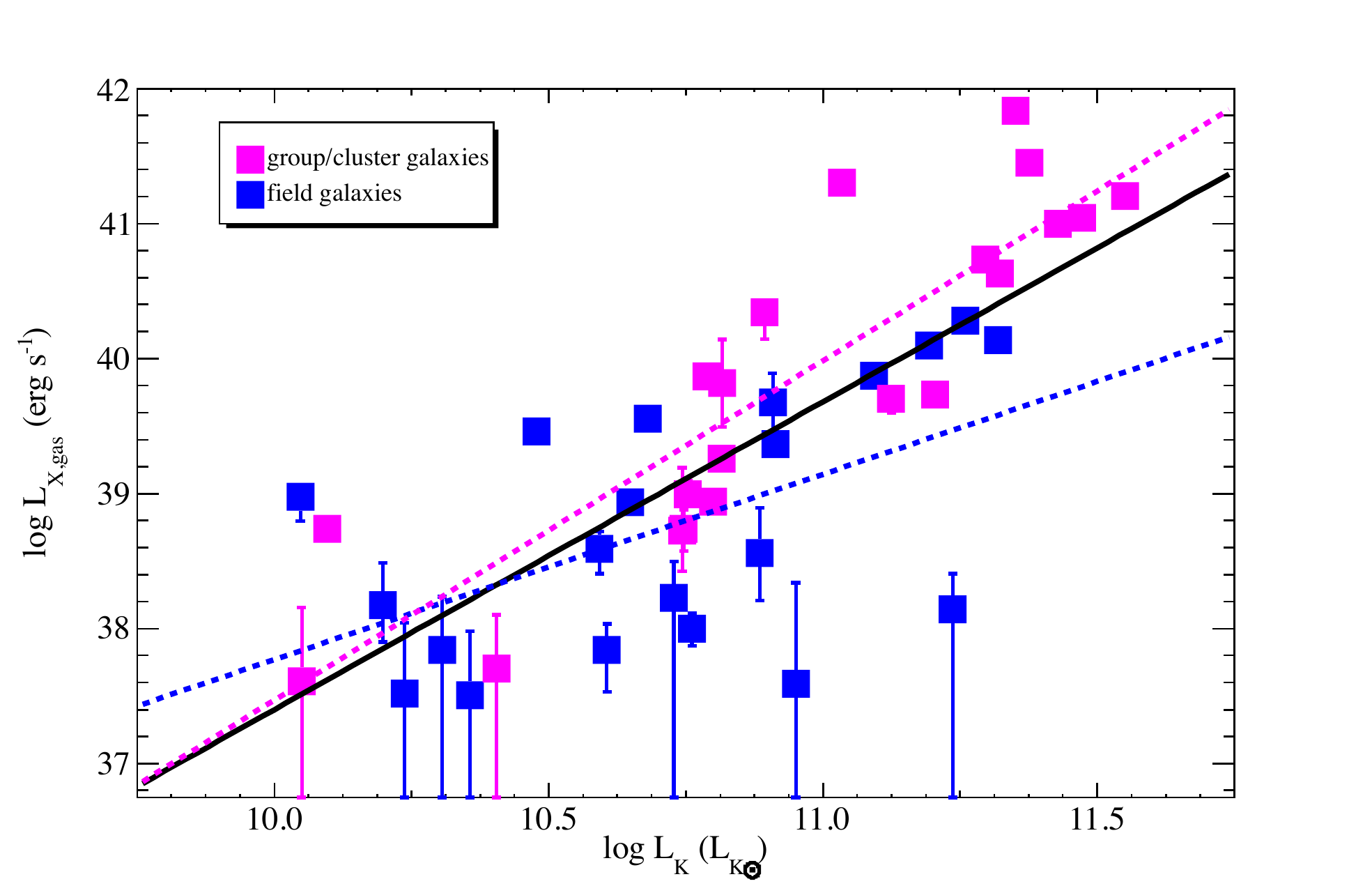}
\figcaption{\label{fig:lxks}  
Gaseous X-ray luminosities as a function of $K$-band luminosities for all galaxies in this work. 
The black solid line is the best-fit $L_{X_{\rm gas}}$--$L_K$ relation 
( ${\rm log}(L_{X_{\rm gas}})=2.3{\rm log}(L_K)+14.6$) for all sample galaxies. Dashed lines indicate the best-fit of each sub-group. {\it top}: ``true lenticular" galaxies (brown) and ``true elliptical" galaxies (green). {\it middle}: high mass galaxies (red) and low mass galaxies (black). {\it bottom}: field galaxies (blue) and galaxies in groups and clusters (magenta).}
\end{figure}

In Figure~\ref{fig:lxks} we show the $L_{X_{\rm gas}}$--$L_K$ relation for each 
of the three sets of sub-groupings (E/S0, mass, environment) described in \S2. 
We find that massive galaxies, ``true elliptical" galaxies, and galaxies in groups and clusters generally belong to a single population (denoted population A), while low-mass galaxies, ``true lenticular" galaxies, and field galaxies generally belong to a separate population (denoted population B).

\begin{figure*} 
   \begin{center}
\epsscale{1.11}
    \leavevmode
\plottwo{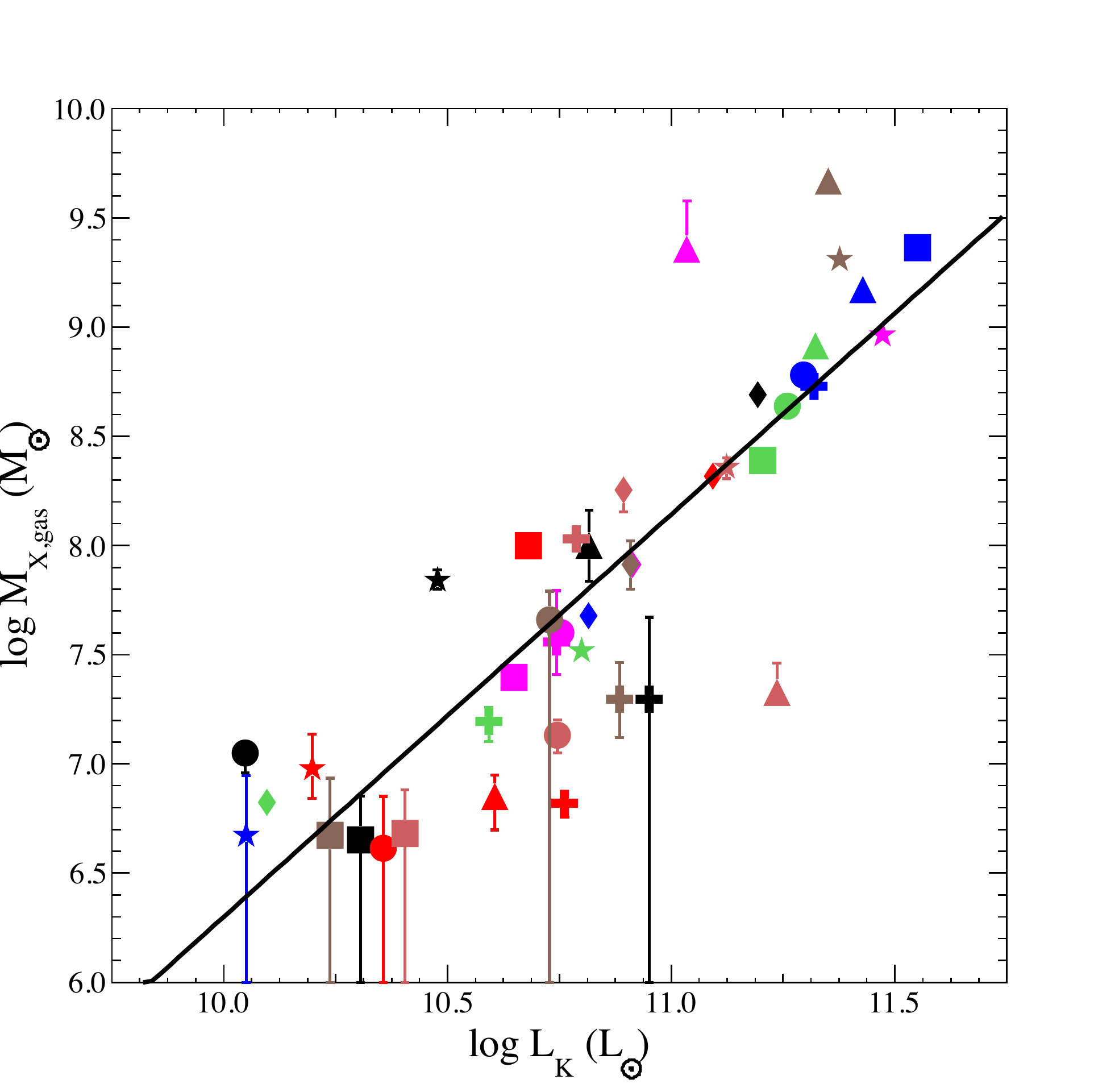}{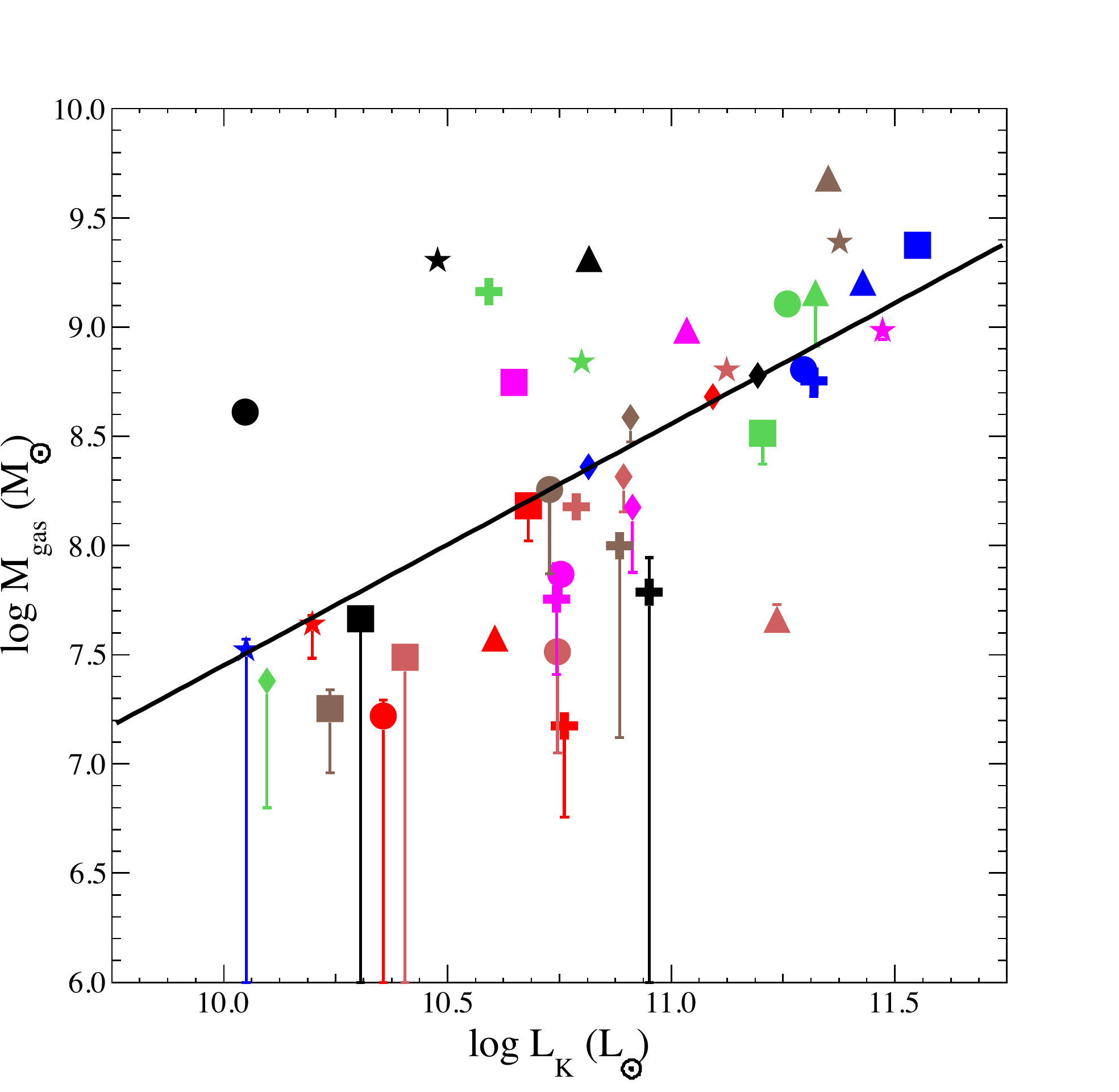}
\caption{{\it left}: hot gas mass as a function of $K$-band luminosities for all galaxies in this work. {\it right}: total gas mass (the sum of hot gas and cold gas) as a function of $K$-band luminosities for all galaxies in this work. The scatter of the total mass -- $L_K$ relation is larger than that of hot gas mass --  $L_K$ relation.
Color code is the same as in Figure~\ref{fig:lxk}.} 
  \label{fig:mxk} 
  \end{center}
\end{figure*}

If ETGs retained all their accumulated stellar-mass loss, the hot gas mass $M_{X_{\rm gas}}$ 
should in principle be linearly related to $L_K$.
Figure~\ref{fig:mxk} shows $M_{X_{\rm gas}}$ as a function of $L_{K}$ for our sample. 
The scatter of the $M_{X_{\rm gas}}$--$L_K$ relation is smaller than the $L_{X_{\rm gas}}$--$L_K$ relation (Figure~\ref{fig:lxk}). 
This is manifested in the smaller uncertainties in its best-fit power fit:
a best-fit slope of $A=1.84\pm 0.17$ and intercept $B=-12.1\pm 1.8$, as indicated by the black solid line in Figure~\ref{fig:mxk}-{\it left}. This result, although steeper than linear, slightly reduces the discrepancy we had between the hot gas content and the optical light of ETGs. 
We further compared the total gas mass $M_{\rm gas}$ 
(hot gas plus cold gas: $M_{\rm gas}=M_{X_{\rm gas}}+M_{\rm HI}+M_{\rm H_2}$) 
as a function of $L_K$ for our sample as shown in Figure~\ref{fig:mxk}-{\it right}. 
We obtained a best-fit slope of $A=1.11\pm 0.21$ and intercept $B=-3.61\pm 2.20$.
The scatter of this $M_{\rm gas}-L_K$ relation is larger than in the relation between the hot gas mass and 
$L_K$.
This suggests that the cold gas content is unlikely to be from
accumulated stellar-mass loss or cooling from hot gas.  However, we note that the total gas mass $M_{\rm gas}$  increases almost linearly with $L_K$, which is exactly what we expect if stellar-mass loss is 
the primary source of all gaseous components.

In subsequent sections (\S4.2--\S4.8), we relate 
the $L_{X_{\rm gas}}-L_K$ relation
to various internal and external factors: galaxy age,  cold gas masses ($M_{\rm HI}$, $M_{\rm H_2}$,  $M_{\rm HI}+M_{\rm H_2}$), 
star formation rate (SFR),  total galaxy mass ($M_{\rm tot}$, $r_e$, $M_{\rm tot}/r_e$), 
hot gas temperature ($T_X$), stellar velocity dispersion ($\sigma$), 
a rotational parameter ($\lambda$), galaxy eccentricity ($\epsilon$), their combination ($\lambda/\sqrt{\epsilon}$),
and the local  density of galaxies in the neighborhood of each galaxy.
We also investigate the deviations in $L_{X_{\rm gas}}$ from the best-fit $L_{X_{\rm gas}}-L_K$  relation 
and assess whether these deviations [defined as $\Delta L_{X_{\rm gas}}=
{\rm log}(L_{X_{\rm gas}})-2.3\,{\rm log}(L_K)-14.6$]
are correlated with the internal and external factors listed above.
We use the Spearman correlation in the Penn State statistical package ASURV\footnote{http://www2.astro.psu.edu/statcodes/asurv} 
 to quantify the correlations of these relations, which allows us to properly treat any upper limits. The correlation coefficient (with uncertainties) and its associated null hypothesis probability 
for any relation between $L_{X_{\rm gas}}/L_K$ and each factor are listed in Table~4. The results for the relation between the residuals $\Delta L_{X_{\rm gas}}$  and each factor are listed in Table~5. 
We also repeated all these analyses using $M_{X_{\rm gas}}/L_K$ and analogous deviations $\Delta M_{X_{\rm gas}}$, instead of $L_{X_{\rm gas}}/L_K$ and $\Delta L_{X_{\rm gas}}$; we obtained very similar results, so they are not presented here.

\subsection{Stellar age}

$L_{X_{\rm gas}}/L_K$ and $\Delta L_{X_{\rm gas}}$ are plotted against the stellar ages 
of our sample galaxies in 
Figures~A1-{\it top-left} and \ref{fig:age}, respectively. 
These figures show that galaxies with older stellar populations
tend to have relatively larger X-ray halos. Older galaxies have a longer time to accumulate hot gaseous halos.
This trend was suggested by Boroson et al.\ (2011) for nearby ETGs. 
In contrast, Civano et al.\ (2014) found in the {\sl Chandra} COSMOS survey that younger galaxies 
tend to have larger $L_{X_{\rm gas}}/L_K$. 
This may be related to younger galaxies experiencing recent major mergers.
Note that galaxies in the {\sl Chandra} COSMOS survey extend out to $z=1.5$, 
so may have different evolutionary histories compared with nearby ETGs. 
At the same time, the galaxy ages listed in Table~3 are taken from heterogenous sources. 
The lack of consistency between these analyses and the fact that many galaxies show 
radial age gradients may further complicate this study. 
We also note that the correlation between age and $\Delta L_{X_{\rm gas}}$ seems to only exist among population A galaxies.

\begin{figure} 
\epsscale{1.2}
\plotone{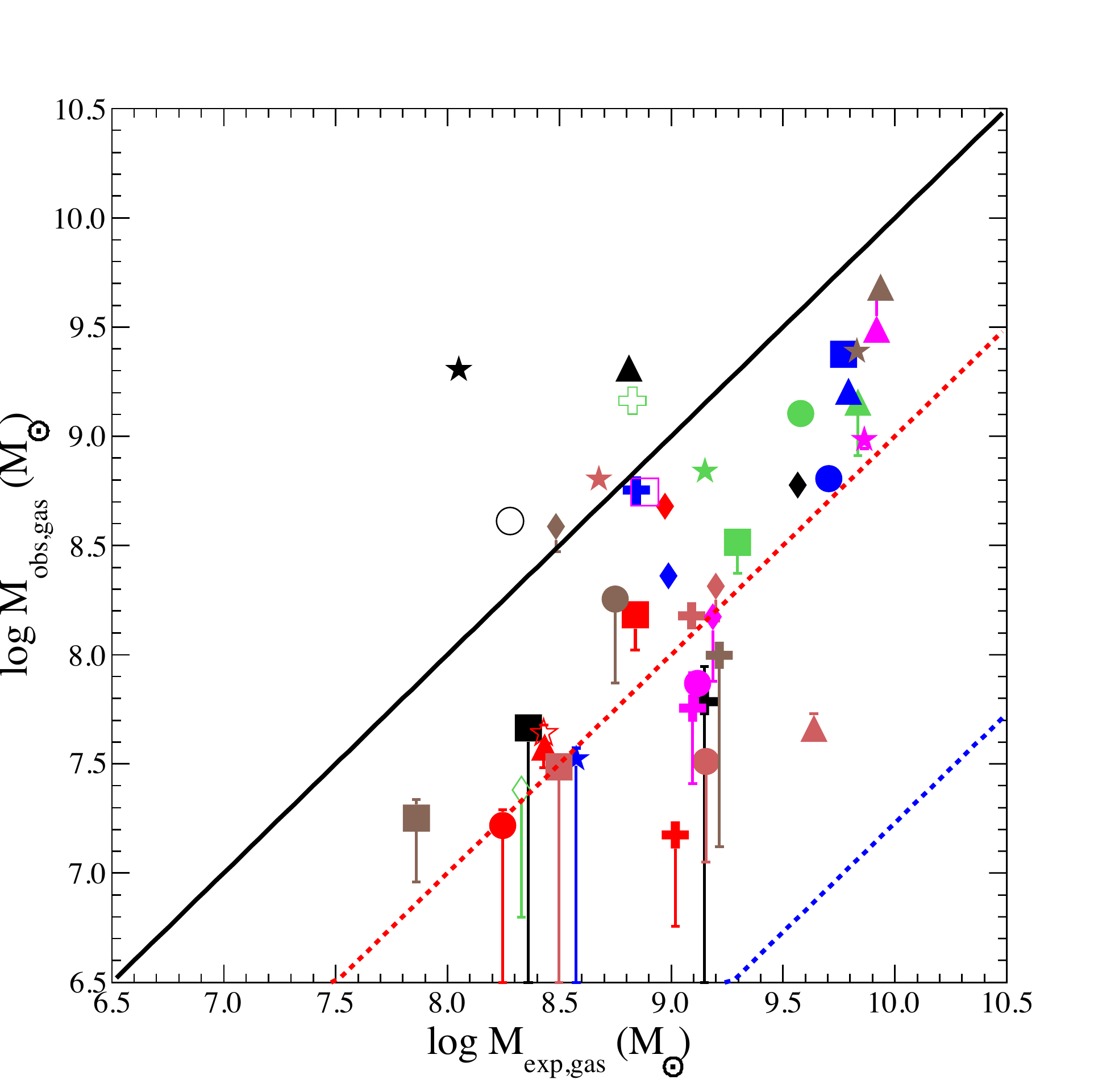}
\figcaption{\label{fig:mass} Observed gas mass (the sum of cold and hot gas mass) as a function of expected gas mass (stellar age times stellar-mass loss rate). 
We use the average age (8.15 Gyr) for galaxies that do not have stellar age available in the literature and are marked by open symbols. 
Color code is the same as in Figure~\ref{fig:lxk}. The solid black line indicates the equality between these two masses. For most galaxies, their observed gas mass is much less than expected. 
Dashed red line: equality between these two masses assuming gas distributes out to 10$r_{\rm e}$ ($\sim$ 45 kpc). Dashed blue line: equality between these two masses assuming gas distributes out to $r_{\rm vir}$ ($\sim$ 300 kpc). 
[{\sl see the electronic edition of the journal for a color version of this figure.}]}
\end{figure}

We estimated the stellar-mass loss rate for each galaxy by using the current value determined 
for elliptical galaxies by Knapp et al.\ (1992):
$\sim$0.0021 $L_K/L_{K\odot}~M_{\odot}$ Gyr$^{-1}$.
This can be regarded as a lower limit to the average stellar-mass loss rate over cosmic time.
We calculated the expected gas mass for each galaxy by multiplying its stellar-mass loss rate by its stellar age.  
In Figure~\ref{fig:mass} we compare the observed total gas mass (hot gas plus cold gas) of each galaxy to the 
expected gas mass (the solid black line indicates equality). 
For most galaxies, the observed gas mass is much less than that expected from accumulated stellar-mass loss. 
This discrepancy suggests that gas has been driven out of these galaxies.   
Alternatively, this may indicate that gas has been distributed to larger radii than 
our extraction apertures and is more extended than the stellar light distributions. 
Humphrey et al.\ (2011) studied the isolated elliptical galaxy NGC~720 in great detail using deep {\sl Chandra} and {\sl Suzaku} observations, the combination of which provides both high spatial resolution and low instrumental background. 
They detected hot ISM emission to beyond 50 kpc and extrapolate its hot gas properties out to the virial radius $r_{\rm vir}$ ($\sim$ 300 kpc). 
Based on its measured and extrapolated gas mass profiles, its enclosed gas mass increases by 10 times (600 times) out to 45 kpc (300 kpc) compared to its gas mass within $2\,r_{\rm e}$ ($\sim$10 kpc).
If we scale the gas mass of galaxies in our sample out to 60 kpc and 300 kpc, based on the gas mass profile of NGC~720, the equality between the observed total mass and the expected gas mass would shift to the red and blue dashed lines, respectively, in Figure~\ref{fig:mass}. 
In principle, we may be seeing the inner parts of the history of accumulated stellar-mass loss,
while the bulk has been pushed beyond the optical extent of the galaxies.

\subsection{Cold gas content and star formation}

\begin{figure*} 
   \begin{center}
    \leavevmode
\epsscale{1.11}
\plottwo{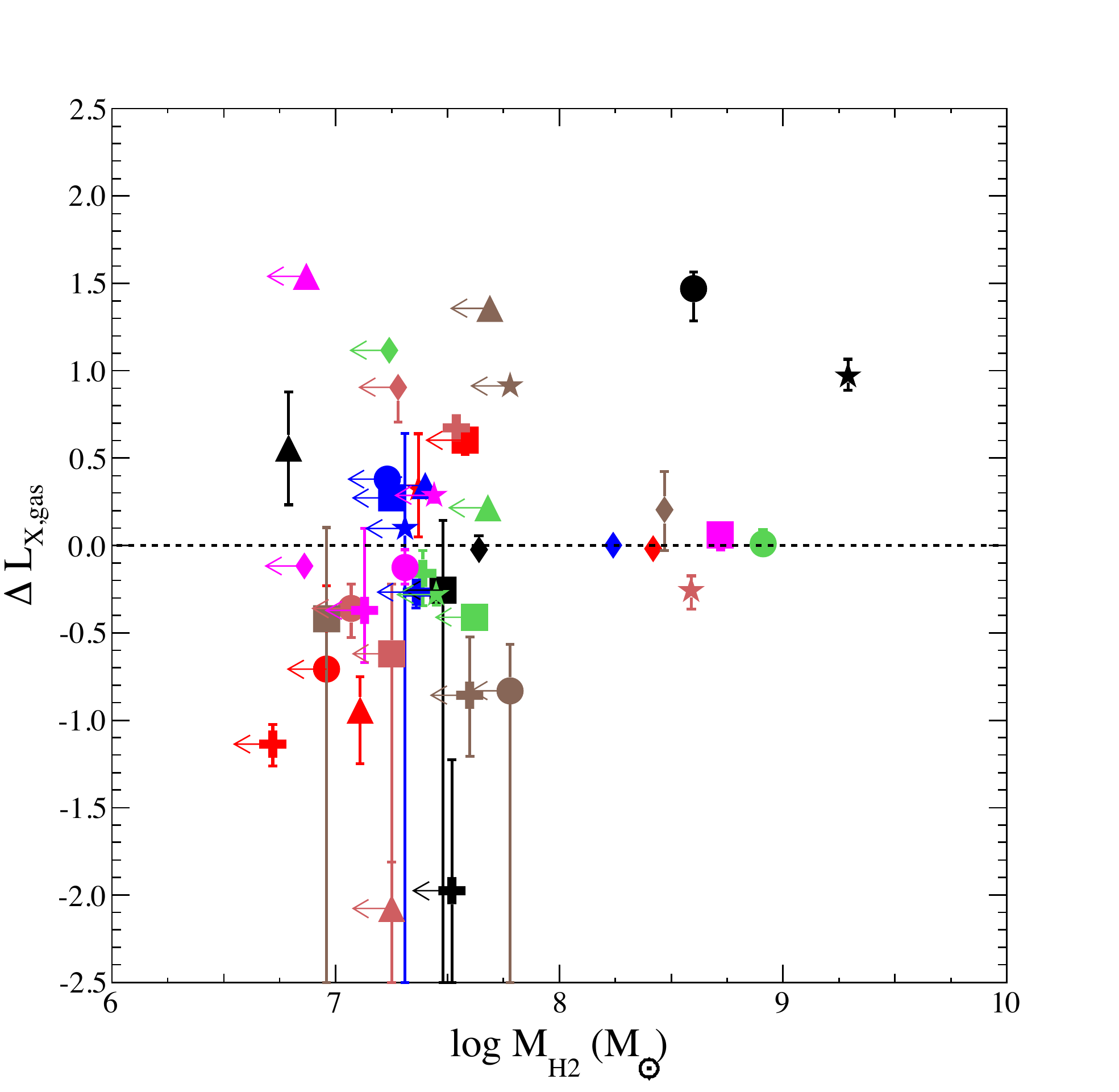}{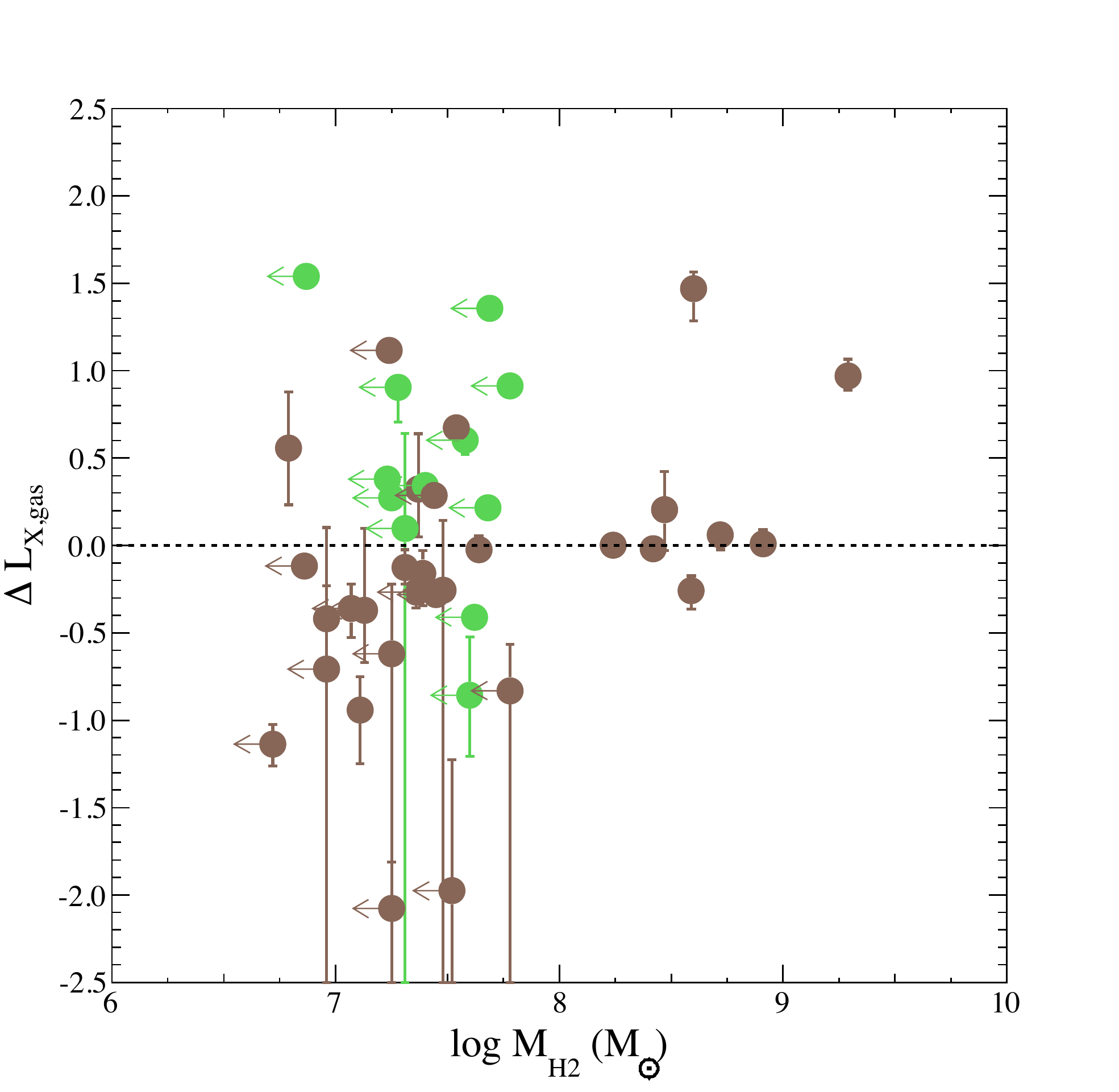}
\plottwo{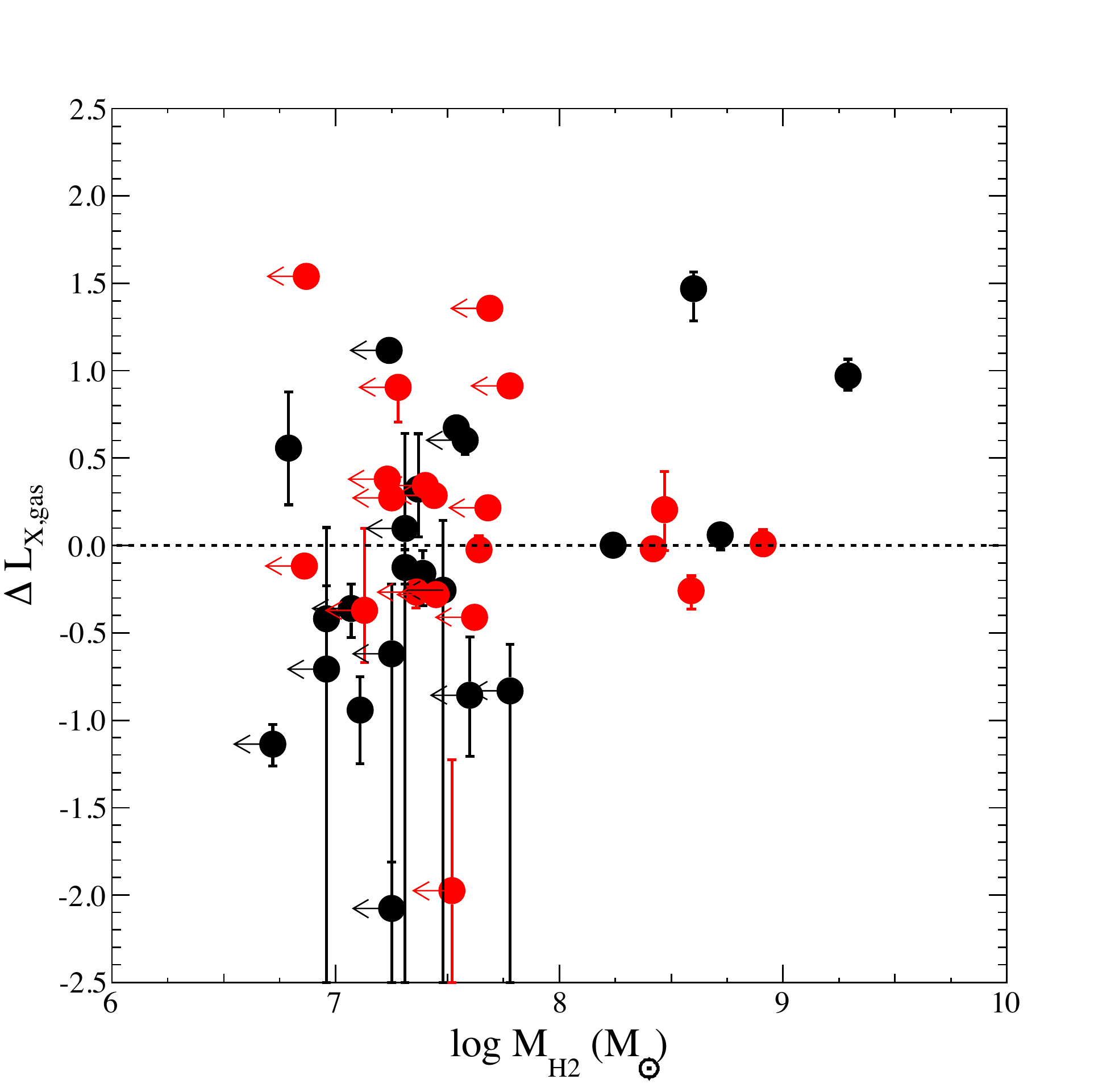}{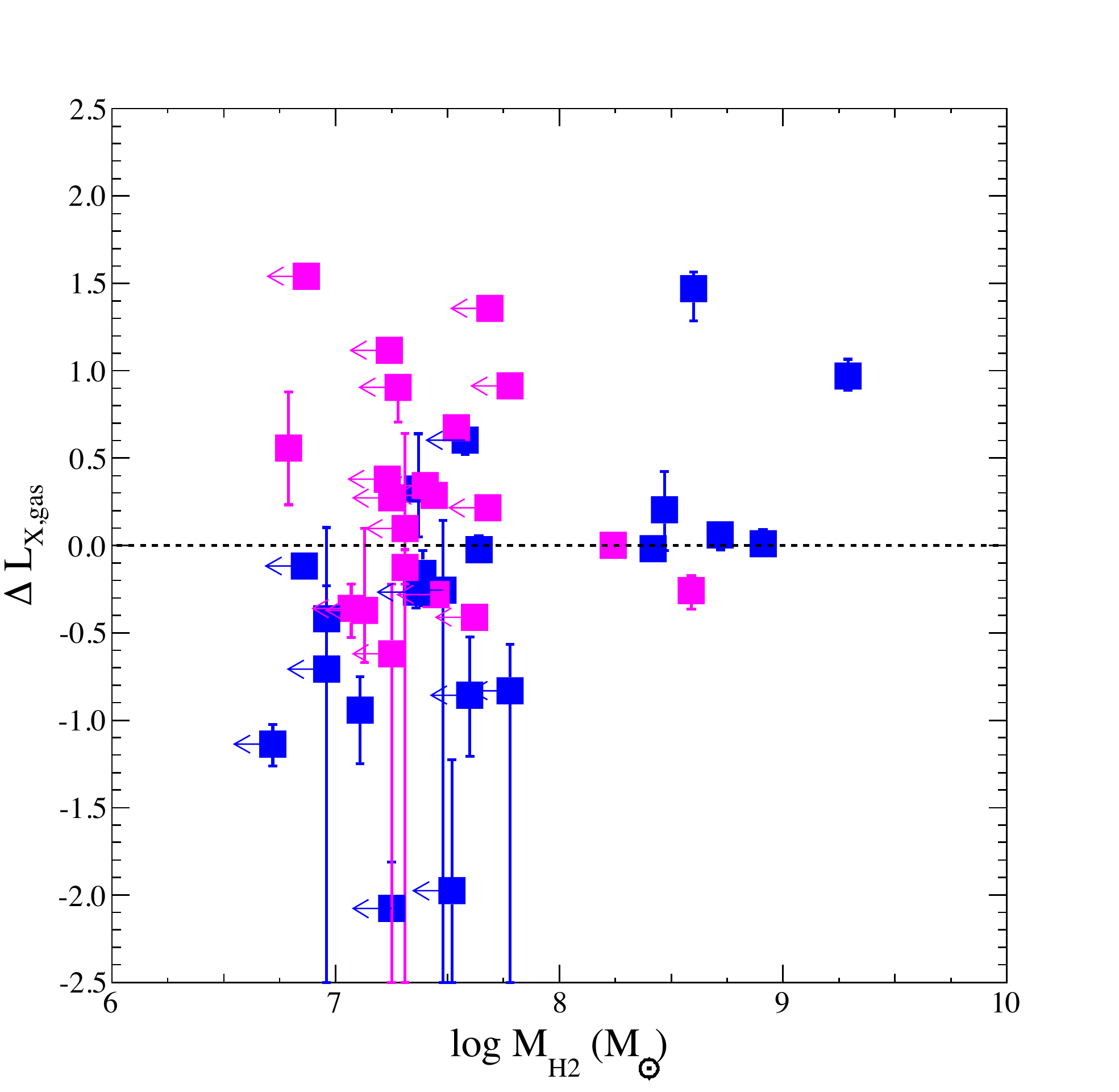}
\caption{$\Delta L_{X_{\rm gas}}$ as a function of molecular gas mass.  {\it top-left}: color code is the same as in Figure~\ref{fig:lxk}. {\it top-right}: ``true lenticular" galaxies (brown) and ``true elliptical" galaxies (green). {\it bottom-left}: high mass galaxies (red) and low mass galaxies (black). {\it bottom-right}: field galaxies (blue) and galaxies in groups and clusters (magenta).}
 \label{fig:h2} 
 \end{center}
\end{figure*}

More than 20\% of the ETGs in the ATLAS$^{\rm 3D}$ multiwavelength survey are observed to contain 
atomic and molecular gas in significant amounts (at least $10^7$--$10^8~M_{\odot}$).  
Su \& Irwin (2013) proposed that the hot ISM of ETGs may have accreted cold gas that was subsequently heated to an X-ray emitting phase. 
We thus expect that the hot gas and cold gas content of ETGs may be  related.
We investigated the relation between the hot gas content in our sample galaxies and their cold gas masses: $M_{\rm HI}$, $M_{\rm H_2}$, and $M_{\rm HI}+M_{\rm H_2}$. 
We found that 
the deviations $\Delta L_{X_{\rm gas}}$ generally increase with the cold gas masses in these galaxies, as shown in Figures~\ref{fig:hi} to \ref{fig:cold}; in particular, the hot gas 
in field galaxies is correlated with their molecular gas masses, as shown in Figure~\ref{fig:h2} ($\rho=0.714\pm0.117$), while there is almost no correlation for population A galaxies. 
Figure~\ref{fig:h2}-{\it top-right} shows that H$_2$ is detected only in ``true lenticular" galaxies
in our sample.
The effect of cold gas on galaxies in groups and clusters may be overshadowed by more complicated interactions 
with the ICM.

\begin{figure*} 
   \begin{center}
    \leavevmode
\epsscale{1.11}
\plottwo{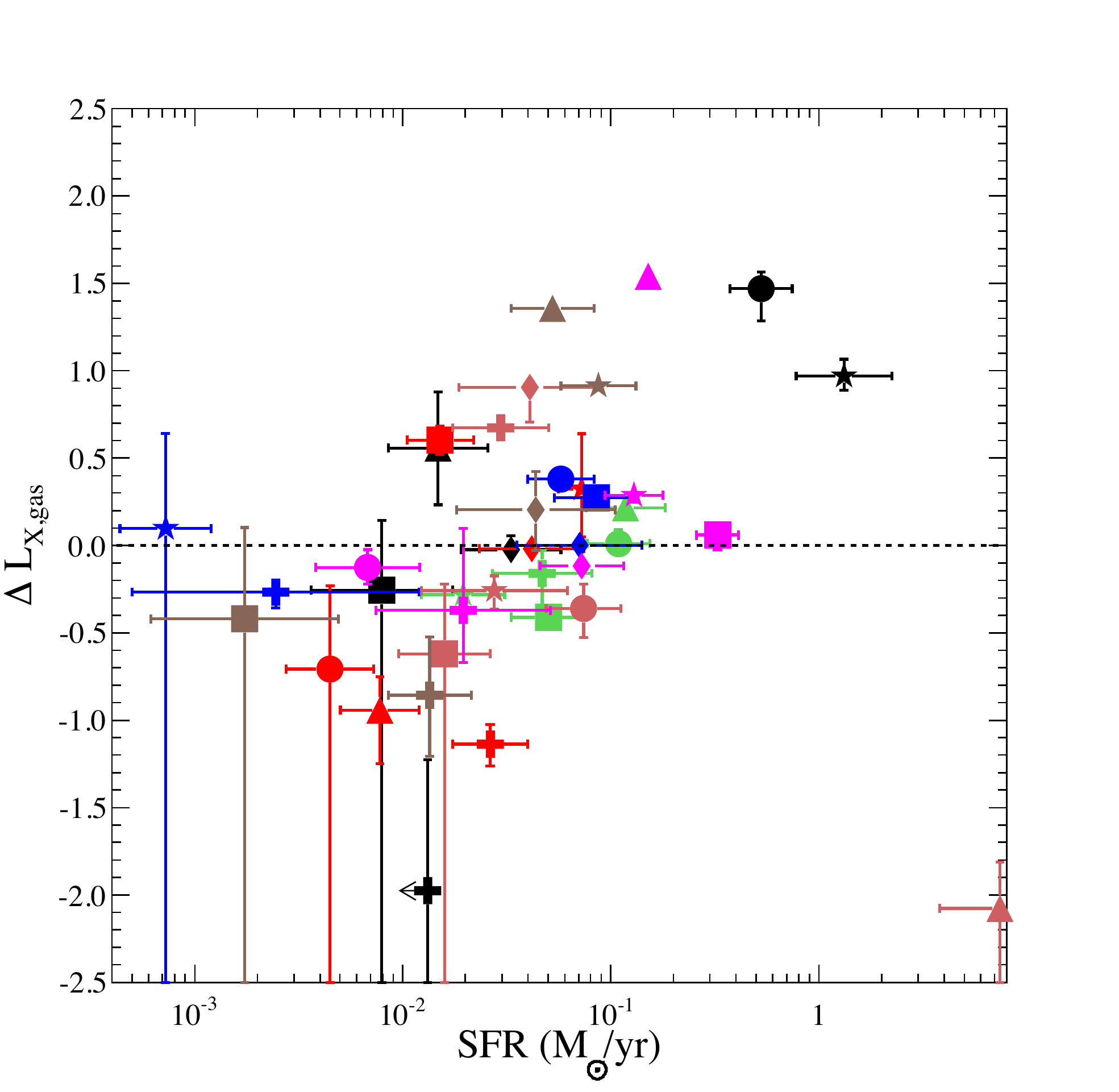}{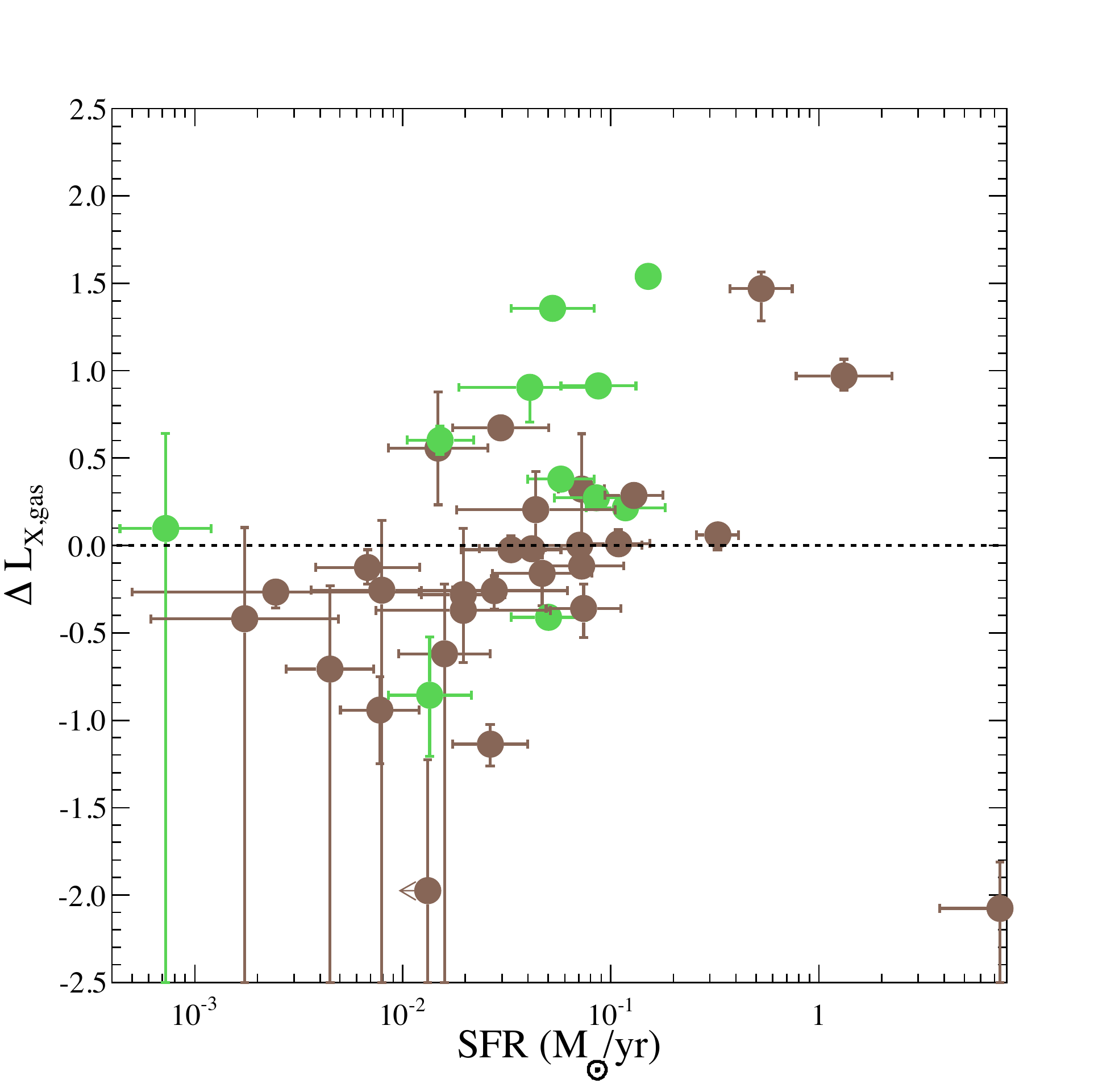}
\plottwo{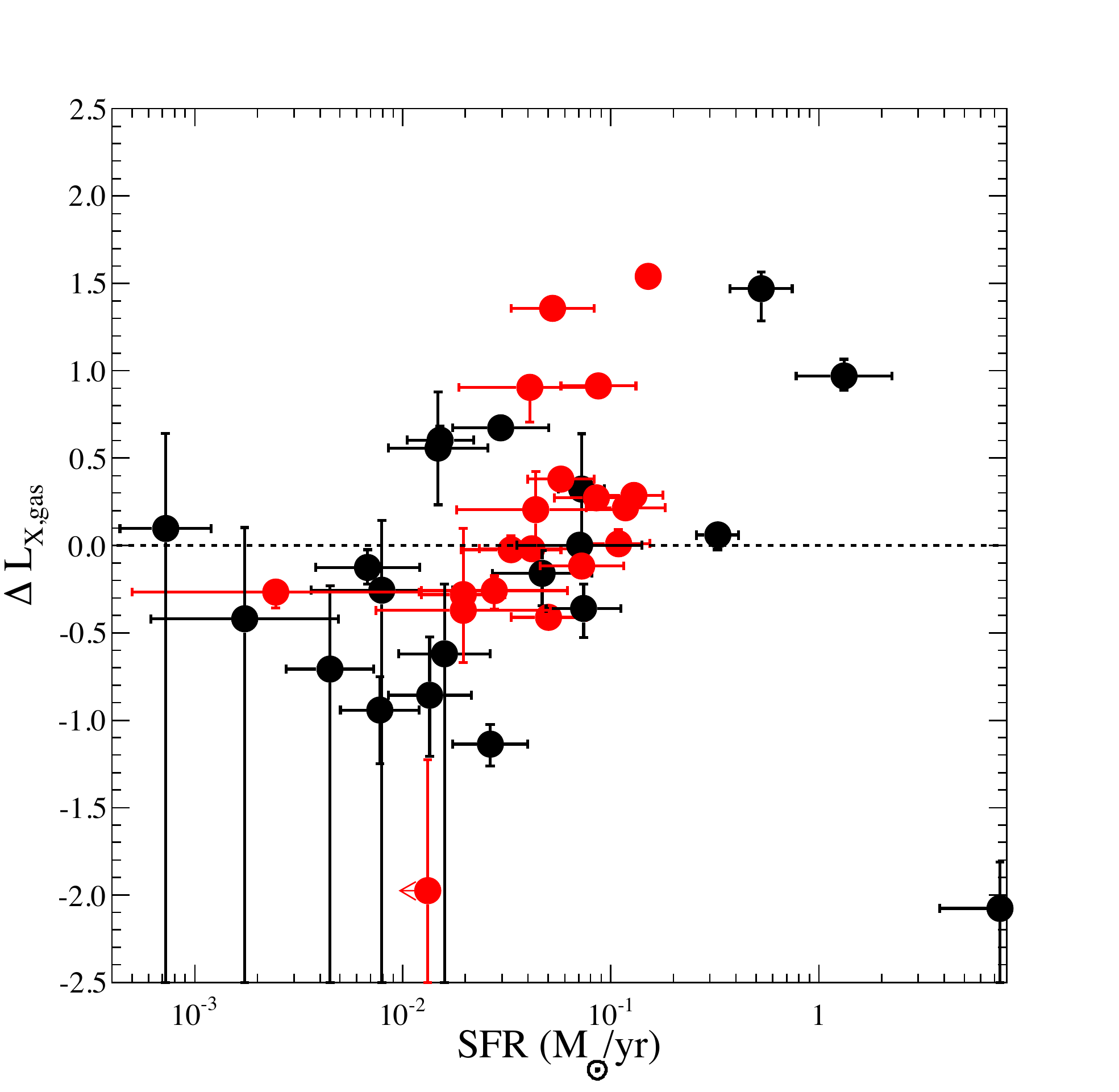}{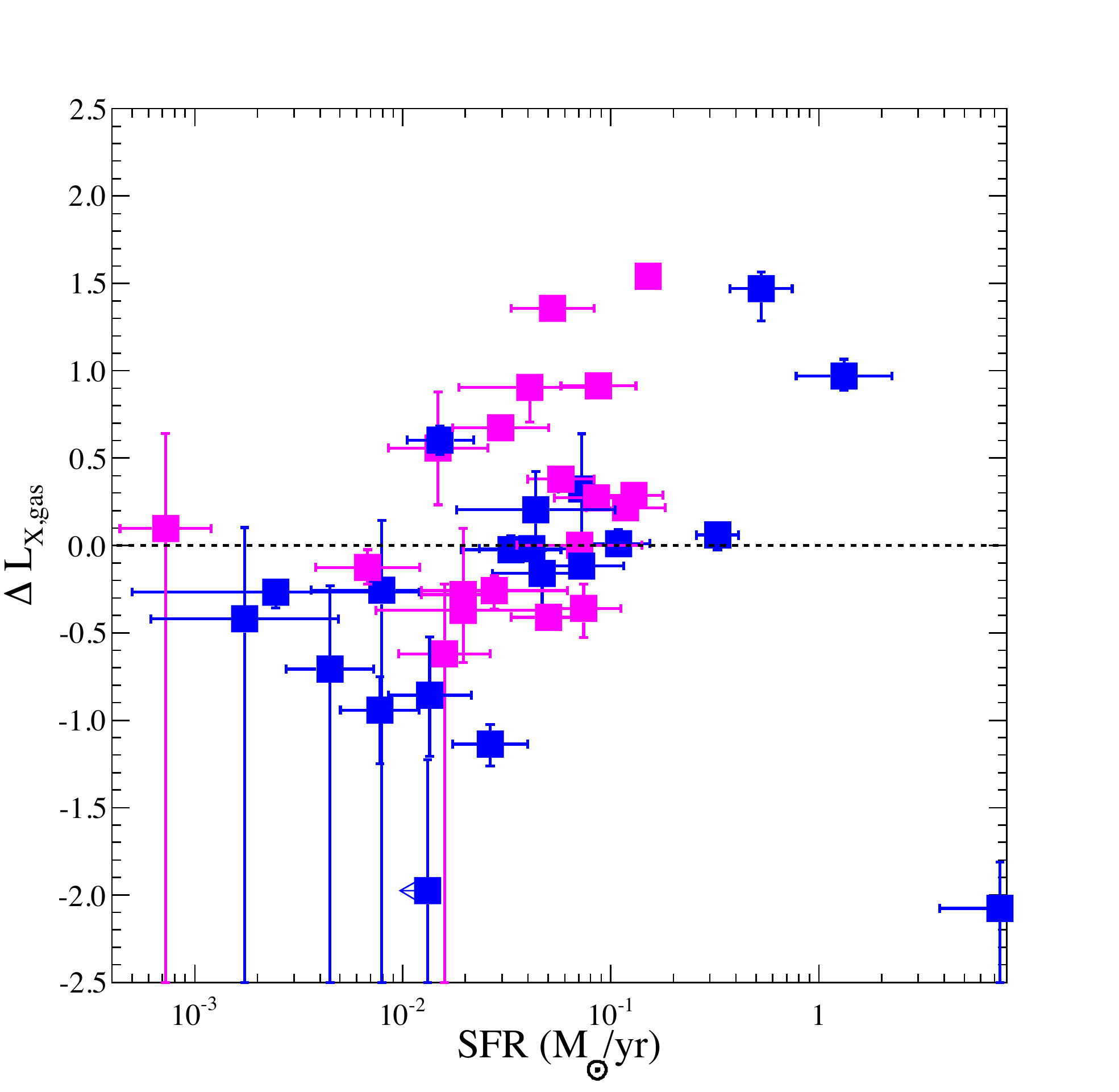}
\caption{$\Delta L_{X_{\rm gas}}$ as a function of star formation rate. {\it top-left}: color code is the same as in Figure~\ref{fig:lxk}. {\it top-right}: ``true lenticular" galaxies (brown) and ``true elliptical" galaxies (green). {\it bottom-left}: high mass galaxies (red) and low mass galaxies (black). {\it bottom-right}: field galaxies (blue) and galaxies in groups and clusters (magenta).}
 \label{fig:sfr} 
 \end{center}
\end{figure*}

These appreciable amounts of cold gas in ETGa are found to be associated with low levels of star formation, with rates ranging up to $\sim1\,M_{\odot}/$yr (e.g.\ Temi et al.\ 2009; Amblard et al.\ 2014). 
Diffuse hot gas can be produced by the energy output associated with star formation 
in late-type galaxies (e.g.\  Mineo et al.\ 2014). 
Stinson et al.\ (2013)  show that hot gaseous halos form at the same time as star formation develops.  
Current star formation in ETGs, even at low levels, could have affected the ISM hot gas content
and contribute to the scatter in the $L_{X_{\rm gas}}$--$L_K$ relation.
We found that $\Delta L_{X_{\rm gas}}$ is positively correlated with the SFR, as shown in Figure~\ref{fig:sfr}. A single obvious outlier is NGC~4494, an X-ray faint galaxy with a very large SFR of $\sim7.0~M_{\odot}/$yr. 
NGC~4494 was also noted to have an unusually high SFR in our source reference, Amblard et al.\ (2014),
who studied the correlation between SFR and dust content;
Amblard et al.\ (2014) suggested that its anomalously high SFR may be related to its AGN
and their SED fitting may have led to an unphysical solution. 
In fact, Wu \& Gao (2006) and Satyapal et al.\ (2005) report a 50 times smaller SFR for NGC~4494,
which is more consistent with its very small molecular hydrogen mass $M_{\rm H_2}$. 
Excluding NGC~4494, we found a very strong correlation between $\Delta L_{X_{\rm gas}}$ and SFR.

\subsection{Gravitational potential}

Galaxy mass has been regarded as one of the most plausible factors in influencing the hot gas content of ETGs (e.g.\  Mathews et al.\ 2006). 
Of all the correlations described above,
we find that
$L_{X_{\rm gas}}/L_K$ is most strongly correlated with $M_{\rm tot}$, the total mass within $1\,r_e$ (see Figure~\ref{fig:lxk2}--{\it bottom left}).
Since  $M_{\rm tot}$ is measured within $1\,r_{\rm e}$, it is dominated by the stellar mass, 
which scales with $L_K$ (so does $r_{\rm e}$). 
The measurement of $M_{\rm tot}$ is not within large enough radii to be dominated by dark matter.
The $L_{X_{\rm gas}}$--$L_K$ relation is very steep, with $L_{X_{\rm gas}}$ depending on $L_K$ 
to a power of at least 2. 
This means the $L_{X_{\rm gas}}/L_K$ ratio has a residual  dependence on $L_K$. 
Thus, the apparent $L_{X_{\rm gas}}/L_K$--$M_{\rm tot}$ correlation could be a by-product of the intrinsic dependence of $L_{X_{\rm gas}}/L_K$ on $L_K$.

We find that the deviations $\Delta L_{X_{\rm gas}}$ also generally increase with $M_{\rm tot}$ (as well as $r_{\rm e}$ and  $M_{\rm tot}/r_{\rm e}$) for our sample galaxies, as demonstrated in Figure~\ref{fig:mtot}. 
This may reflect that it is easier for more massive galaxies to retain their hot gas. 
A closer look shows that  $\Delta L_{X_{\rm gas}}$ increases with these factors only in more massive galaxies, while the deviations $\Delta L_{X_{\rm gas}}$ are actually anti-correlated with $M_{\rm tot}$ in low mass galaxies. 
We show in the Appendix that this latter trend does not survive tests of systematic errors. 

\subsection{Hot gas temperature and stellar velocity dispersion}

\begin{figure*} 
   \begin{center}
    \leavevmode
\epsscale{1.11}
\plottwo{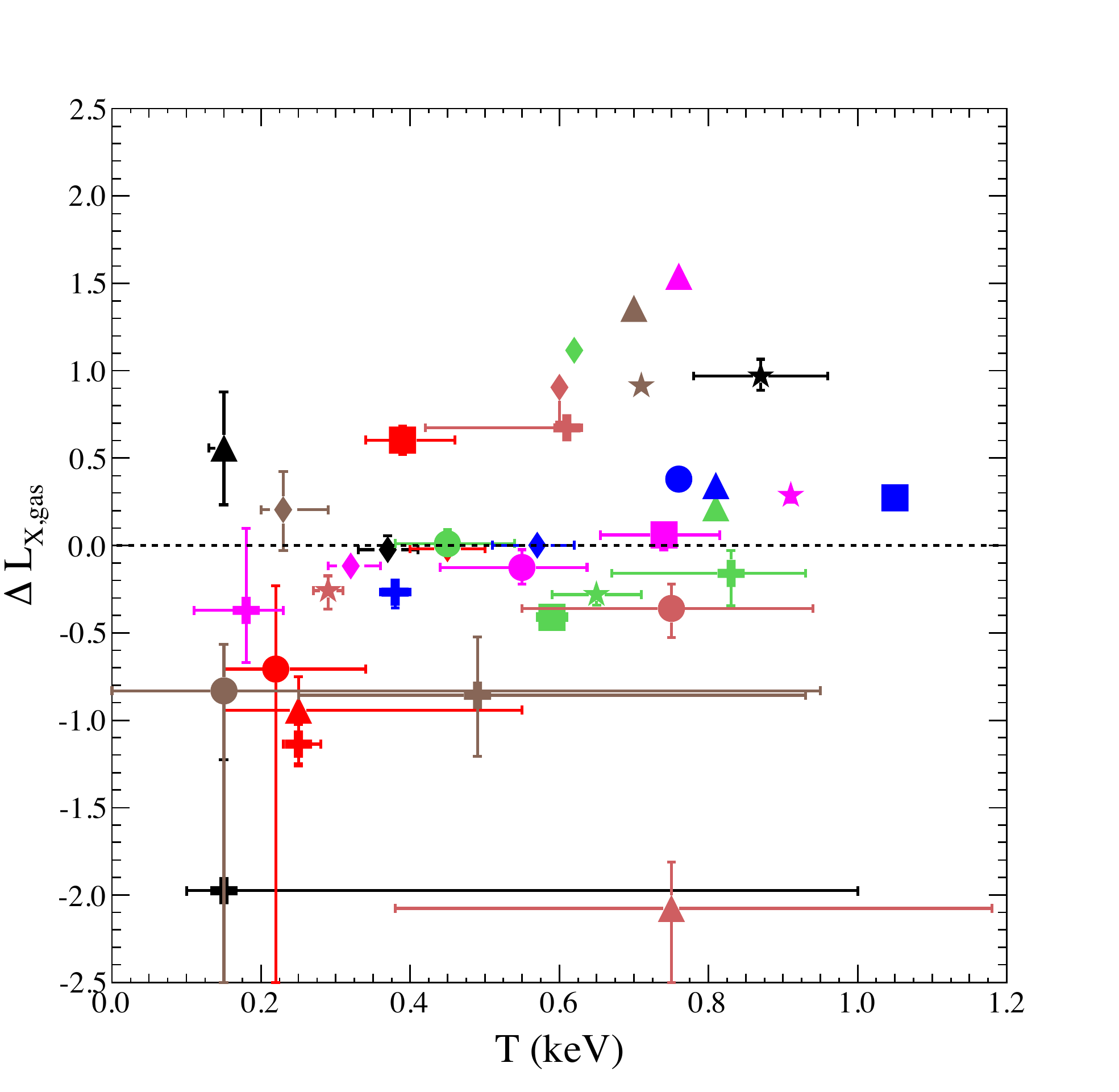}{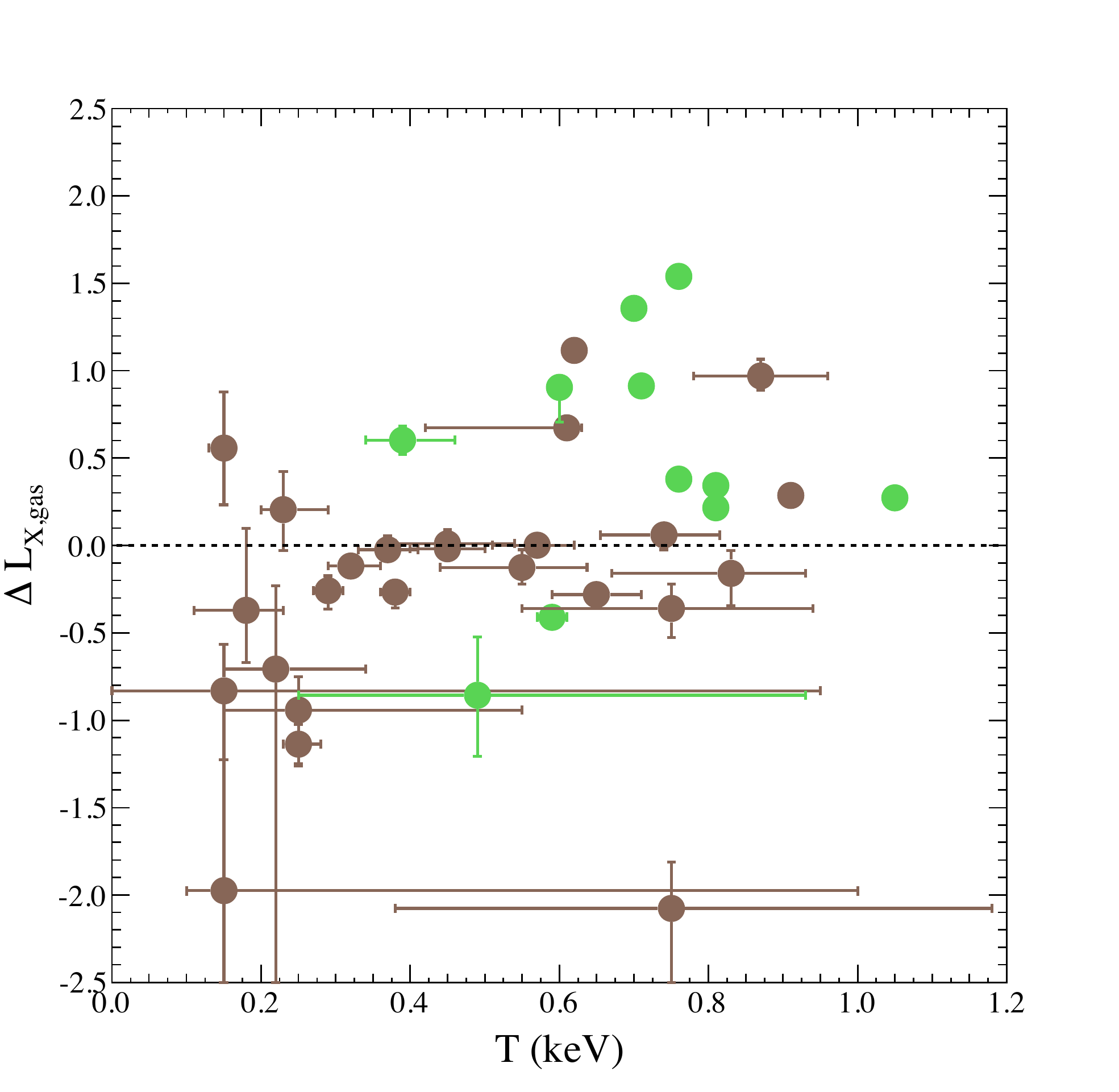}
\plottwo{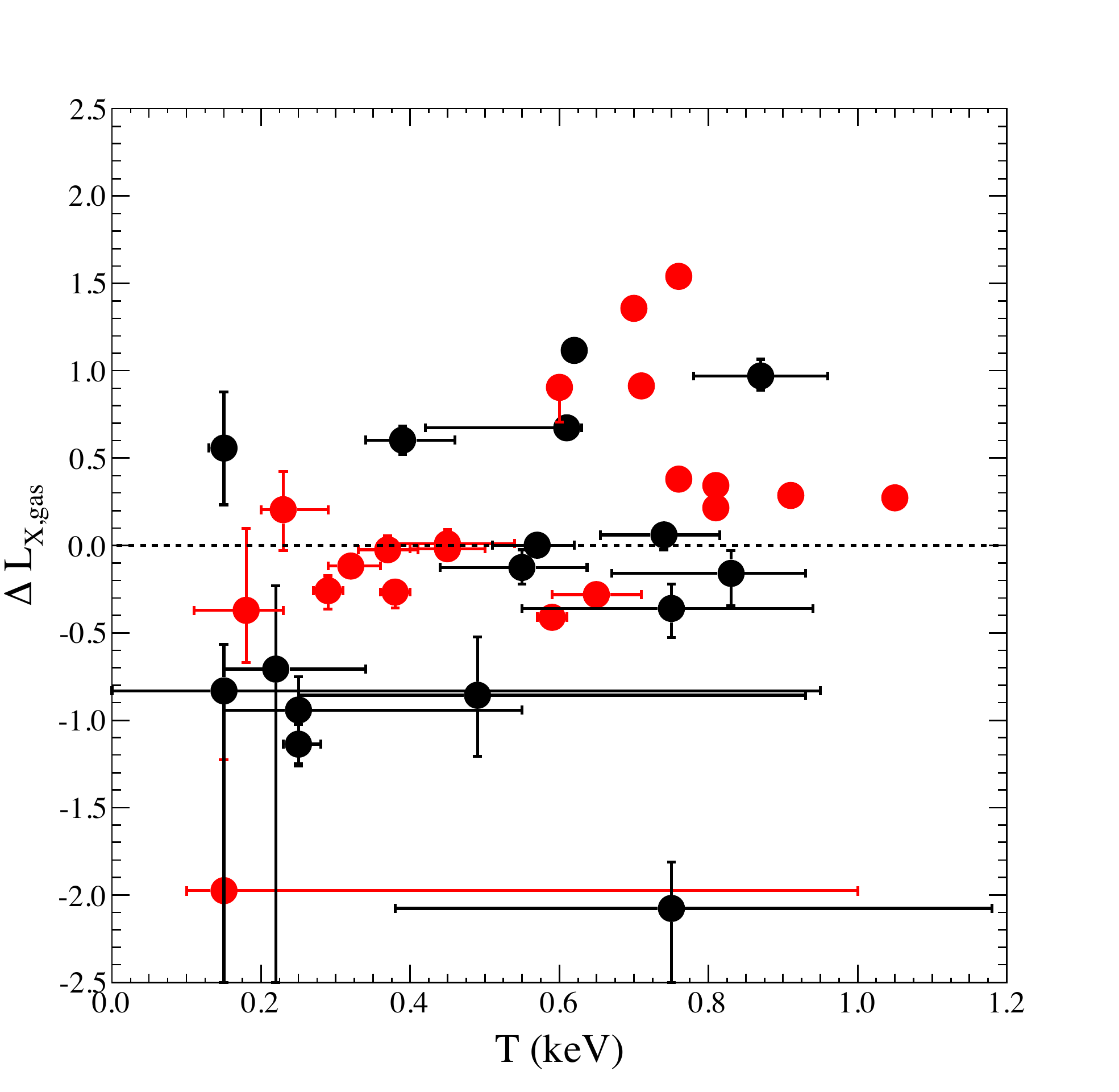}{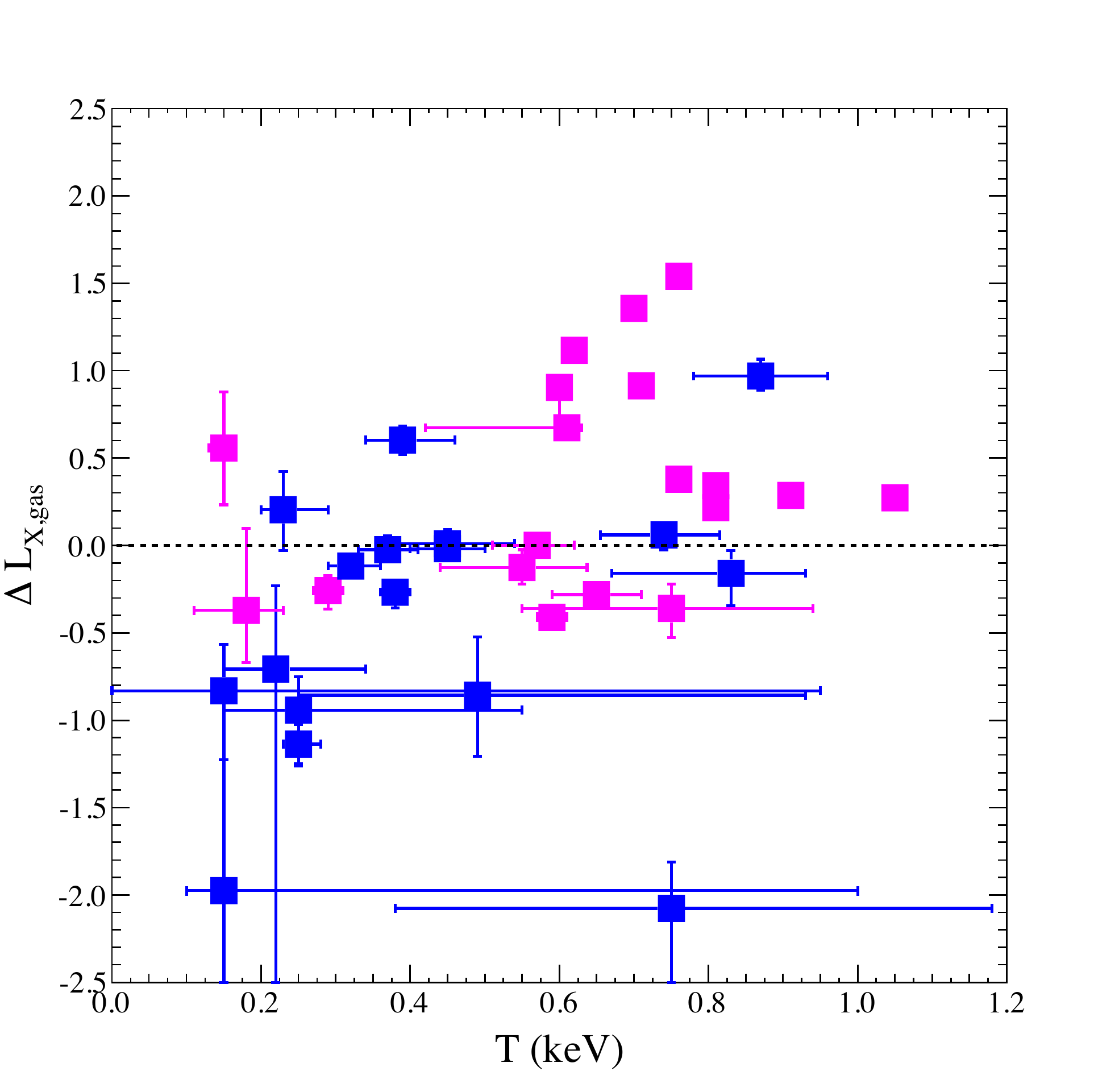}
\caption{$\Delta L_{X_{\rm gas}}$ as a function of hot ISM temperature.  {\it top-left}: color code is the same as in Figure~\ref{fig:lxk}. {\it top-right}: ``true lenticular" galaxies (brown) and ``true elliptical" galaxies (green). {\it bottom-left}: high mass galaxies (red) and low mass galaxies (black). {\it bottom-right}: field galaxies (blue) and galaxies in groups and clusters (magenta).}
 \label{fig:t}
  \end{center}
\end{figure*}

The stellar velocity dispersion and the temperature of a hydrostatic gaseous halo in an ETG 
are partial measures of
the depth of its gravitational potential in a thermalized system (the gas temperature 
can also be a sensitive indicator of non-gravitational energy feedback).
Figure~\ref{fig:t} indicates that those 
population A galaxies with hotter atmospheres generally
have larger $\Delta L_{X_{\rm gas}}$. It is difficult to obtain accurate hot ISM temperatures in low mass galaxies\footnote{The temperatures of NGC~821, NGC~3377, and NGC~3379 were taken from Boroson et al.\ (2011), while we could not have their temperatures constrained in our fitting. This may be because Boroson et al.\ (2011) use a smaller $L_X/L_K$ for CV+AB component which leaves more counts for the hot gas components for these extremely faint galaxies.}. 
Previous studies show that X-ray luminosities increase with ISM temperature in ETGs (Boroson et al.\ 2011; Matthew et al.\ 2006; and Kim \& Fabbiano 2013).
Our study focuses on the hot gas content per unit stellar light, rather than the X-ray luminosity itself. 
We find that $\Delta L_{X_{\rm gas}}$ and $T_X$ are positively correlated. 
The relation between $\Delta L_{X_{\rm gas}}$ and $\sigma$ is similar to the  $\Delta L_{X_{\rm gas}}$--$M_{\rm tot}$ relation.

\subsection{Rotation and ellipticity}

\begin{figure}
\epsscale{1.2}
 \plotone{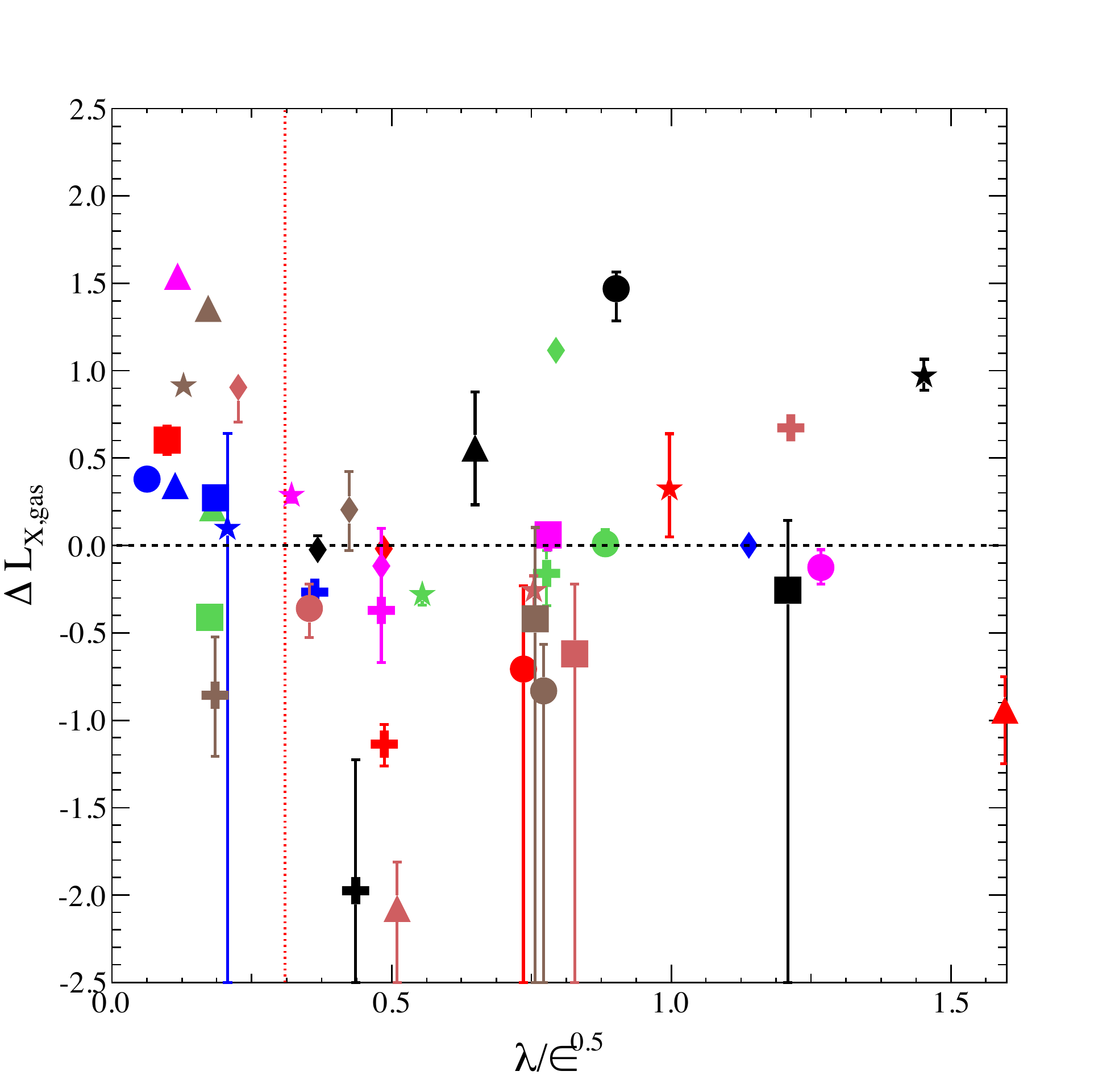}
 \caption{$\Delta L_{X_{\rm gas}}$ as a function of the parameter $
\lambda/\sqrt{\epsilon}$. Color code is the same as in Figure~\ref{fig:lxk}. The red dashed line indicates $\lambda/\sqrt{\epsilon}=0.31$. We call galaxies with $\lambda/\sqrt{\epsilon} < 0.31$ true elliptical galaxies and galaxies with $\lambda/\sqrt{\epsilon} > 0.31$ true lenticular galaxies.} 
 \label{fig:kfs}
\end{figure}

The ATLAS$^{\rm 3D}$  project found that a parameter $\lambda/\sqrt{\epsilon}$, combining a galaxy rotational parameter $\lambda$ with 
a galactic eccentricity $\epsilon$, distinguishes slowly-rotating ``true elliptical"  from fast-rotating  ``true lenticular" galaxies
(Emsellem et al.\ 2011) .
The parameter $\lambda$ is  robust measure of a galaxy's angular momentum and is defined as
$$
\lambda \equiv \frac{\langle R \lvert V\rvert \rangle}{\langle R \sqrt{V^2+\sigma^2}\rangle}
$$
(Jesseit et al.\ 2009),
where $R$, $V$, and $\sigma$ are the circular radius (1\,$r_{\rm e}$), circular velocity and stellar velocity dispersion. 
The values of $\lambda$ and $\epsilon$ of galaxies in our sample are taken from the ATLAS$^{\rm 3D}$  survey
(Emsellem et al.\ 2011) and listed in Table~2.
We show $\Delta L_{X_{\rm gas}}$ as a function of the parameter $\lambda/\sqrt{\epsilon}$ in Figure~\ref{fig:kfs}. We find that true elliptical galaxies ($\lambda/\sqrt{\epsilon} < 0.31$) have a larger X-ray excess than true lenticular galaxies ($\lambda/\sqrt{\epsilon} > 0.31$).

We also compared the hot gas content to the rotation parameter $\lambda$ and ellipticity $\epsilon$ separately. 
We find that the deviations $\Delta L_{X_{\rm gas}}$ are anti-correlated with rotation and flatness, as shown in Figures~\ref{fig:rot} and \ref{fig:flat}, respectively; however, this only holds for population A galaxies. 
The situation is further complicated by the fact
that flatness and rotation are usually highly correlated, in the sense that only flat galaxies can support fast rotation. This is the case for galaxies in our sample. It is difficult to tell whether the above two correlations have real physical origins or one correlation is simply a byproduct of the other. 
The best way to test this is to study the correlation of hot gas and rotation for galaxies that have similar shapes or to study the correlation between hot gas and ellipticity for galaxies that have similar rotation. 

\begin{figure*}
  \begin{center}
    \leavevmode
    \epsscale{1.11}
 \plottwo{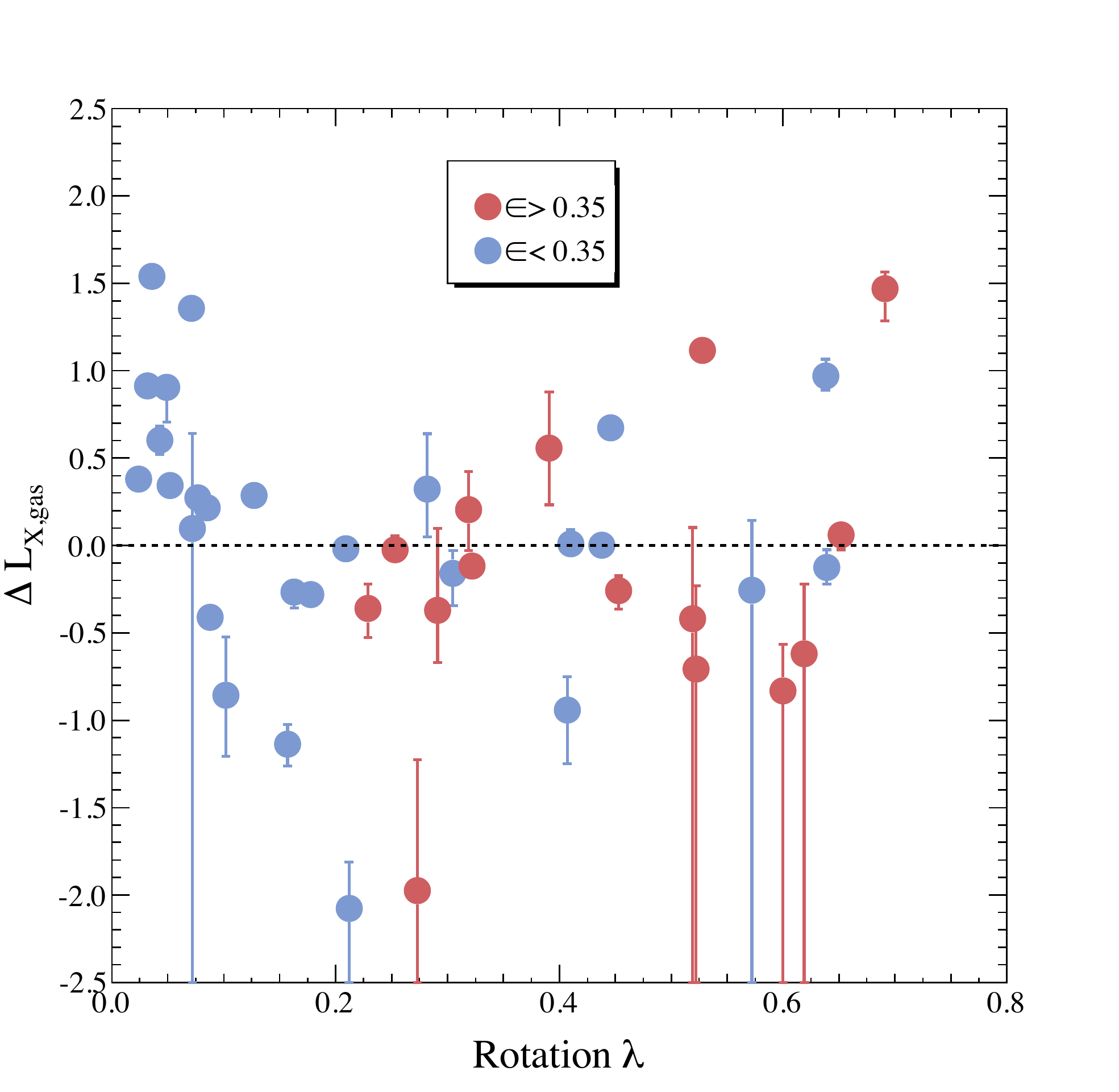}{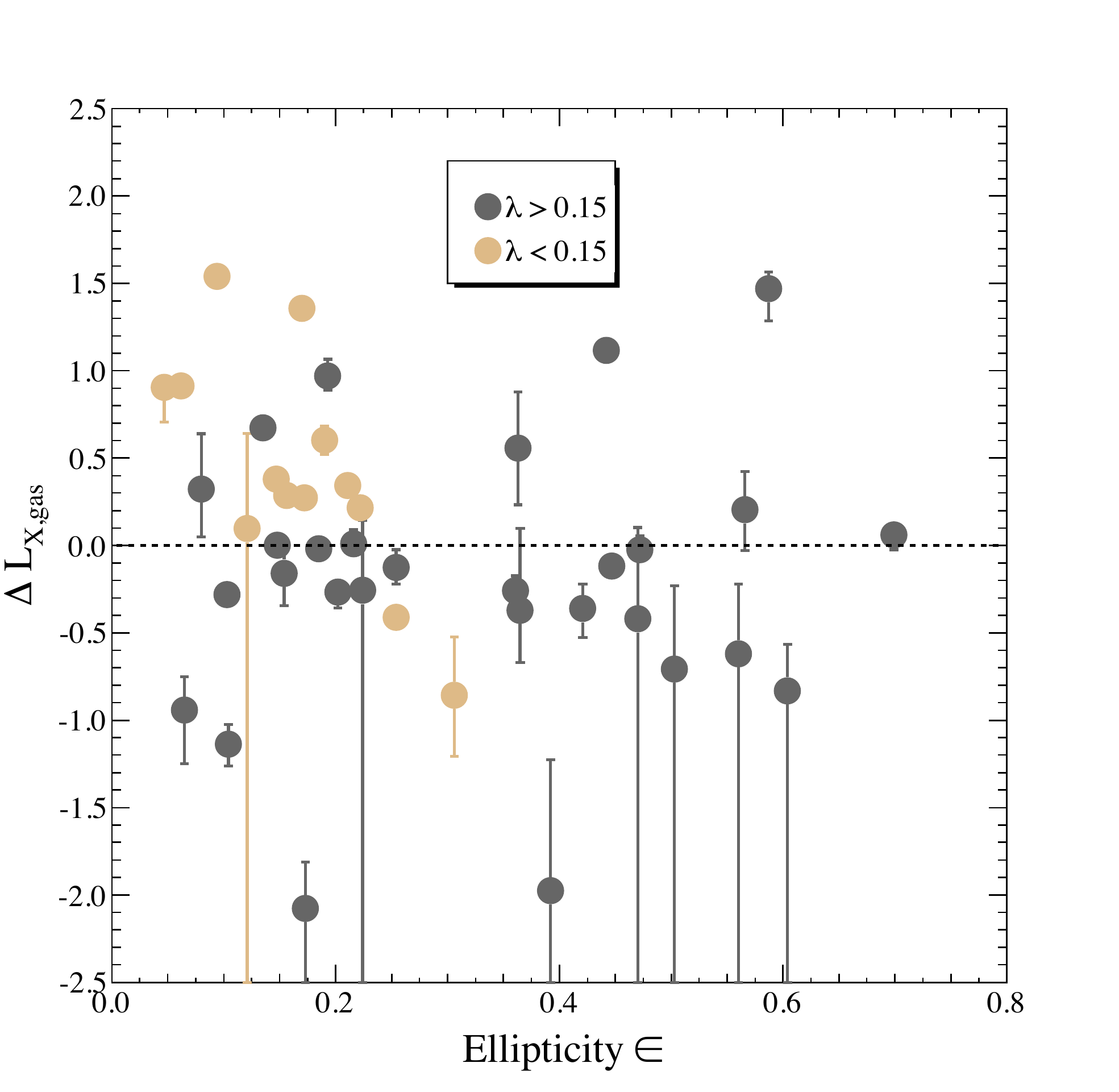}
 \caption{{\it left}: $\Delta L_{X_{\rm gas}}$ as a function of rotation $\lambda$ for flat galaxies (red) and round galaxies (blue). {\it right}: $\Delta L_{X_{\rm gas}}$ as a function of ellipticity $\epsilon$ for fast-rotating galaxies (gray) and slow-rotating galaxies (yellow).}
  \label{fig:rotflat} 
  \end{center}
\end{figure*}

Based on the data in Table~2 and Figures~\ref{fig:rot} and \ref{fig:flat}, we find that $\lambda=0.15$ provides a natural cut  separating
slow-rotating and fast-rotating galaxies and $\epsilon=0.35$ provides a natural separation between flat and round galaxies. 
We divide our sample into flat ($\epsilon > 0.35$) and round ($\epsilon < 0.35$) galaxies in the $\Delta L_{X_{\rm gas}}$--rotation relation, as shown in Figure~\ref{fig:rotflat}-{\it left}. Neither sub-group exhibits a significant correlation. As shown in Figure~\ref{fig:rotflat}-{\it right}, we divide our sample into fast-rotating ($\lambda > 0.15$) and slow-rotating ($\lambda < 0.15$) galaxies in the $\Delta L_{X_{\rm gas}}$--ellipticity relation. Slow-rotating galaxies show a significant correlation between the negative residuals of $\Delta L_{X_{\rm gas}}$ and ellipticity.
This result indicates the relatively important role of ellipticity.

\subsection{ Environmental density}

It is debated whether intracluster gas has a net positive or negative effect on the X-ray luminosities of cluster galaxies (Sun et al.\ 2007; Mulchaey \& Jeltema 2010). For galaxies in dense environments, gaseous outflows can be suppressed by ICM pressure confinement.
In more isolated environments, gaseous outflows could be responsible for the scatter in  $L_{X_{\rm gas}}/L_K$ for ETGs. 
From the SDSS we obtained the number of  neighboring galaxies for 38 out of 42 galaxies in our sample, as listed in Table~3. Figure~\ref{fig:sdss} shows  $L_{X_{\rm gas}}/L_K$ and $\Delta L_{X_{\rm gas}}$ as a function of the number of nearby galaxies. Galaxies in higher galaxy density environments contain more hot gas per unit stellar light (although this trend seems very mild), as found by Brown \& Bregman (2000) using {\sl ROSAT} and {\sl ASCA}. 
This may be a result of pressure confinement of the ICM, or the accretion of the surrounding ICM into ISM (Pinino et al.\ 2005), or they may be massive slow-rotators,

\subsection{ Ram pressure stripping in the Virgo Cluster} 

The Virgo Cluster is the nearest relaxed galaxy cluster, residing at a distance of only 16\,Mpc. 
Cen et al.\ (2014) show through simulations that ram pressure stripping of galaxies starts to become effective within 3\,Mpc of the cluster center.  
In our sample, 15 galaxies have their deprojected radii within 3\,Mpc of M87, the central 
galaxy of the Virgo Cluster. This allows us to study the effects of ram pressure
on many galaxies in a single cluster.
A quantitative description of the strength of ram pressure stripping was first presented by Gunn \& Gott (1972). Based on a static force argument, a stripping condition can be derived for a gaseous disc moving face-on through the ICM. 
Hot gas in the ISM will be stripped when the ram pressure ($P_{\rm ram}=\rho_{_{\rm ICM}}v^2$) exceeds the gravitational restoring force per unit area, $P_{\rm grav}$.
McCarthy et al.\ (2008) developed an analogous model for the ram pressure stripping of galaxies with a spherically-symmetric gas distribution. Their model, which is more suitable for ETGs with an extended hot gas halo rather than a cold gaseous disk, yields 

\begin{equation}
P_{\rm ram} >\frac{\pi}{2}\frac{GM_{\rm tot}\rho_{_{\rm ISM}}}{R_{_{\rm ISM}}},
\end{equation}
where $M_{\rm tot}$ is the total mass of the galaxy, $\rho_{_{\rm ISM}}$ is the galaxy ISM gas density, and $R_{_{\rm ISM}}$ is the radius of the galaxy at which the stripping occurs.
We consequently obtained the ratio of ram pressure to gravitational restoring pressure:

\begin{equation}
{ P_{\rm ram} \over P_{\rm grav} } = {2\over\pi} { \rho_{_{\rm ICM}} \over \rho_{_{\rm ISM}}} {{ v^2 R_{_{\rm ISM}} } \over 
{GM_{\rm tot} }}.
\end{equation}

\begin{figure*}
  \begin{center}
    \leavevmode
\epsscale{1.11}
\plottwo{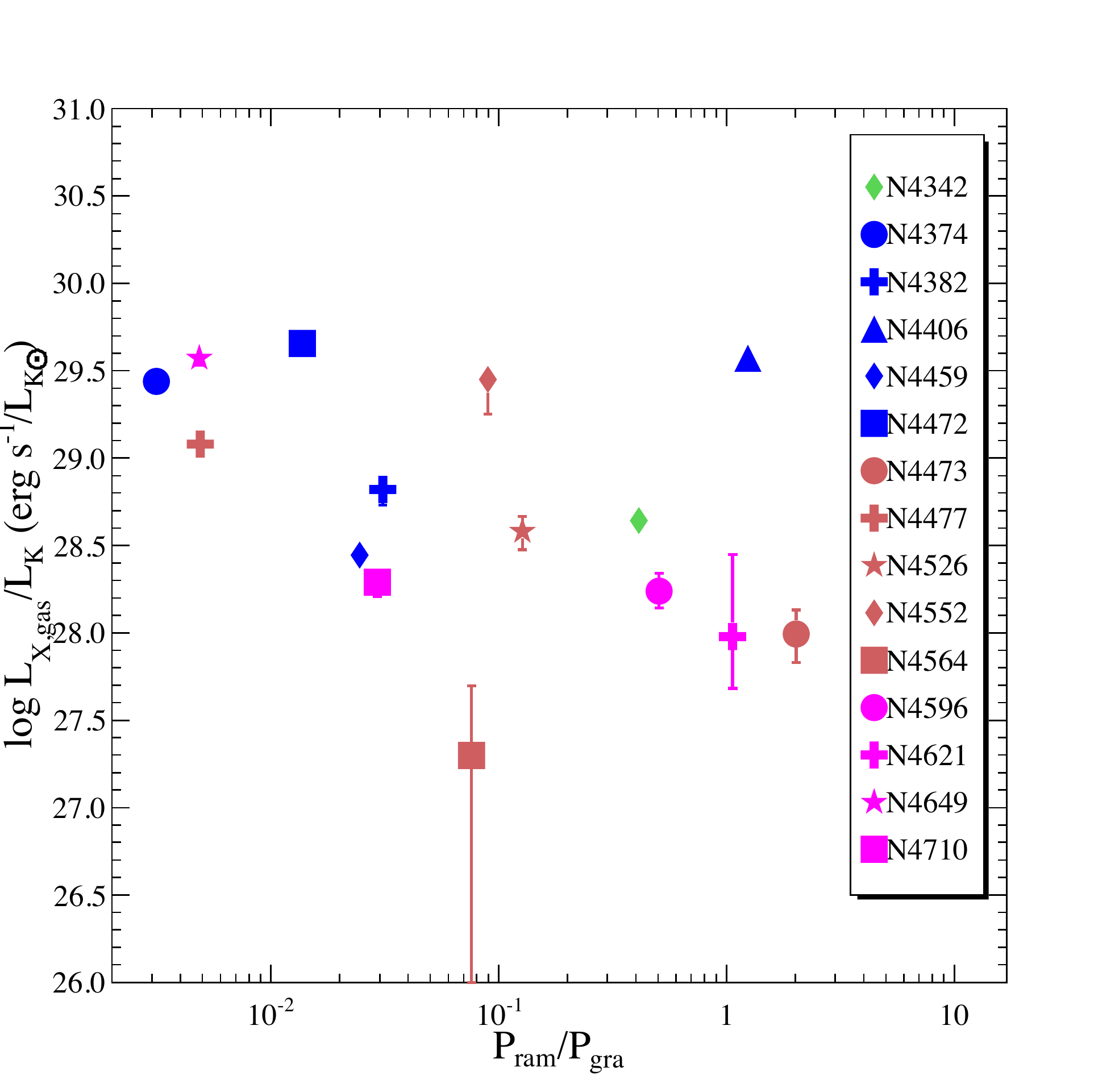}{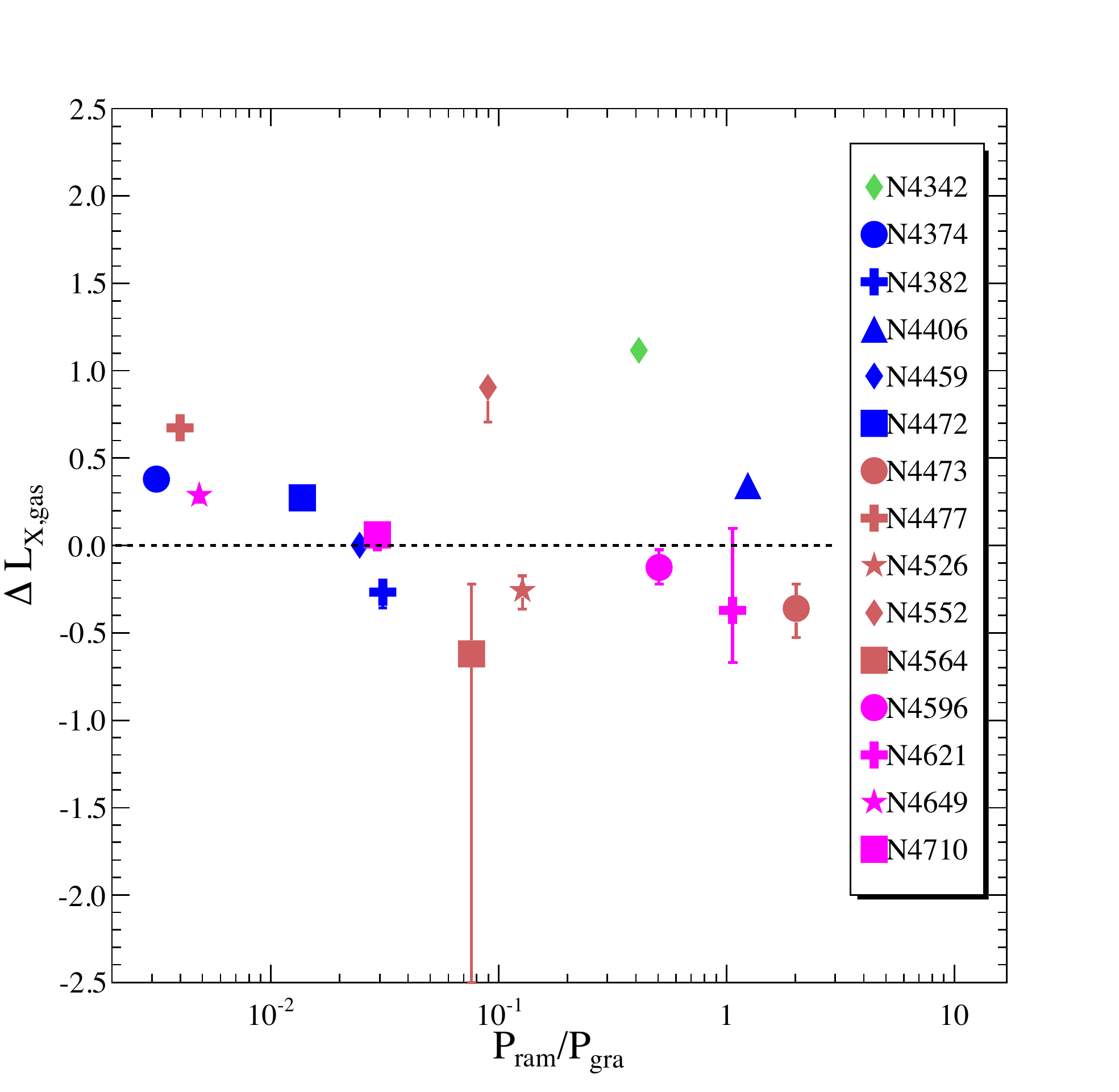}
\caption{$L_{X_{\rm gas}}/L_K$ and $\Delta L_{X_{\rm gas}}$ of galaxies in the Virgo Cluster in this work as a function of the ratio of ram pressure to gravitational restoring pressure. Color code is the same as in Figure~\ref{fig:lxk}. [{\sl see the electronic edition of the journal for a color version of this figure.}]}
 \label{fig:p}
  \end{center}
\end{figure*}

We adopt eq.(2) to calculate the instantaneous strength of ram pressure stripping relative to the
gravitational restoring force.  
We estimate $\rho_{_{\rm ICM}}$ from the deprojected gas density profile of the Virgo Cluster in the radial range of 0.3--1\,$r_{\rm vir}$, as derived from 13 {\sl XMM-Newton} observations out to $r_{\rm vir}$ (1.2 Mpc).   
This results in a deprojected density profile of the form 
$\rho_{_{\rm ICM}}\propto r^{-1.2}$
 (Urban et al.\ 2011), where $r$ is the 3D separation between a member galaxy and M87, as taken from ATLAS$^{\rm 3D}$. 
 The relative velocity $v$ is chosen to be $\sqrt{3}$ times the observed radial velocity relative to M87, obtained from NED, and
$R_{_{\rm ISM}}$ is chosen to be $2\,r_e$.
We obtained values of 
$\rho_{_{\rm ISM}}$ through the best-fit norms of the {\tt apec} component in our spectral analysis. 
Figure~\ref{fig:p} shows the ratio $L_{X_{\rm gas}}/L_K$ and $\Delta L_{X_{\rm gas}}$  as a function of the ratio $P_{\rm ram}/P_{\rm grav}$ for the 15  Virgo Cluster ETGs in our sample, as described in \S2. 
X-ray faint galaxies tend to have larger $P_{\rm ram}/P_{\rm grav}$ than X-ray luminous galaxies. 
This trend implies that X-ray luminous galaxies have retained a larger fraction of their gas against ram pressure,
while X-ray faint galaxies have lost a larger fraction of their gas to ram pressure.

\begin{table*}
  \caption{Summary of correlation of $L_{X_{\rm gas}}/L_K$ with each factor}
  \begin{center}
    \leavevmode
    \begin{tabular}{lccccc} \hline \hline       
{Relations}&{Correlation coefficient $\rho$} &Errors (STA)&{Null hypothesis probability}&{Number of galaxies}&{Correlation coefficient $\tau$}\\ \hline 
Age &0.439&$\pm0.034$&0.0084&37&0.215\\
$M$(H~{\sc i}) &0.412&$\pm0.033$&0.0100&39&0.182\\
$M(\rm H_2)$&0.152&$\pm0.036$&0.3310&42&0.098\\
$M(\rm H_2)+M$(H~{\sc i})&0.336&$\pm0.030$&0.0314&42&0.224\\
SFR$^{*}$&0.568&$\pm0.068$&0.0006&38&0.425\\
$T_X$&0.515&$\pm0.184$&0.0023&36&0.352\\
$\sigma$&0.553&$\pm0.025$&0.0004&42&0.317\\
 $M_{\rm tot}$ ($r<r_e$)&{\bf 0.639}&$\pm0.070$&0.0000&42&0.457\\
$r_e$&{\bf 0.622}&$\pm0.024$&0.0001&42&0.460\\
$M_{\rm tot}/r_e$&0.611&$\pm0.027$&0.0001&42&0.171\\
$\lambda/\sqrt{\epsilon}$&-0.441&$\pm0.026$&0.0047&42&-0.285\\
Rotation $\lambda$& -0.439&$\pm0.028$& 0.0050&42&-0.282\\
Ellipticity $\epsilon$&-0.280&$\pm0.034$&0.0732&42&-0.193\\
SDSS &0.346&$\pm0.033$&0.0352&38&0.185\\
$P_{\rm ram}/P_{\rm gra}$&-0.479&$\pm0.041$&0.0733&15&-0.333\\ \hline
    \end{tabular}
  \end{center}
  \tablecomments{$^{*}$ not including NGC~4494.}
\end{table*}

\begin{table*}
  \caption{Summary of correlation of $\Delta L_{X_{\rm gas}}$  with each factor}
  \label{tab:distance}
  \begin{center}
    \leavevmode
    \begin{tabular}{lccccc} \hline \hline    
{Relations}&{Coefficient $\rho$$^{\dagger}$}&Errors (STA, SYS$^{\sharp}$)&{Null hypothesis prob.}&{No. of galaxies}&{Correlation coefficient $\tau$$^{\ddagger}$}\\ \hline
Age &0.283&$\pm0.080$, ${-0.004}$, ${+0.009}$, $-0.041$&0.0892&37&0.182\\
$M$(H~{\sc i}) &0.319&$\pm0.067$, $+0.049$, $-0.067$, $-0.028$&0.0495&39&0.153\\
$M(\rm H_2)$&0.213&$\pm0.058$, $+0.020$, $-0.026$, $+0.008$&0.1717&42&0.128\\
$M(\rm H_2)+M$(H~{\sc i})&0.296&$\pm0.060$, $+0.032$, $-0.041$, $+0.057$&0.0584&42&0.227\\
SFR$^{*}$&{\bf 0.536}&$\pm0.103$, $+0.025$, $-0.065$, $-0.026$&0.0011&38&0.428\\
$T_X$&{\bf 0.476}&$\pm 0.186$, ${-0.024}$, ${+0.004}$, $-0.098$&0.0049&36&0.337\\
$\sigma$&0.220&$\pm0.063$, $+0.106$, $-0.131$, $-0.088$&0.1595&42&0.064\\
$M_{\rm tot}$ ($r<r_e$)&0.175&$\pm0.065$, $+0.127$, $-0.149$, $-0.062$&0.2631&42&0.120\\
$r_e$&0.218&$\pm0.064$, $+0.114$, $-0.141$, $-0.070$&0.1631&42&0.151\\
$M_{\rm tot}/r_e$&0.231&$\pm0.070$, $+0.117$, $-0.139$, $-0.076$&0.1396&42&0.171\\
$\lambda/\sqrt{\epsilon}$&-0.217&$\pm0.070$, $-0.060$, $+0.061$, $+0.101$&0.1656&42&-0.138\\
Rotation $\lambda$& -0.247&$\pm0.071$, $-0.064$, $+0.064$, $+0.106$& 0.1140&42&-0.159\\
Ellipticity $\epsilon$&-0.272&$\pm0.063$, $+0.009$, $+0.011$, $+0.060$&0.0821&42&-0.261\\
SDSS &0.362&$\pm0.063$, $-0.011$, $+0.006$, $-0.086$&0.0276&38&0.202\\
$P_{\rm ram}/P_{\rm gra}$&-0.343&$\pm0.073$, $-0.032$, $+0.100$, $+0.168$&0.1995&15&-0.295\\ \hline
    \end{tabular}
  \end{center}
\tablecomments{
$^{*}$ not including NGC~4494. $^{\sharp}$ Systematic errors obtained using dashed red, blue, and black (broken power law) lines respectively in Figure~\ref{fig:lxk_sys} as the benchmark of the $L_{X_{\rm gas}}/L_K$ relation. $^{\dagger}$ determined by Spearman's correlation. $^{\ddagger}$ determined by Kendall's tau correlation.}
\end{table*}

\section{\bf Discussion}

The large scatter (up to a factor of 1000) in the $L_{X_{\rm gas}}$--$L_K$ relation for ETGs has been a 
long-standing puzzle in extragalactic astronomy. 
We use $L_{X_{\rm gas}}/L_K$ and the residuals $\Delta L_{X_{\rm gas}}$ to describe the hot gas content per unit stellar light in our sample of 42 galaxies and 
compare them
to various internal and external factors. 
We find that the X-ray luminosity deviations $\Delta L_{X_{\rm gas}}$  are most strongly correlated with the galaxies' star formation rates and hot gas temperatures. 
We also find two distinct populations of ETGs.
These results point towards some important implications that are discussed in the following subsections.

\subsection{Population A and population B galaxies}

Galaxies in our sample tend to belong to two different populations, as shown in Figure~\ref{fig:lxks}. 
Population A galaxies are generally X-ray luminous, 
massive, slowly rotating, hot, big, and tend to reside in 
high galaxy density environments. 
Population B galaxies tend to be X-ray sub-luminous, low mass, fast-rotating, cool, small, and reside in the low galaxy density environments (the field). 
These factors are highly degenerate and it is difficult to pin down the driving factor. 
Perhaps galaxies in groups and clusters have experienced more mergers,
which could slow down their rotation and produce more massive and bigger galaxies with 
deeper gravitational potentials.

\begin{table*}
\caption{Best-fits of the $L_{X_{\rm gas}}$--$L_K$ to a single power law for each subgroup and their dispersions}
  \begin{center}
    \leavevmode
    \begin{tabular}{lccccc}\hline \hline
{Subgroup}&{Slope}&{Difference in slopes}&{Dispersion}&{Number of galaxies}&{Slope-Kelly$^{\ast}$} \\ \hline 
True elliptical &2.545$\pm$0.508&\multirow{2}{*}{1.8\,$\sigma$}&0.646&12&$2.459\pm0.772$\\%1.8
True lenticular &1.619$\pm$0.373&&0.717&30&$1.591\pm0.342$\\
\hline
High mass &3.222$\pm$0.700&\multirow{2}{*}{3.1\,$\sigma$}&0.690&20&$2.913\pm0.681$\\%3.15
Low mass &0.843$\pm$0.394&&0.685&22&$0.690\pm0.611$\\
\hline
Goups \& Clusters &2.515$\pm$0.326&\multirow{2}{*}{2.4\,$\sigma$}&0.589&21&$2.393\pm0.412$\\%2.25
Field &1.374$\pm$0.481&&0.762&21&$1.422\pm0.494$ \\ \hline 
    \end{tabular}
  \end{center}
  \tablecomments{
$^{\ast}$ Slopes determined by the maximum likelihood procedure of Kelly (2007).}
\end{table*}

The best-fit power law characterizing each sub-group is listed in Table~6. 
The slopes of the sub-group $L_{X_{\rm gas}}$--$L_K$ relations are 
somewhat  steeper for population A galaxies than for population B galaxies; however, the slopes are significantly different for only the high-mass / low-mass sub-groups (which differ by  more than $3\sigma$).
We define the dispersion in $L_{X_{\rm gas}}-L_K$ for each sub-group as 
$\sqrt{N^{-1}\sum{\Delta L_{X_{\rm gas}}}^2}$,
where $\Delta L_{X_{\rm gas}}$ is the difference between the observed $L_{X_{\rm gas}}$ and the best-fit $L_{X_{\rm gas}}$--$L_K$ relation for each sub-group and $N$ is the number of galaxies in each sub-group, as listed in Table~6.
Galaxies in the field have significantly larger dispersion than galaxies in groups and clusters (by 20\%).
This is consistent with our investigation of the environments of 209 ETGs detected by {\sl ROSAT}, as shown in Figure~\ref{fig:osullivan}. 
%This suggests that the large scatter in the $L_{X_{\rm gas}}$-$L_K$ relation has a largely internal origin. 

Galaxies in groups and clusters experience complicated environmental effects. Relatively lower mass galaxies with a lower ISM temperature (thus a small ISM pressure) may suffer from ram pressure stripping and their atmospheres may evaporate due to the heating of the surrounding ICM. The atmospheres of massive galaxies are relatively hot with a high gas pressure. The surrounding ICM may provide pressure confinement. These processes make faint galaxies fainter and luminous galaxies more luminous, leading to a steeper slope in the $L_{X_{\rm gas}}$--$L_K$
relation for Population A galaxies.

\subsection{Cold gas and stellar feedback}

\subsubsection{Feedback from mild star formation}

In our study, we find that the hot gas content is most strongly correlated with SFR compared to all other factors we have considered. This applies to both population A galaxies and population B galaxies. There are three explanations for this correlation. 
One explanation is that the hot gas transforms into cold gas through radiative cooling and forms into stars. 
We have three lines of evidence to rule out this scenario: (1). we find that the radiative cooling rate is smaller than the SFR for more than half of the galaxies. Thus, radiative cooling is unable to sustain the star formation. (2). $\Delta L_{X_{\rm gas}}$ is positively correlated with hot gas temperature and the SFR is mildly positively correlated with the hot gas entropy. Hot gas is not cooling fast. (3). the correlation between  $\Delta L_{X_{\rm gas}}$ and SFR is stronger than that between $\Delta L_{X_{\rm gas}}$ and $M_{\rm H_2}$.
The second explanation is that stellar feedback energizes the ISM. 
Given that the star formation rates in our sample galaxies are very low (mostly $<0.1\,M_{\odot}/$yr) 
it seems unlikely that such star formation
could drive significant amounts of gas out of the host galaxies;
associated supernova explosions (1 SN per 100 years per $M_{\odot}/$yr of star formation) 
would just heat up more gas into an X-ray emitting phase. 
Energetic feedback is expected to easily sweep out the gaseous component in dwarf galaxies (Hopkins et al.\ 2012). ETGs in our sample are at least as massive as the Milky Way. Their gravitational potentials may be deep enough to retain their hot atmospheres. 
Thus, our result provides observational support of star formation feedback in early-type galaxies.
The third explanation is that both low level star formation and enhanced X-ray luminosities may both
be consequences of recent wet minor mergers, as suggested by Civano et al.\ (2014).
These processes need to be studied in more detail (especially via simulations) in the future.

\subsubsection{Origins of cold gas}

\begin{figure} 
\epsscale{1.2}
\plotone{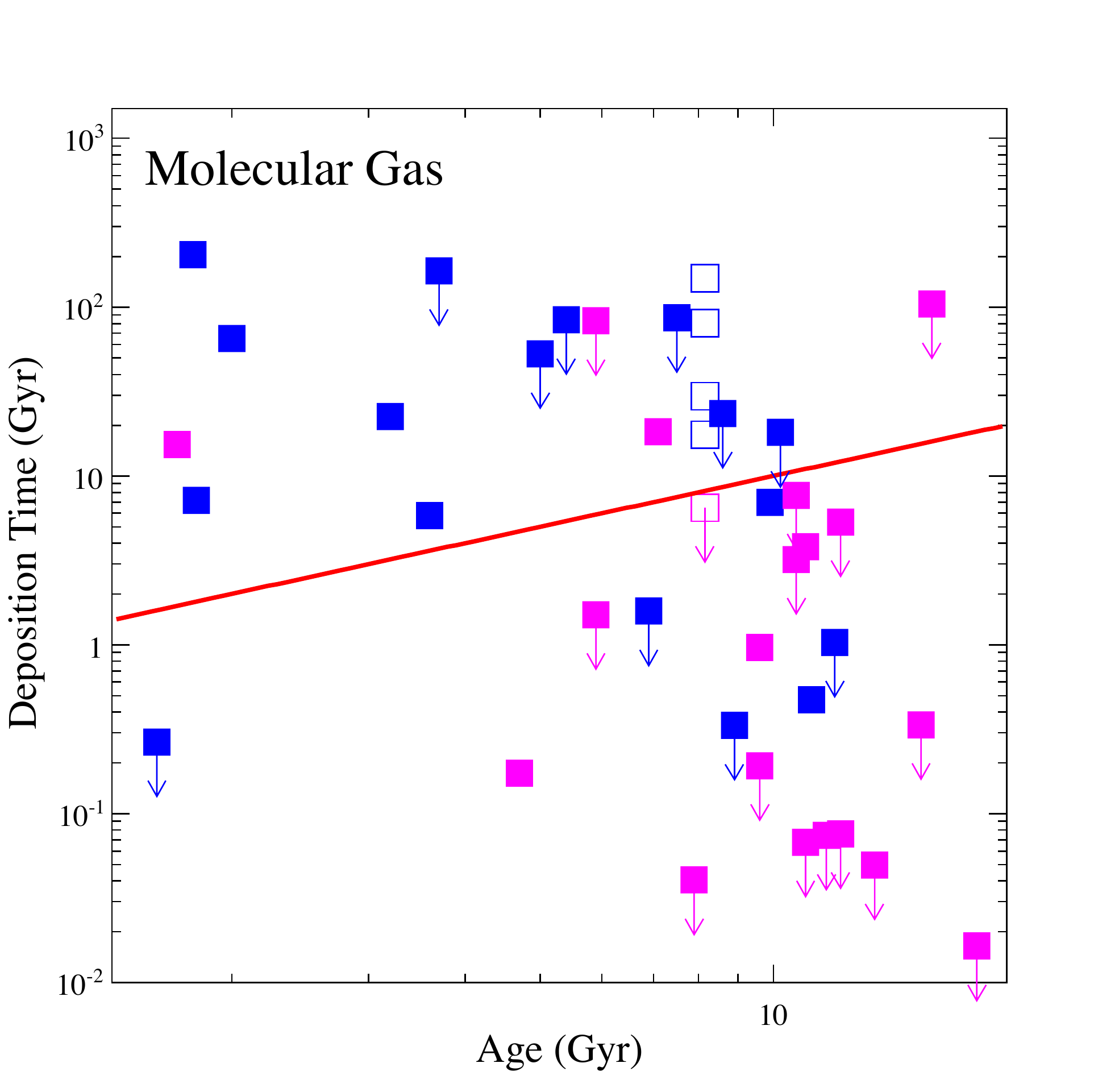}
\figcaption{\label{fig:timeage} Time needed to form the observed amount of molecular gas from radiative cooling of hot gas as a function of stellar age for field galaxies (blue) and galaxies in groups and clusters (magenta). The solid red line indicates the equality between these two time scales. 
We use the average age (8.15 Gyr) for galaxies that do not have stellar age available in the literature and are marked by open symbols. We use the average temperature (0.56 keV) for galaxies that do not have a constrained ISM temperature measurement and are marked by open symbols.
[{\sl see the electronic edition of the journal for a color version of this figure.}]}
\end{figure}

In this work we find that the hot gas content of population B galaxies is significantly correlated with their neutral gas, particularly molecular gas mass. 
If the radiative cooling of hot gas is the main supplier of the cold gas, 
this would naturally explain why
the cold gas content is observed to increase with the hot gas content.
In order to probe this scenario, we calculated the classical mass deposition rate for galaxies in our sample: $\dot{M}_c = 2\mu {\rm m_p} L_X/5 kT$ (David et al.\ 2014). Then, we obtained the time required for each galaxy to form their cold gas contents through this deposition. In Figure~\ref{fig:timeage}, we compared the time needed to form the observed amount of molecular gas and the stellar age of each galaxy. 
There is not enough time for field galaxies in our sample to have their cold gas supplied by hot gas cooling. Their cold gas may be acquired externally from mergers/accretions. 
However, it is possible for galaxies in groups and clusters to have most of their cold gas supplied by hot gas cooling. Sun et al.\ (2007) observed a connection between the hot coronae and the radio emission in cluster galaxies, which suggests that radiative cooling of the coronal gas may fuel the central black holes in galaxies in a cluster environment where galactic cold gas is difficult to acquire.

Su \& Irwin (2013) demonstrated an anti-correlation between molecular gas mass and the hot gas metallicity. 
One explanation for this trend would be that hot ISM is diluted by mixing with less metal-enriched molecular gas. This process would also add more hot gas to the ISM and lead to the $\Delta L_{X_{\rm gas}}$--$M_{\rm H_2}$ correlation observed in this study. 
This is only effective in population B galaxies. Small mass systems have a smaller hot gas halo and they are easier to dilute.
Galaxies in the field may have acquired cold gas through external processes such as {\sl recent} mergers; these mergers would increase hot gas content. In contrast, galaxies in groups and clusters are more likely to have molecular gas supplied by stellar-mass loss or condensation from hot gas and their hot gas content may also be influenced by environmental factors such as ram pressure stripping. 

\subsubsection{Star formation efficiency}

\begin{figure*} 
   \begin{center}
    \leavevmode
\epsscale{1.11}
\plottwo{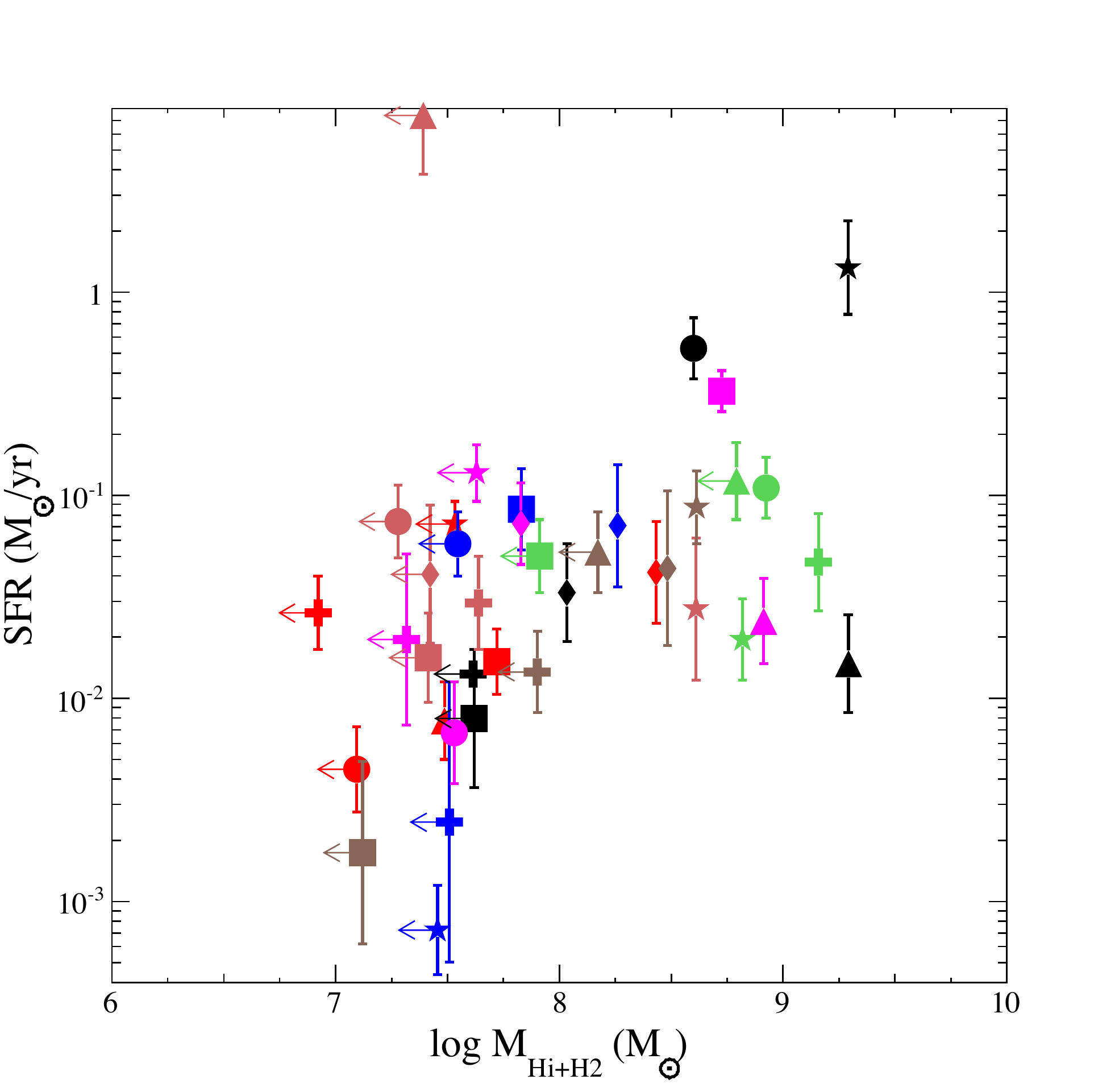}{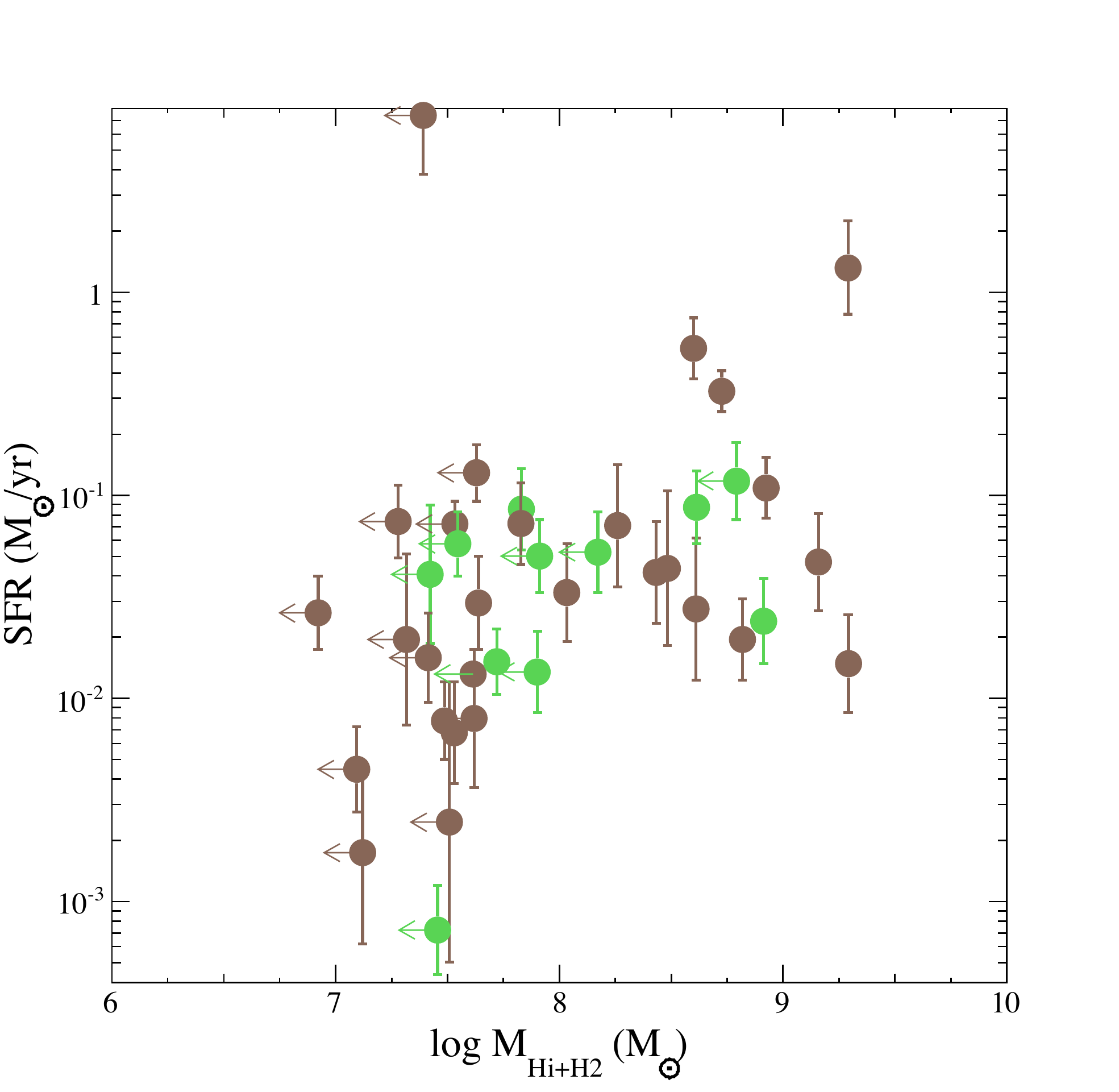}
\plottwo{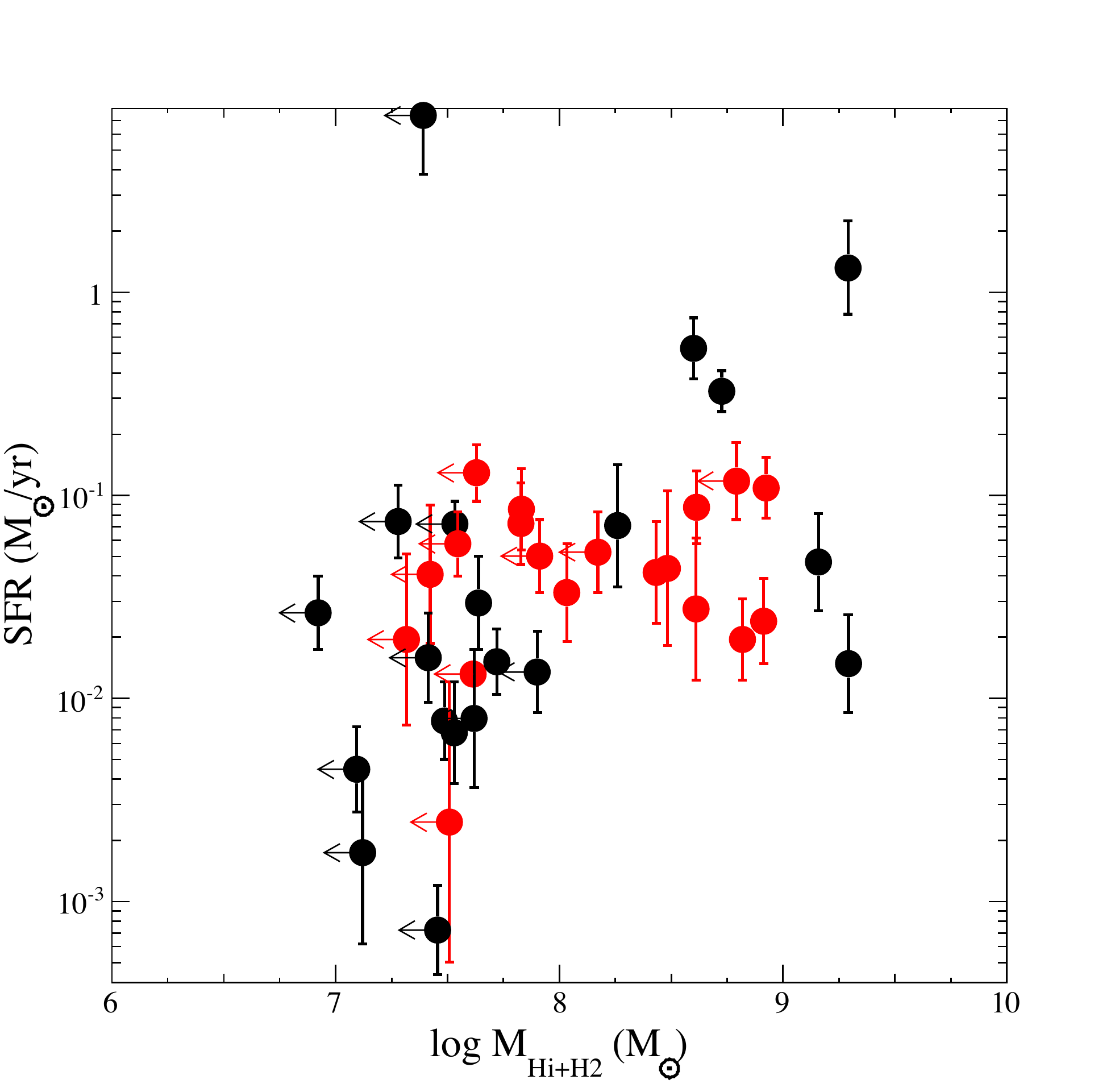}{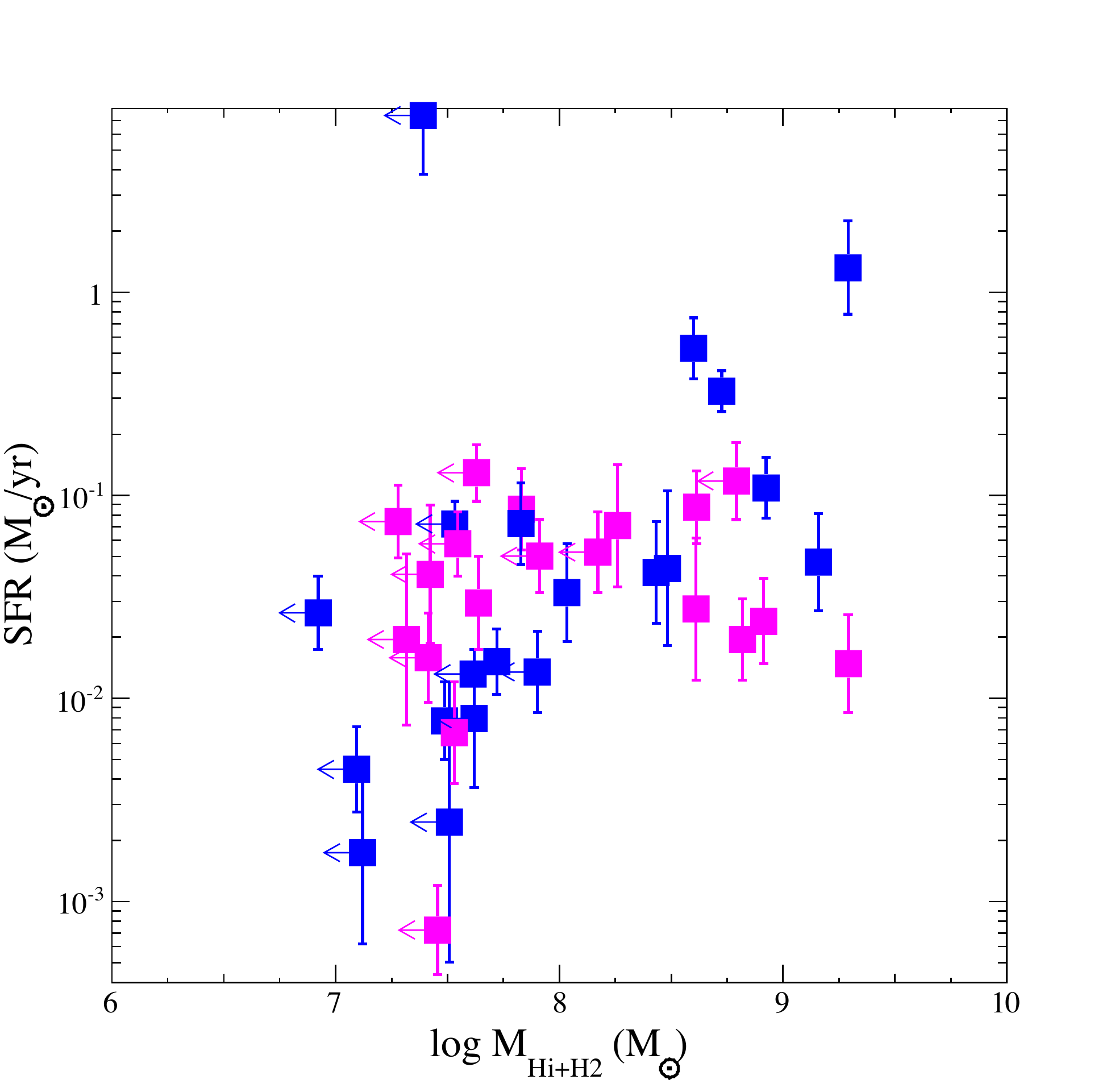}
\caption{Star formation rate as a function of cold gas mass.  {\it top-left}: color code is the same as in Figure~\ref{fig:lxk}. {\it top-right}: ``true lenticular" galaxies (brown) and ``true elliptical" galaxies (green). {\it bottom-left}: high mass galaxies (red) and low mass galaxies (black). {\it bottom-right}: field galaxies (blue) and galaxies in groups and clusters (magenta).}
 \label{fig:coldsfr} 
  \end{center}
\end{figure*}

Star formation is observed to qualitatively correlate with its fuel, the cold gas. However, the exact process of star formation is still unclear. Various factors, such as environment, depth of potential well, metallicity, and AGN feedback, may have played noticeable roles in regulating the star formation. The star formation efficiency (the ratio of SFR to cold gas mass) can vary dramatically among galaxies. In Figure~\ref{fig:coldsfr}, we show the relation between SFR and cold gas mass for galaxies in our sample. SFR increases with cold gas mass for population B galaxies, while population A galaxies tend to have constant SFR regardless of their cold gas mass. Various mechanisms may have interfered with the star formation process in population A galaxies. On the positive side, the ICM confinement may increase the gas density of cluster galaxies, which would boost star formation.
Meanwhile, cluster galaxies may have their cold gas removed quickly through ram pressure stripping or heat conduction and appear to have a large SFR per unit gas. Moreover, star formation efficiency is found to be positively correlated with gas metallicity (Yates \& Kauffmann 2014). We demonstrated that cold gas in cluster galaxies and field galaxies may have different origins. Cold gas in cluster galaxies, formed through hot gas cooling, is probably contaminated, while cold gas in field galaxies, acquired externally, is more likely to be pristine. This could enhance the star formation efficiency in cluster galaxies.
On the negative side, Davis et al.\ (2014) demonstrated that the depletion time of cold gas positively correlates with the mean shear rate (A$/\Omega$; A is the first Oort constant and $\Omega$ is the angular velocity). 
``True elliptical" galaxies rotate slower (high shear rates) and thereby may take longer time to form stars. 
%galaxies with a fast rising rotation curve tend to have a shorter gas depletion time and form stars more efficiently. Note that the derivative of the galaxy rotation curve is related to angular velocity ($\Omega$): $$\Omega=2A+dV/dR,$$ where $V$ is circular velocity, $R$ is radius, and A is the first Oort constant. ``True elliptical" galaxies rotate slower and thereby take longer time to form stars. 
Overall, the expected SFR and cold gas mass relation may have been washed out for population A galaxies due to various mechanisms.

\subsection{Galaxy masses}

Galaxy masses, dominated by dark matter halos, have been regarded as the most 
crucial factor in regulating
the hot gas content of ETGs. 
Galactic winds supply gas to the hot gas halo; the explosion of supernovae provides energy. These processes would make massive galaxies brighter by adding more hot gas. However, these same processes may instead make low mass galaxies fainter if the ejected gas becomes energized enough to escape the galaxy due to its shallower gravitational potential well. 
Not only strongly motivated in theory, the role of dark matter halos in retaining hot gas has been supported by observations of ETGs. Kim \& Fabbiano (2013) demonstrated that $L_{X_{\rm gas}}$ is highly correlated with the total mass for a small sample of ETGs, although the absolute values of $L_{X_{\rm gas}}$ is not our primary interest. More relevantly, Mathews et al.\ (2006) found that $L_X/L_K$ and even $\Delta L_{X_{\rm gas}}$ by their definition (which is similar to $\Delta L_{X_{\rm gas}}$ defined in our work) are correlated with their total masses. 
Galaxies in their sample are mostly bright elliptical galaxies with $L_K > 2\times10^{11}~L_{K\odot}$ and $L_X > 3\times10^{40}$ erg\,s$^{-1}$, with most residing at group centers. 

We do not observe a significant relation between $\Delta L_{X_{\rm gas}}$  and $M_{\rm tot}$ for galaxies in our sample (Figure~\ref{fig:mtot}). 
The total mass used in Mathews et al.\ (2006) is the X-ray hydrostatic mass derived under the assumption of hydrostatic equilibrium and spherical symmetry, extending out to fairly large radii. 
The total mass used by Kim \& Fabbiano (2013) is taken from Deason et al.\ (2012) which is the dynamical mass within  $5\,r_{\rm e}$.
Both of these are very good proxies for dark matter mass.
Unfortunately, the X-ray technique requires a large number of counts and can be very observationally expensive, in particular for relatively faint galaxies. The total mass estimated through stellar dynamics, extending to large radii, is not readily available since 
accurate mass estimates from stellar dynamics are generally restricted to the very central regions of galaxies.
The total mass used in this study is measured within 1\,$r_{\rm e}$ by the ATLAS$^{\rm 3D}$ survey, so this total mass is mostly made up of stellar mass and can not fairly reflect dark matter content. 
In order to extend the total mass within 1\,$r_{\rm e}$ out to 5\,$r_{\rm e}$, we parameterize the dynamical mass of 11 ATLAS$^{\rm 3D}$ galaxies studied by Deason et al.\ (2012) out to 5\,$r_{\rm e}$ as a function of their total mass within 1\,$r_{\rm e}$, as estimated in the ATLAS$^{\rm 3D}$ survey.
Applying this parameterization to all galaxies in our sample
we found that 
the correlation between this scaled mass and $\Delta L_{X_{\rm gas}}$ is still not significant.
This does not rule out the role of a dark matter halo in regulating the hot gas content in ETGs.
In some sense, hot gas temperature may serve as a proxy of potential well and we did observe that $\Delta L_{X_{\rm gas}}$ increases with the ISM temperature in spite of the larger scatter.

The lack of significant correlation may also be related to a complicated feedback process, in particular related to AGN activities. 
It has been well established that the bulge mass of a galaxy is highly correlated with the mass of the central supermassive black hole (e.g.\  Bentz et al.\ 2008). Thus, more massive galaxies may have more active and stronger AGN outbursts, leading to stronger gaseous outflow. As a result, the gas content of more massive galaxies cannot be simply correlated with the depth of its gravitational potential.

\subsection{Intrinsic dynamics}

Our results demonstrate an anti-correlation between the residuals $\Delta L_{X_{\rm gas}}$ and some intrinsic kinematic factors such as rotation $\lambda$, ellipticity $\epsilon$, and the parameter $\lambda/\sqrt{\epsilon}$ for population A galaxies (see Figures~\ref{fig:lxk3}, \ref{fig:rot} and \ref{fig:flat}). 
Such an empirical pattern has been noted by previous studies (e.g.\  Sarzi et al.\ 2010, 2013). Rotation could have taken away part of the kinetic energy of stellar motions and slowed down thermalization processes. 
Sarzi et al.\ (2010) found that the hot gas temperature of slow rotators met the stellar-thermalization kinetic energy expectation while fast rotators fell short.
Fast rotating systems are able to expel hot gas to larger radii which leads to a more diffusely distributed halo (Negri et al.\ 2014). 
Meanwhile, it has been noted since the time of the {\sl Einstein} that elliptical galaxies have larger $L_X/L_{\rm opt}$ than lenticular galaxies (Eskridge et al.\ 2005).
Independent of the galaxy kinematical support, and apart from any evolutionary phase, flattening alone would reduce the binding energy of galaxies of a given total mass. 
Yet two correlations, $\Delta L_{X_{\rm gas}}$-- rotation and $\Delta L_{X_{\rm gas}}$-- ellipticity, are likely to be byproducts of one another. This is because rotation and flatness are inherently related. Rotational systems eventually become flatter along the major axis while flatter systems are able to possess higher rational support (Binney \& Tremaine 1987).
We find a significant anti-correlation between $\Delta L_{X_{\rm gas}}$ and ellipticity for slow-rotating galaxies, which favors the role of flattening.
Sarzi et al.\ (2013) also found that a few $L_X$-deficient slow rotating galaxies appear to be flat. 
However,
through two-dimensional simulations, Negri et al.\ (2013) demonstrated that flat galaxies and round galaxies can contain the same amount of hot gas if they have the same rotation, and a fast-rotating flat galaxy could also contain less hot gas than a slowly-rotating flat galaxy, which favors the role of rotation. 
Interestingly, we find that both rotation and flattening are only relevant for population A galaxies. 
This mass-dependence has been noted by some simulation work (Negri et al.\ 2014, Posacki et al.\ 2013). More massive galaxies are more sensitive to flattening and rotation.

\subsection {Ram pressure stripping and stellar-mass loss}

In this work we investigated whether ram pressure stripping can account for the
observed scatter in $L_{X_{\rm gas}}/L_K$.
We compare the hot gas content per stellar mass ($L_{X_{\rm gas}}/L_K$ as well as $\Delta L_{X_{\rm gas}}$) and the strength of the instantaneous ram pressure relative to the gravitational restoring pressure ($P_{\rm ram}/P_{\rm grav}$), for a subsample of the 15 ETGs in the Virgo Cluster in Figure~\ref{fig:p}. We observed an apparent correlation with a correlation coefficient of $\rho=-0.343\pm0.073$, reflecting the role of ram pressure stripping.

Another important stripping process is due to the Kelvin-Helmholtz instability, which allows 
a shear force to be generated as intracluster gas passes by the outer edge of the galaxy;
this shear force allows material to be continually stripped from the galaxy
(Close et al.\ 2013). 
It is referred to as turbulence viscous stripping by some studies, as one of the transport processes (Nulsen 1982) which may 
cause substantial stripping of cluster galaxies at a stripping rate sometimes even greater than that due to instantaneous stripping alone.

 \begin{figure} 
\epsscale{1.2}
\plotone{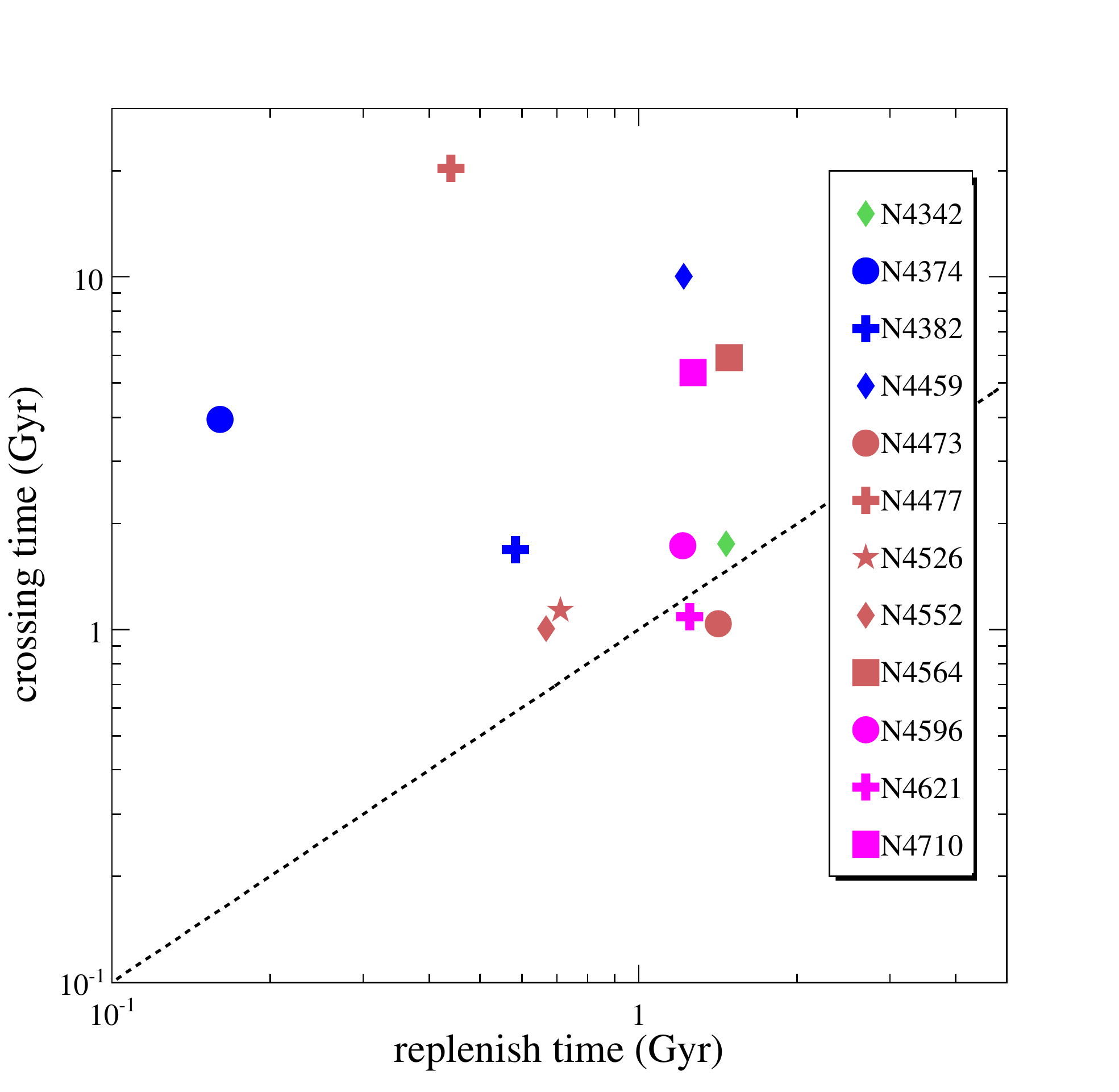}
\figcaption{\label{fig:time} Crossing times as a function of times since gas removal for faint galaxies to reach average $L_X/L_K$ through replenishment to new stellar-mass loss. The dashed black line indicates the equality between these two time scales. [{\sl see the electronic edition of the journal for a color version of this figure.}]}
\end{figure}

Galaxies move through the ICM in roughly elliptical orbits with one focus on the cluster center.
As a galaxy approaches its pericenter, entering
the inner (and denser) region of the cluster, it gains velocity towards the bottom of the gravitational potential well of the cluster. During this time it experiences stronger ram pressure stripping such that it has its hot gas stripped instantaneously. 
When approaching its apocenter, reaching the larger radii of the cluster with a smaller velocity,  
it experiences weaker ram pressure stripping and stops losing gas. During this time, 
gas generated from new stellar-mass loss may be able to rebuild the atmosphere.
While orbiting within the cluster,
galaxies go through gas mass loss and gas mass replenishment phases alternately. 
NGC~4649 is an X-ray luminous galaxy in the Virgo Cluster that is not the center of a subgroup. 
We take NGC~4649 as a typical X-ray luminous early-type galaxy that has a benchmark $L_{X_{\rm gas}}/L_K$. 
Twelve Virgo galaxies in this sample that have smaller $M_{X_{\rm gas}}/L_K$ than NGC~4649 
may have lost some of their hot gas due to ram pressure.
We estimate the time needed for these X-ray sub-luminous galaxies to gain enough gas to reach the average $M_{X_{\rm gas}}/L_K$ through stellar-mass loss.
We calculate their crossing time using the radial velocity and the virial radius ($t_{\rm cr}={r_{\rm vir}}/{v_{\rm rad}}$). Figure~\ref{fig:time} compares the time required for these X-ray sub-luminous galaxies to recover their gas mass with their crossing time. 
Ten out of 12 galaxies have a replenishment time shorter than their crossing time. 
These two time scales are comparable for the other two galaxies. 
This implies that most X-ray sub-luminous galaxies may regain their gas mass and become X-ray luminous galaxies at some point.

\section{\bf Summary}

This work presents our investigation of 42 ETGs in the ATLAS$^{\rm 3D}$ survey that have sufficiently deep {\sl Chandra} observations to constrain their hot gas properties.
In order to find the origin of the large scatter in the $L_X-L_K$ relation of ETGs, 
we measure their hot gas content and relate it to other observables: stellar age, total galaxy mass, effective radius, depth of gravitational potential well, hot gas temperature, stellar velocity dispersion, atomic gas mass, molecular gas mass, cold gas mass, star formation rate, angular momentum $\lambda$, ellipticity $\epsilon$,  $\lambda/\sqrt{\epsilon}$, environmental density, and relative strength of ram pressure (only for 15 Virgo galaxies). These summarize our main conclusions:

$\bullet$ We obtain a scatter of up to a factor of 1000 in the $L_{X_{\rm gas}}$--$L_K$ relation for galaxies in our sample, consistent with the results of previous studies.
The scatter in the $M_{X_{\rm gas}}$--$L_K$ relation is slightly smaller than in the 
$L_{X_{\rm gas}}$--$L_K$ relation.

$\bullet$ We find two populations of early-type galaxies:
galaxies with larger $L_K$ are generally massive, slow-rotating, hot, big, and they tend to reside in high galaxy density environments, while galaxies 
with a smaller $L_K$ tend to be low mass, fast-rotating, cool, and small galaxies, which tend to reside in the low galaxy density environments.

$\bullet$ The hot gas residuals ($\Delta L_{X_{\rm gas}}$) are most strongly correlated with the (rather low) star formation rates  and hot gas temperatures; this indicates that stellar feedback in early-type galaxies heats up the gaseous component into the X-ray emitting phase and these galaxies are massive enough to keep such hot gas bound. Alternatively, both low level star formation and enhanced X-ray luminosities may be consequences
of recent wet minor mergers.  

$\bullet$ We find that star formation rate increases with cold gas mass for low-mass, fast-rotating, and field galaxies, while massive, slow-rotating, and cluster galaxies tend to have constant SFR regardless of their cold gas mass. Various mechanisms may have interfered with the star formation process in the latter population. 

$\bullet$ Early-type galaxies which contain significant amounts of cold gas, especially molecular gas, tend to have more hot gas content. However, in our sample this trend only applies to ``true-lenticular" galaxies, fast-rotating galaxies, low mass galaxies, and/or galaxies in low density environments. This could result from hot gas in the ISM mixing with the molecular gas.

$\bullet$ Slower-rotating galaxies and rounder galaxies tend to have larger hot gas content. This trend only applies to slow-rotating galaxies, high mass galaxies, and/or galaxies in groups and clusters. Flatness may play a bigger role than rotation.

$\bullet$ While cold gas in cluster galaxies may be formed through hot gas cooling or stellar mass loss, the deposition time of the hot gas is longer than the stellar age of field galaxies. Cold gas in field galaxies is unlikely to be supplied by the radiative cooling of hot gas. Adding cold gas mass to $M_{X_{\rm gas}}$, the scatter in the $M_{X_{\rm gas}}$--$L_K$ relation for galaxies in this sample becomes even larger. This further suggests that the cold gas may be supplied by external mergers/accretions. However, the total gas mass $M_{\rm gas}$ increases almost linearly with $L_K$, which is exactly what we expect if stellar-mass loss is the primary source of all gaseous components.

$\bullet$ For the 15 galaxies in our sample that are members of the Virgo Cluster, 
we established a correlation between the deficit of their hot gas content and the
strength of ram pressure, measured by the ratio of the instantaneous ram pressure to the gravitational restoring pressure. We demonstrate that stellar-mass loss is able to replenish stripped gas within the cluster crossing time for most galaxies. 

We conclude that the hot gas content per stellar light for a galaxy is the result of various mechanisms and feedback mechanisms. The situation varies for different galaxies with different evolutionary histories and environments.

\section{Acknowledgments}
We thank the anonymous referee for constructive suggestions.
We are grateful to Liyi Gu, Greg Stinson, David Buote, and James Bullock for helpful discussions. We thank Zhiyuan Li and  Fabrizio Brighenti for reading an early draft and valuable comments. 
To all of the associated PIs, we would like to express our appreciation for the availability of the {\sl Chandra} observations that we use in this work.
We gratefully acknowledge partial support from SAO/NASA {\sl Chandra/XMM} grant GO2-13163X.
This publication makes use of data products from the Two Micron All Sky Survey, which is a joint project of the University of Massachusetts and the Infrared Processing and Analysis Center/California Institute of Technology, funded by the National Aeronautics and Space Administration and the National Science Foundation. Funding for the SDSS and SDSS-II has been provided by the Alfred P. Sloan Foundation, the Participating Institutions, the National Science Foundation, the U.S. Department of Energy, the National Aeronautics and Space Administration, the Japanese Monbukagakusho, the Max Planck Society, and the Higher Education Funding Council for England. The SDSS Web Site is {\url http://www.sdss.org/}.
The SDSS is managed by the Astrophysical Research Consortium for the Participating Institutions.

\appendix

\section{\bf A. Additional Figures}
\renewcommand{\thefigure}{A\arabic{figure}}
\setcounter{figure}{0}

%\newpage
%\clearpage
 \begin{figure} 
\epsscale{1.11}
\plottwo{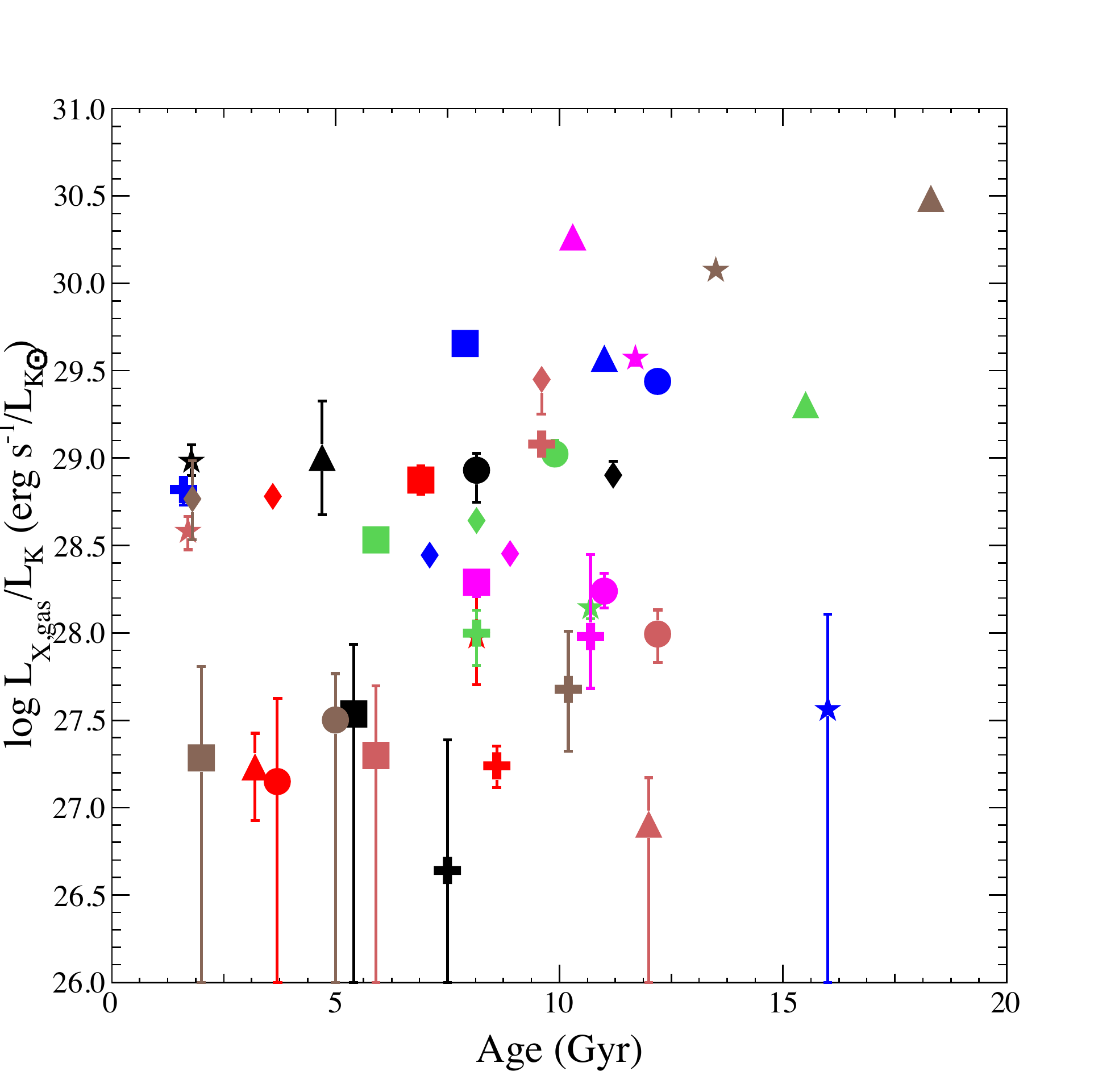}{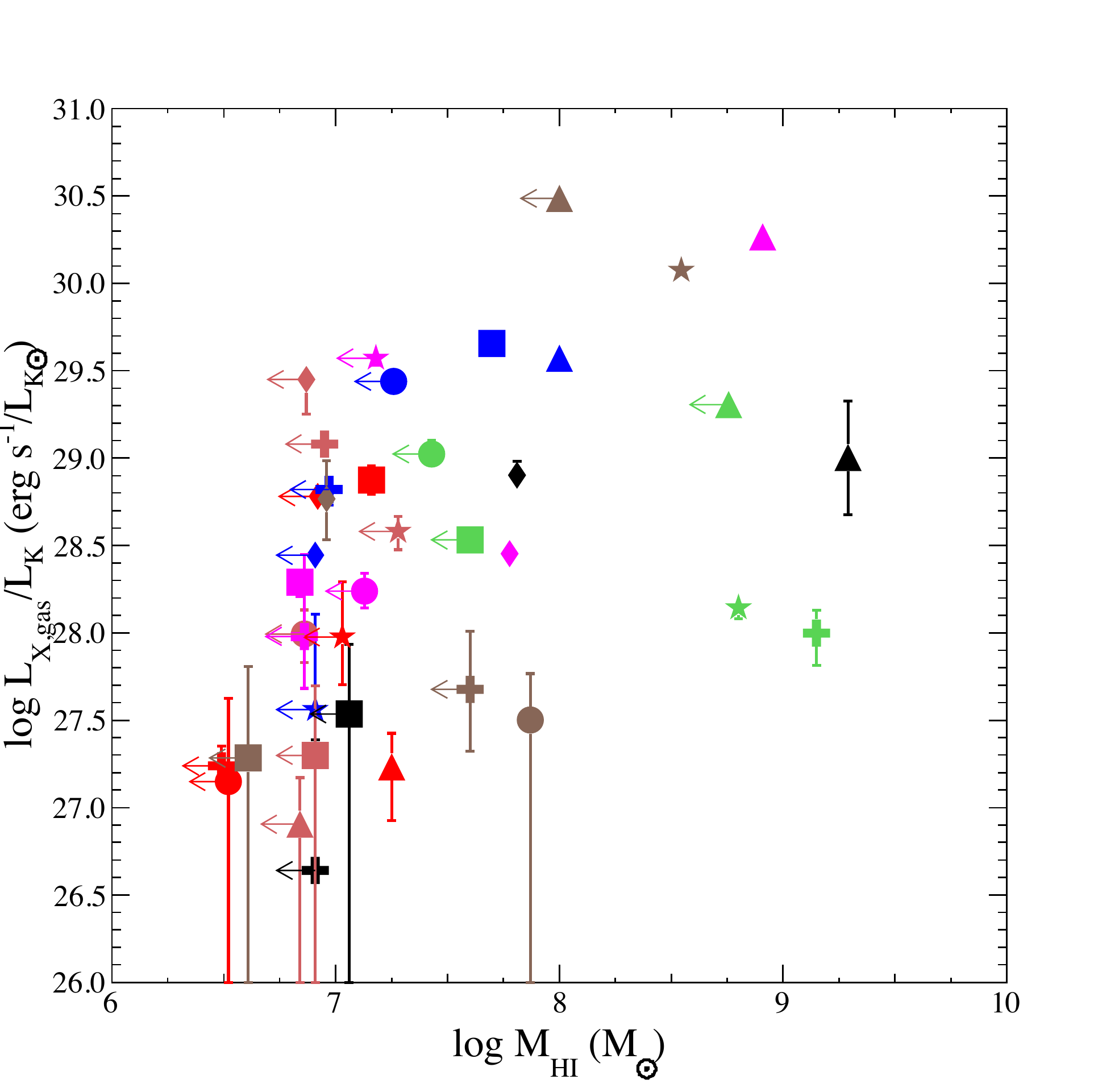}
\plottwo{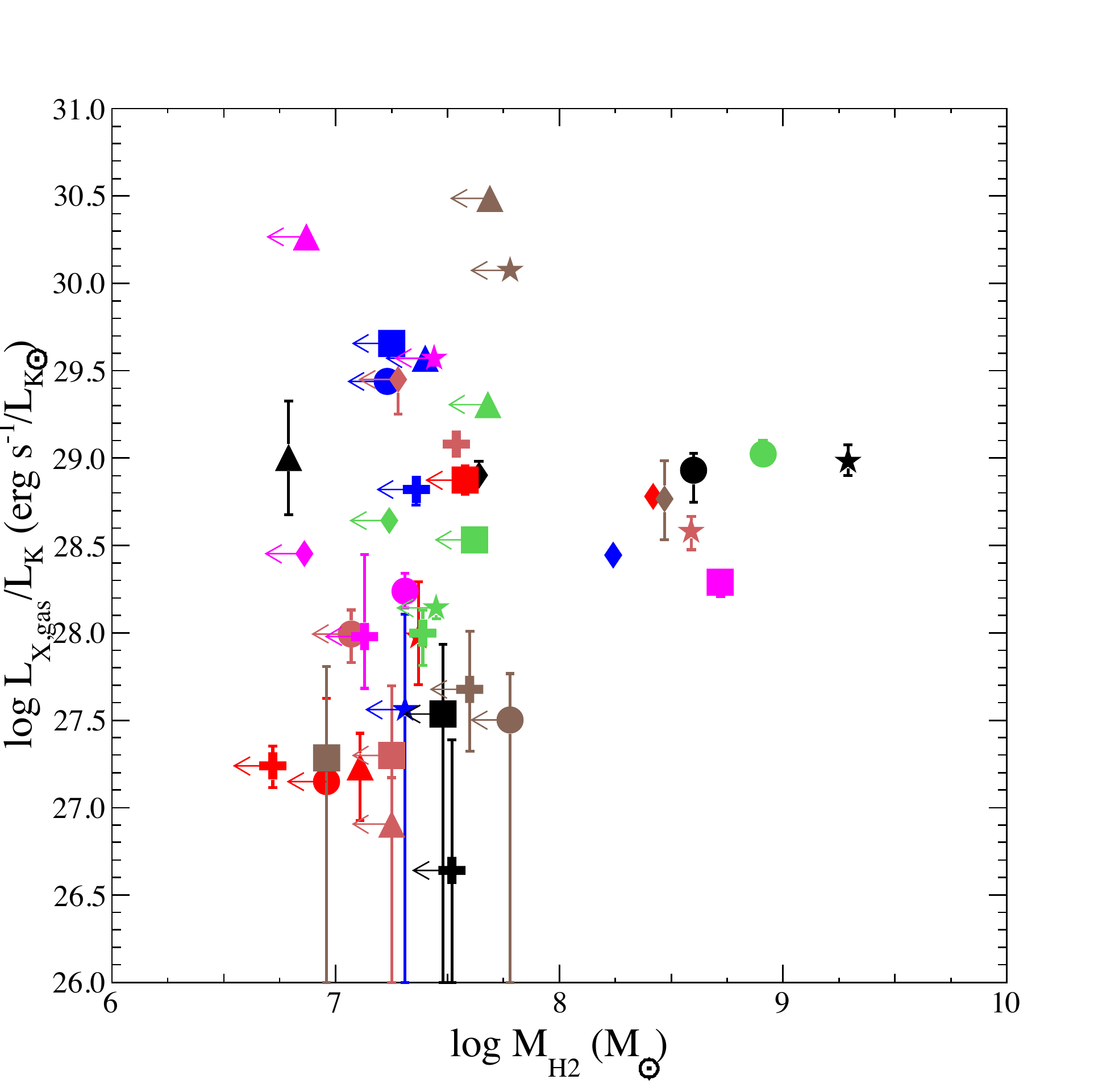}{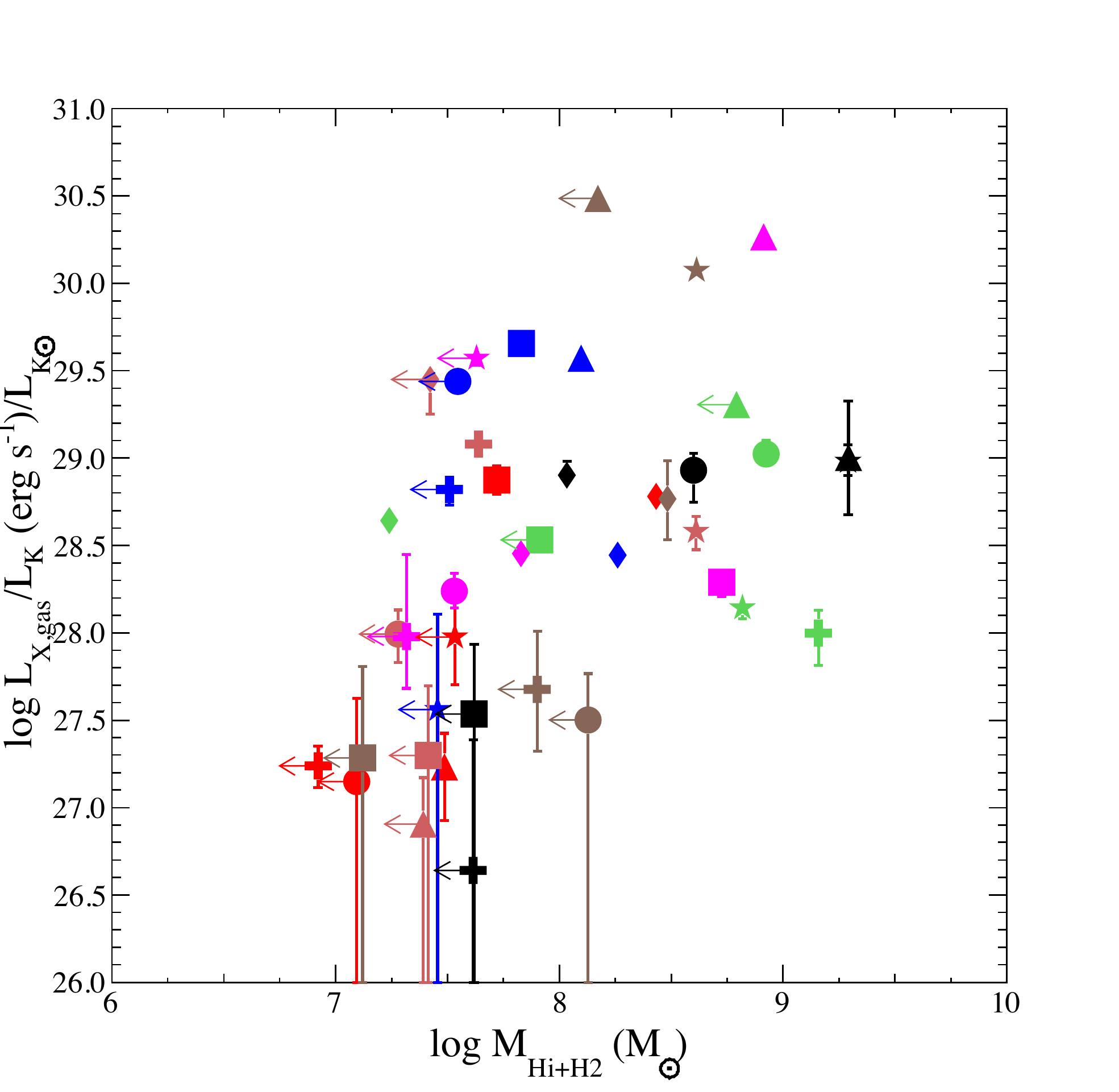}
\figcaption{\label{fig:lxk1} $L_{X_{\rm gas}}/L_K$ as a function of stellar age ({\it top-left}), atomic gas mass ({\it top-right}), molecular gas mass ({\it bottom-left}), and cold gas mass ({\it bottom-right}). Color code is the same as in Figure~\ref{fig:lxk}. }
\end{figure}

 \begin{figure} 
\epsscale{1.11}
\plottwo{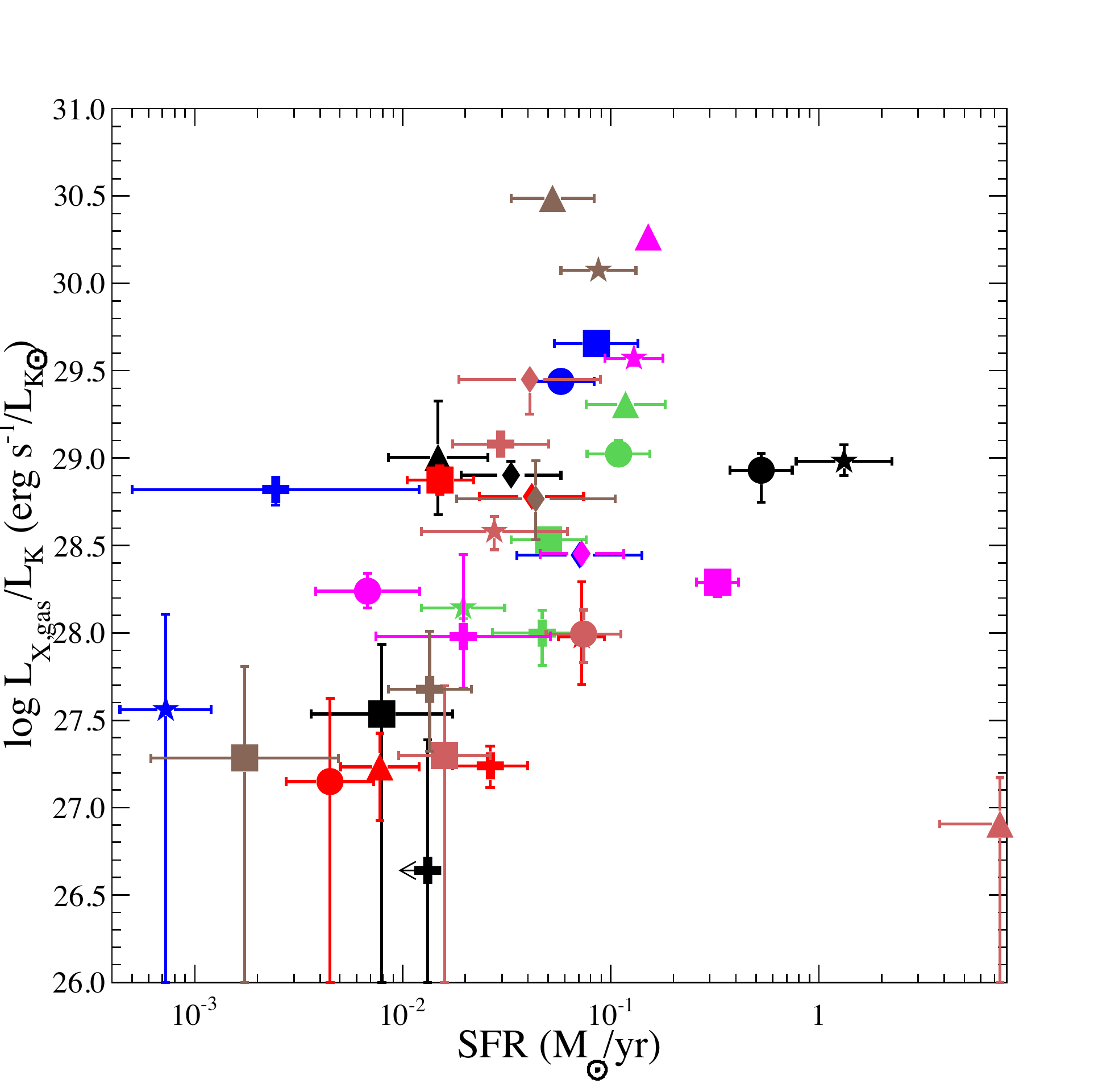}{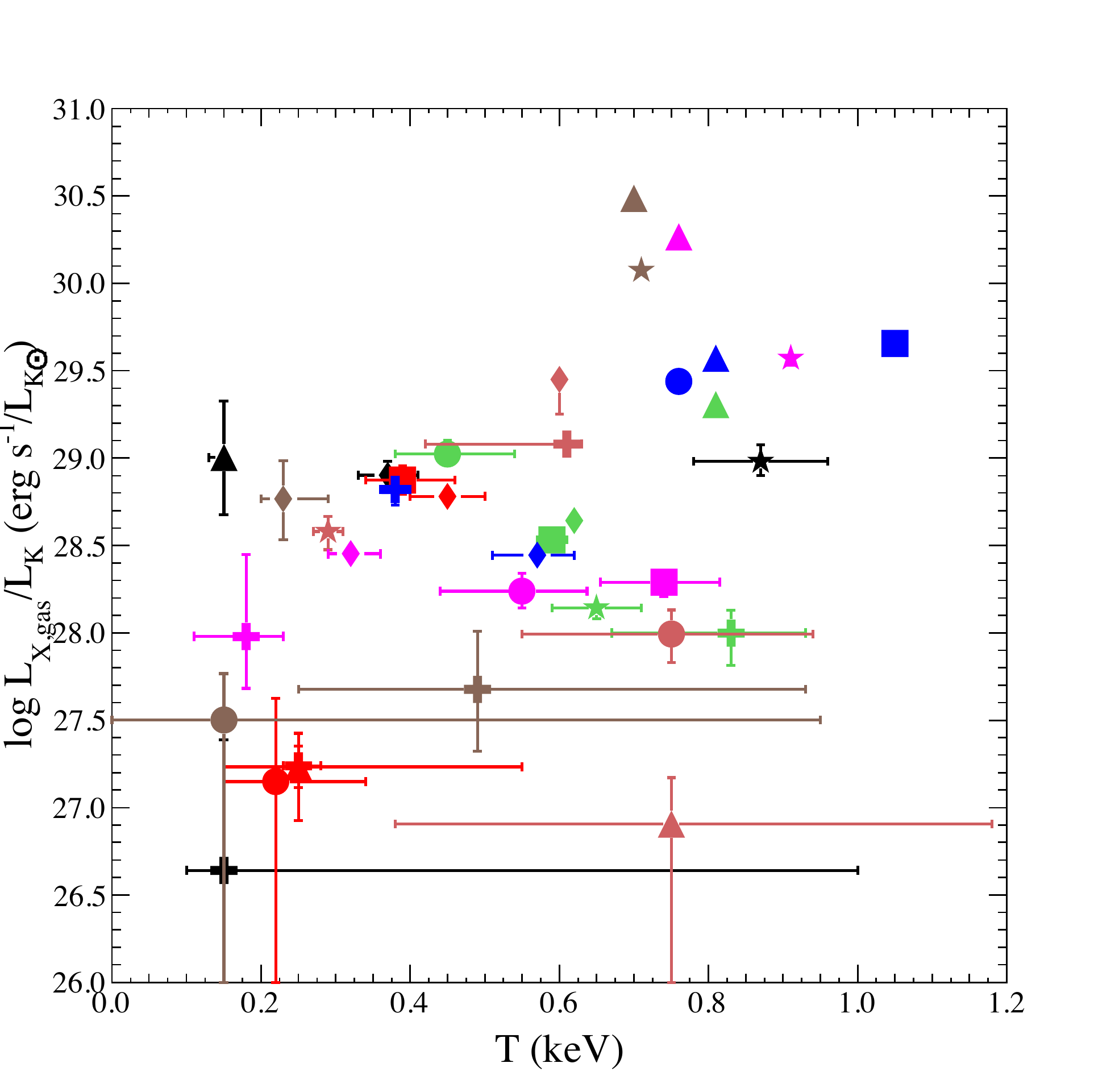}
\plottwo{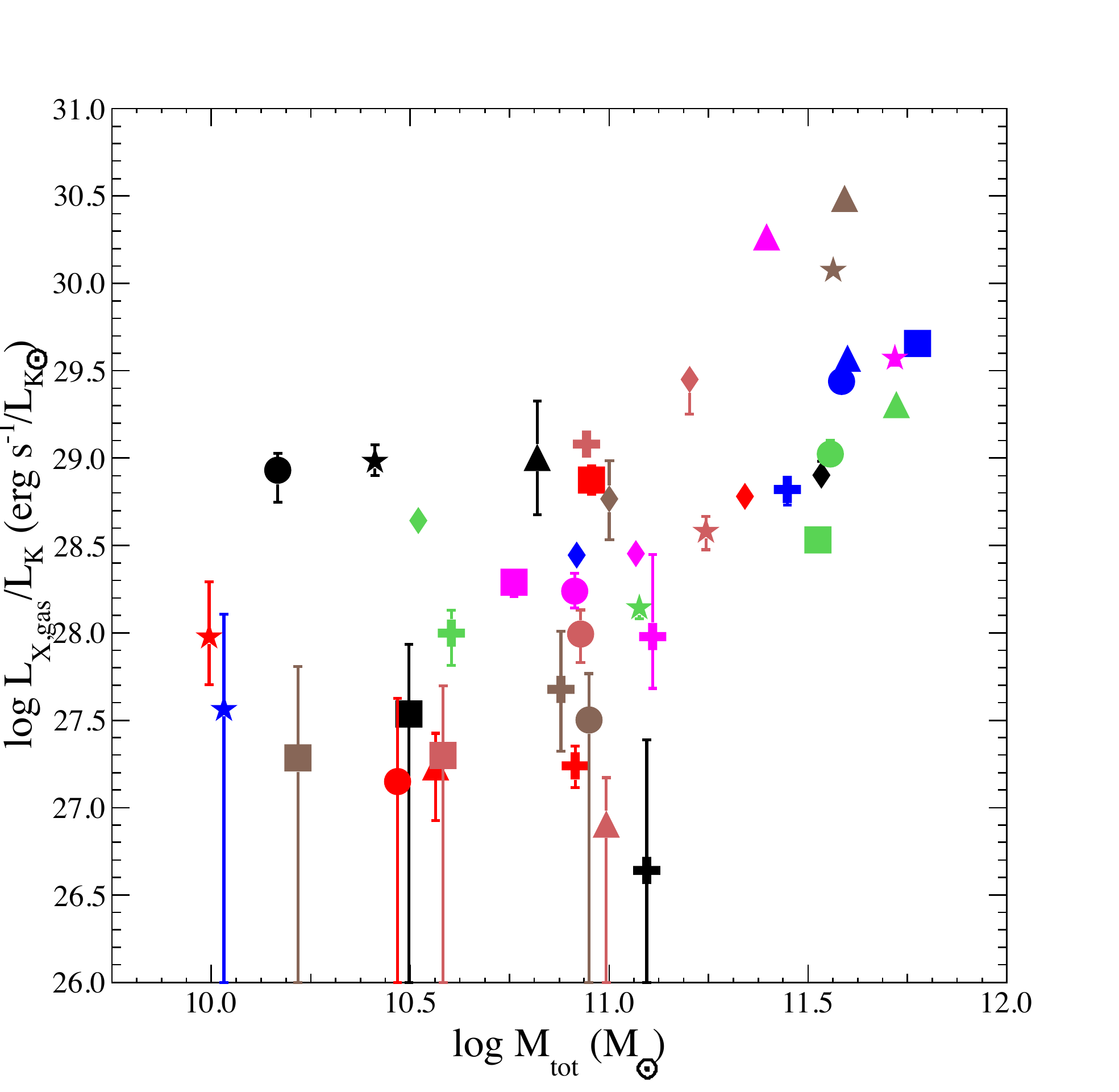}{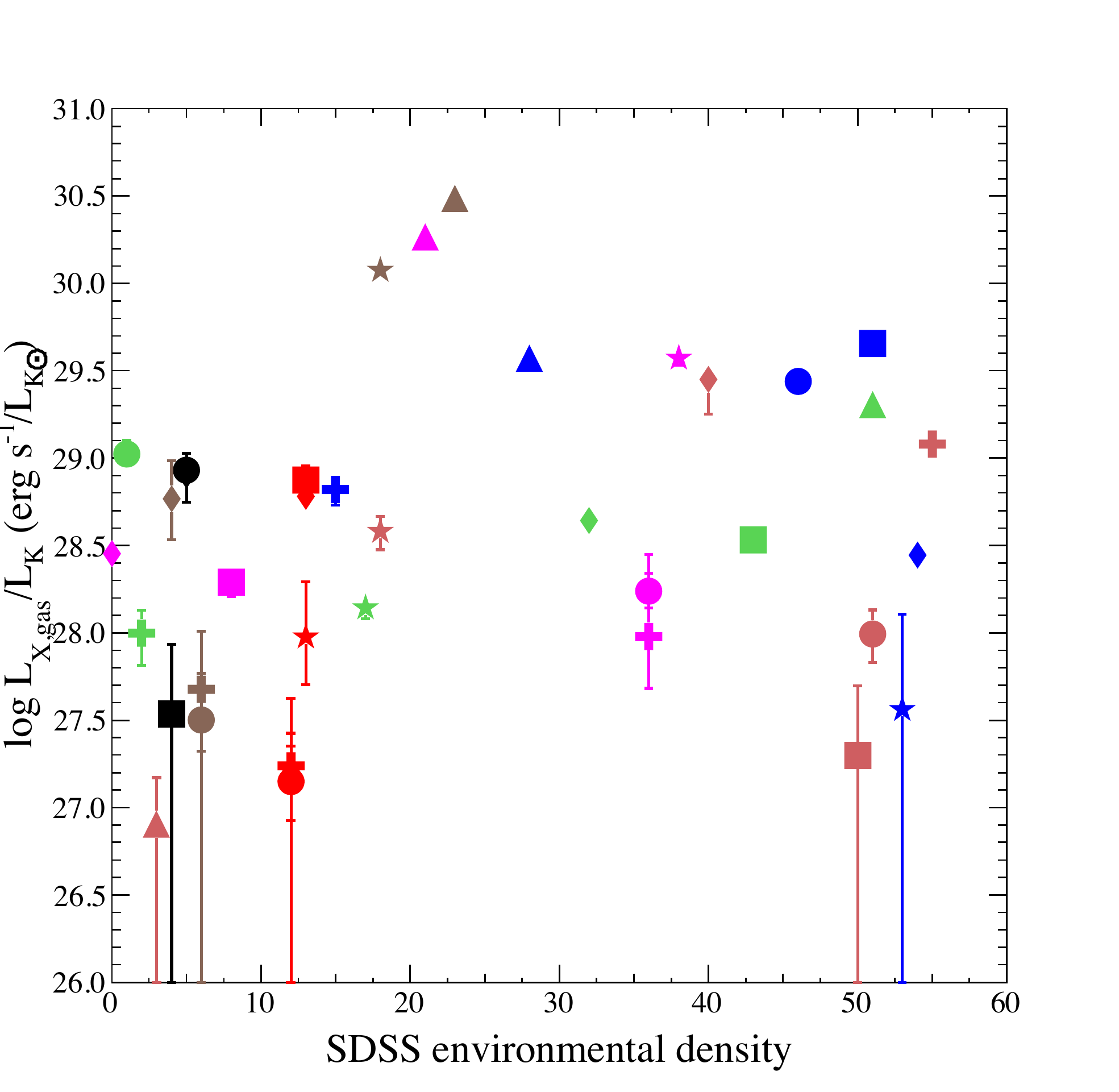}
\figcaption{\label{fig:lxk2} $L_{X_{\rm gas}}/L_K$ as a function of star formation rate ({\it top-left}), hot gas temperature ({\it top-right}), total mass ({\it bottom-left}), and environmental galaxy density ({\it bottom-right}). Color code is the same as in Figure~\ref{fig:lxk}. }
\end{figure}

 \begin{figure} 
\epsscale{1.11}
\plottwo{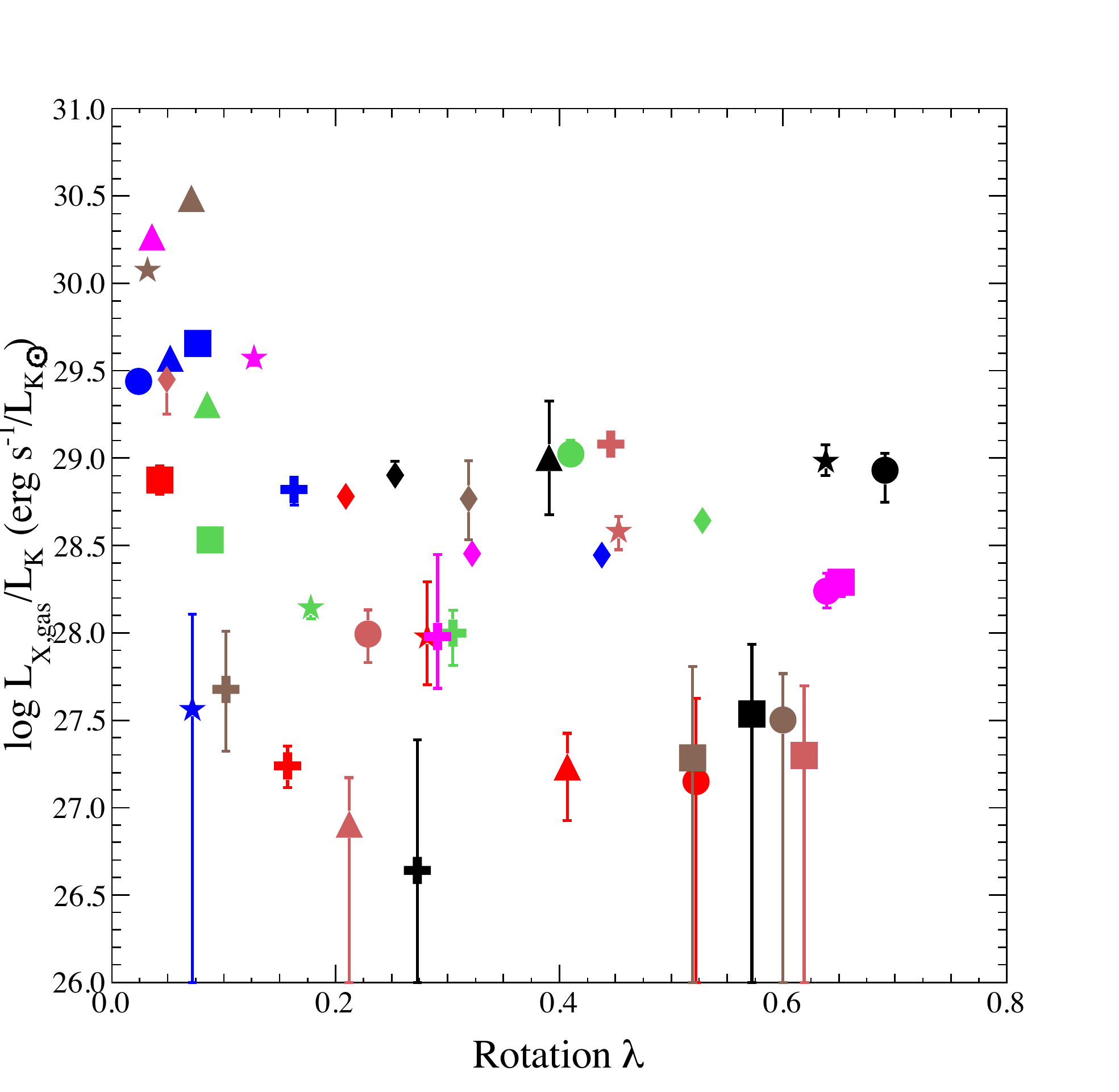}{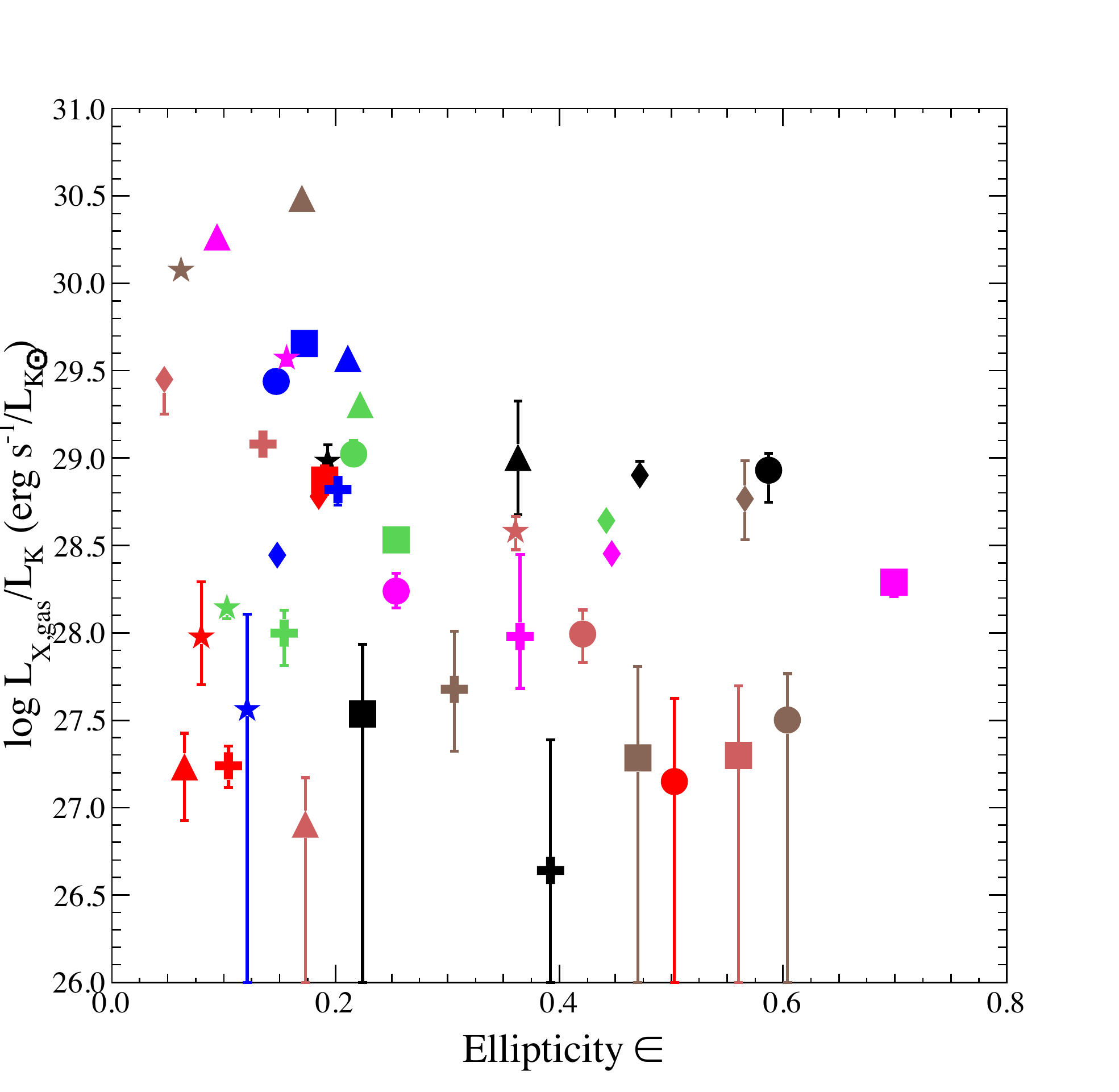}
\figcaption{\label{fig:lxk3} $L_{X_{\rm gas}}/L_K$ as a function of rotation ({\it left}) and ellipticity ({\it right}). Color code is the same as in Figure~\ref{fig:lxk}. }
\end{figure}

 \begin{figure} 
\epsscale{1.11}
\plottwo{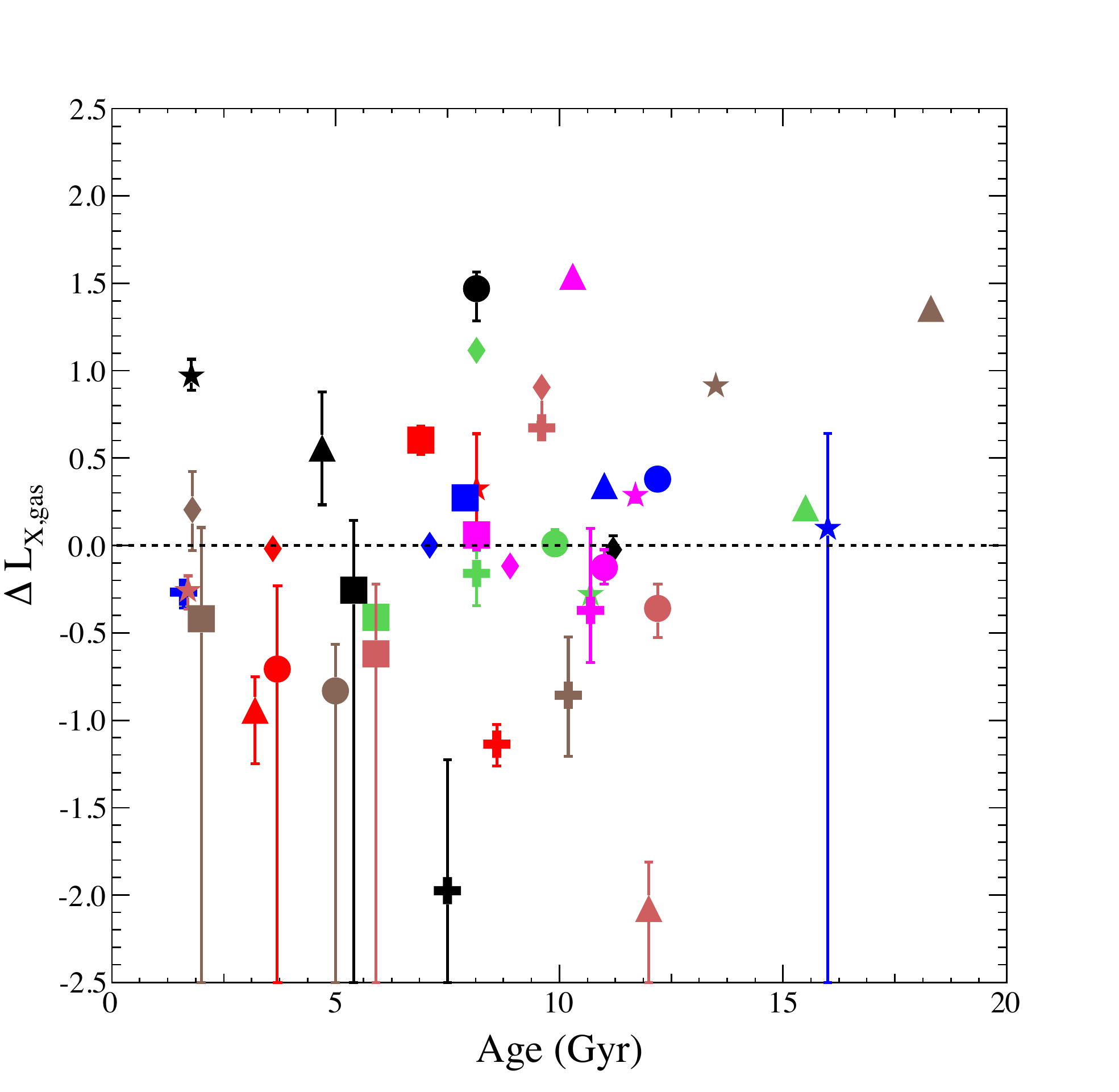}{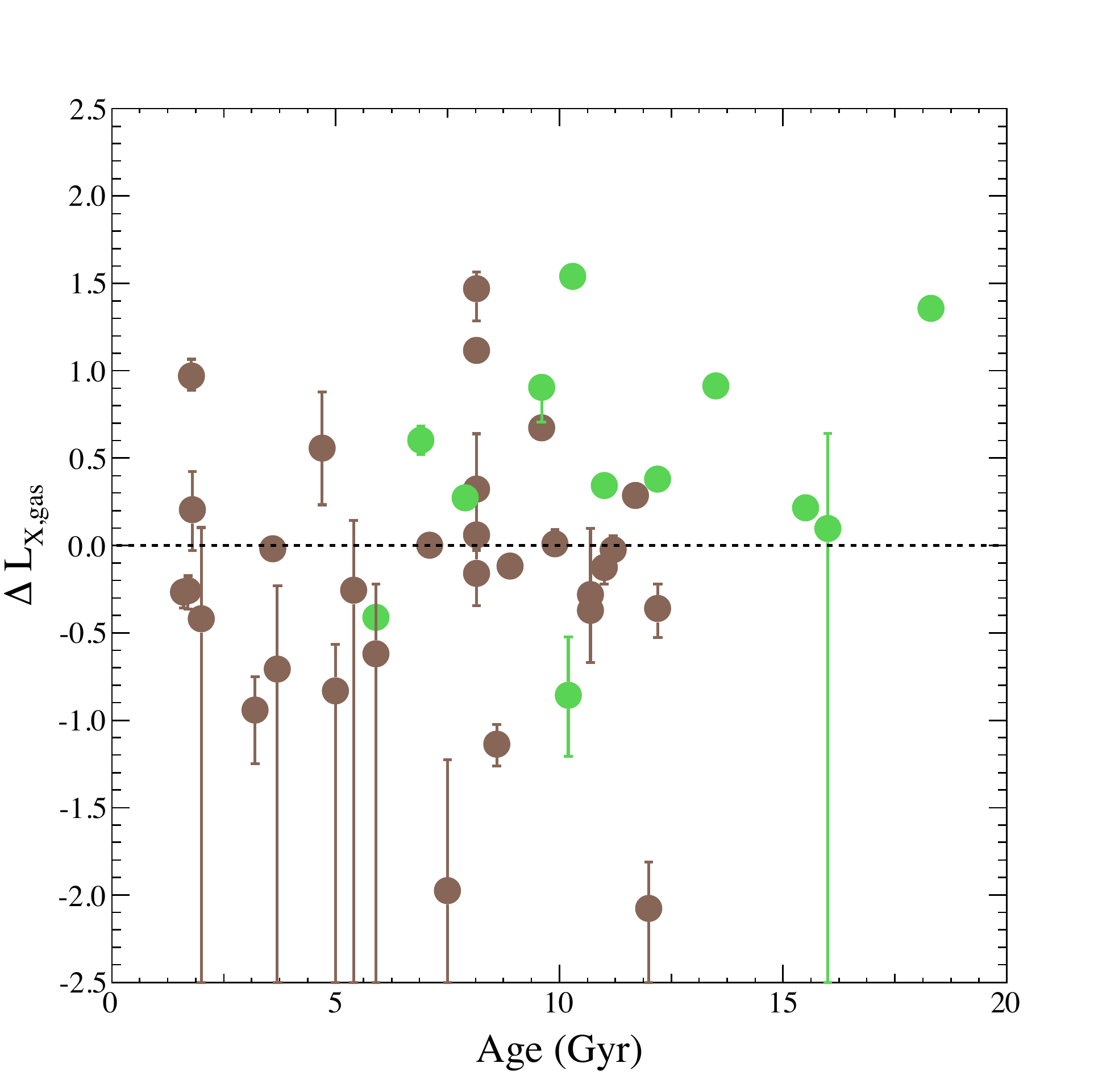}
\plottwo{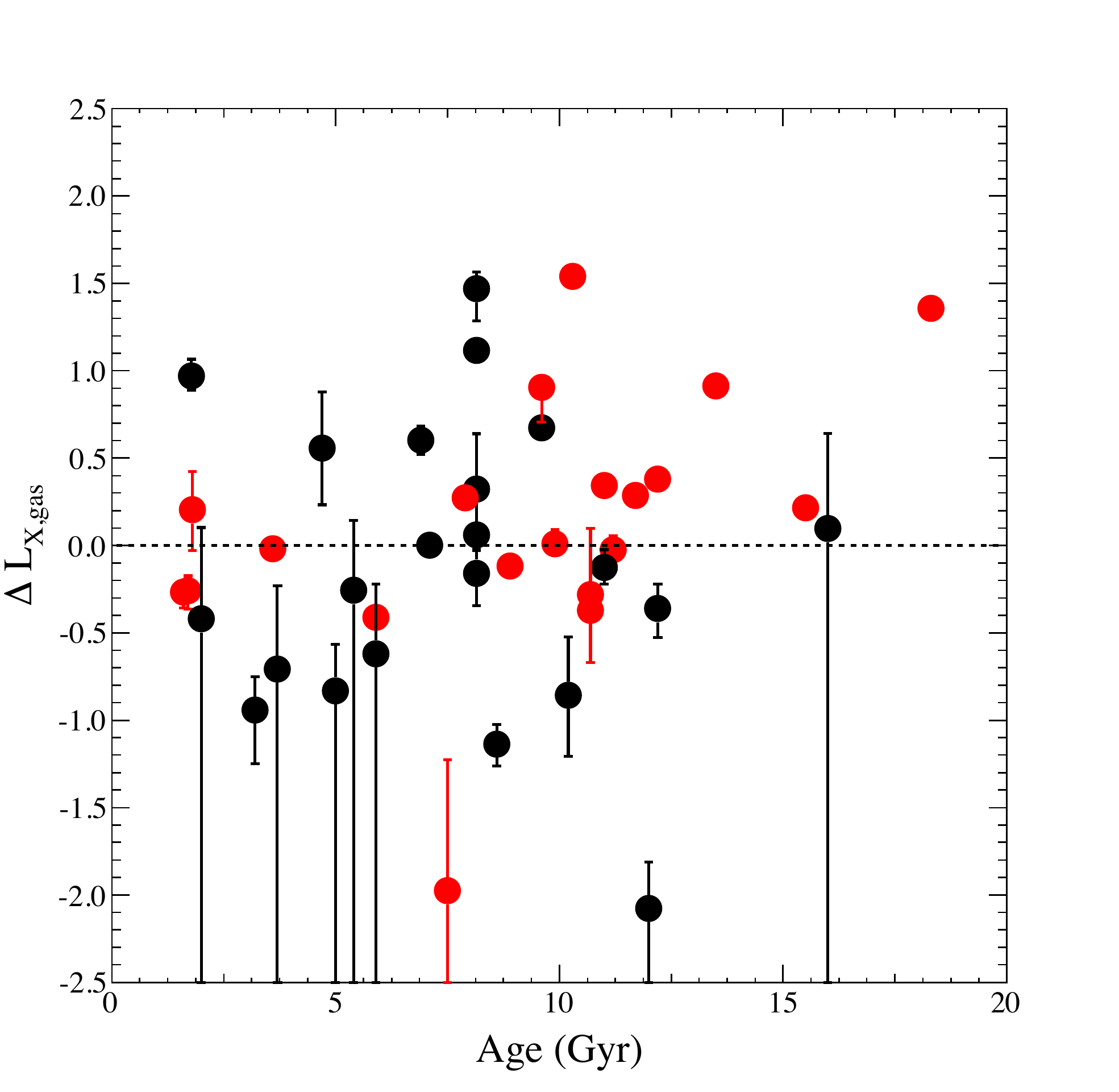}{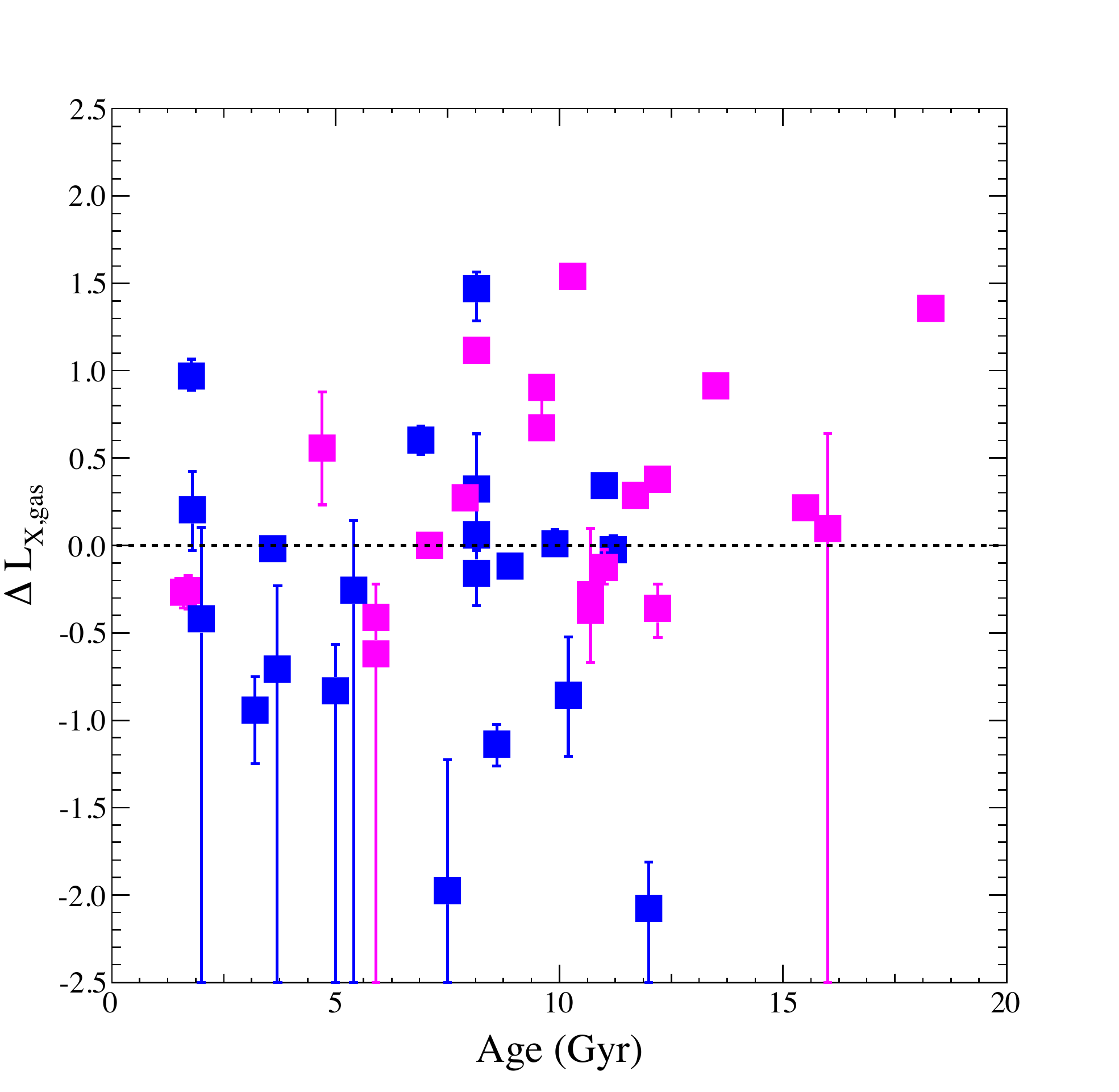}
\figcaption{\label{fig:age} $\Delta L_{X_{\rm gas}}$ as a function of stellar age. {\it top-left}: color code is the same as in Figure~\ref{fig:lxk}. {\it top-right}: ``true lenticular" galaxies (brown) and ``true elliptical" galaxies (green). {\it bottom-left}: high mass galaxies (red) and low mass galaxies (black). {\it bottom-right}: field galaxies (blue) and galaxies in groups and clusters (magenta).}
\end{figure}

 \begin{figure} 
\epsscale{1.11}
\plottwo{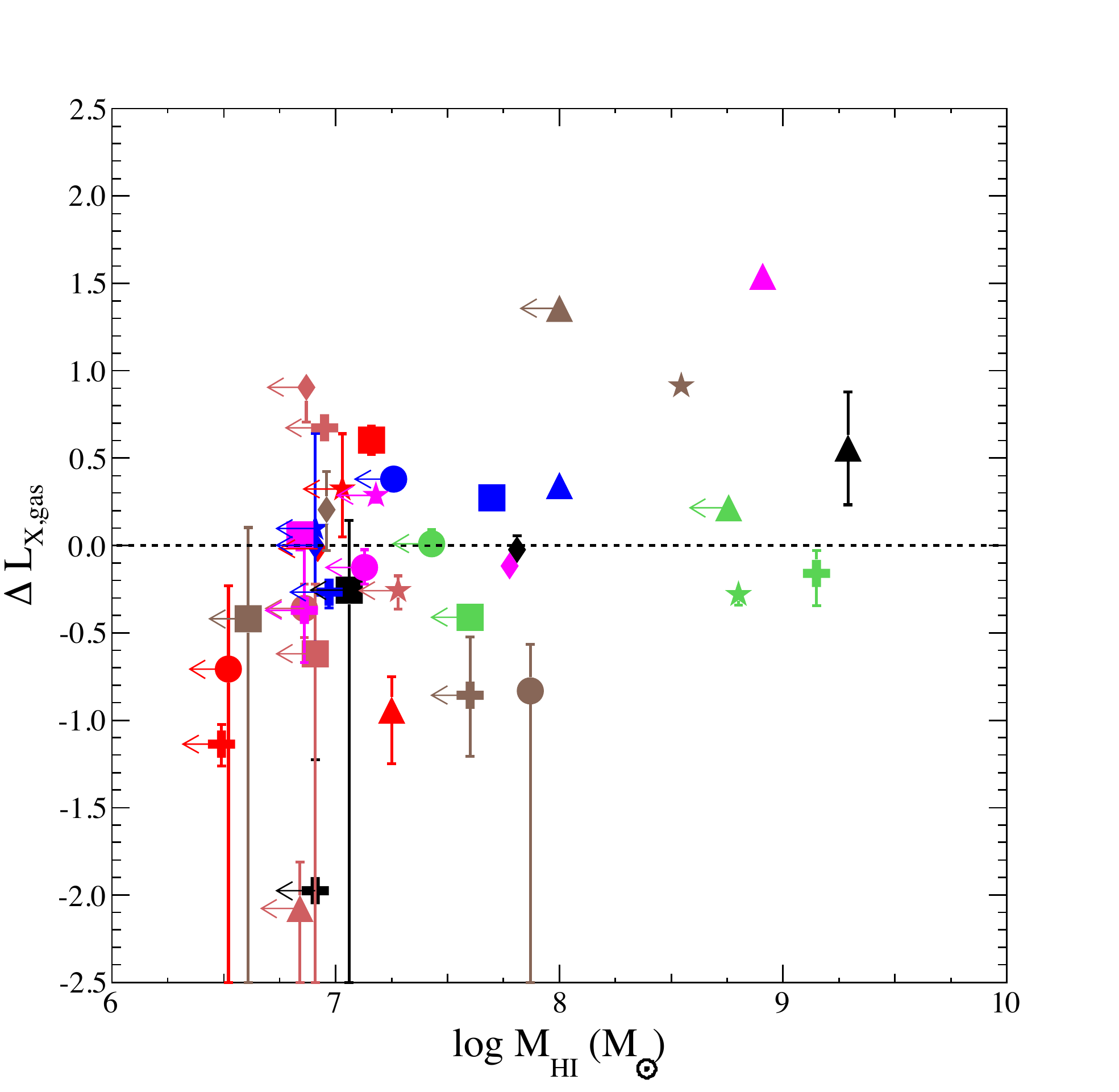}{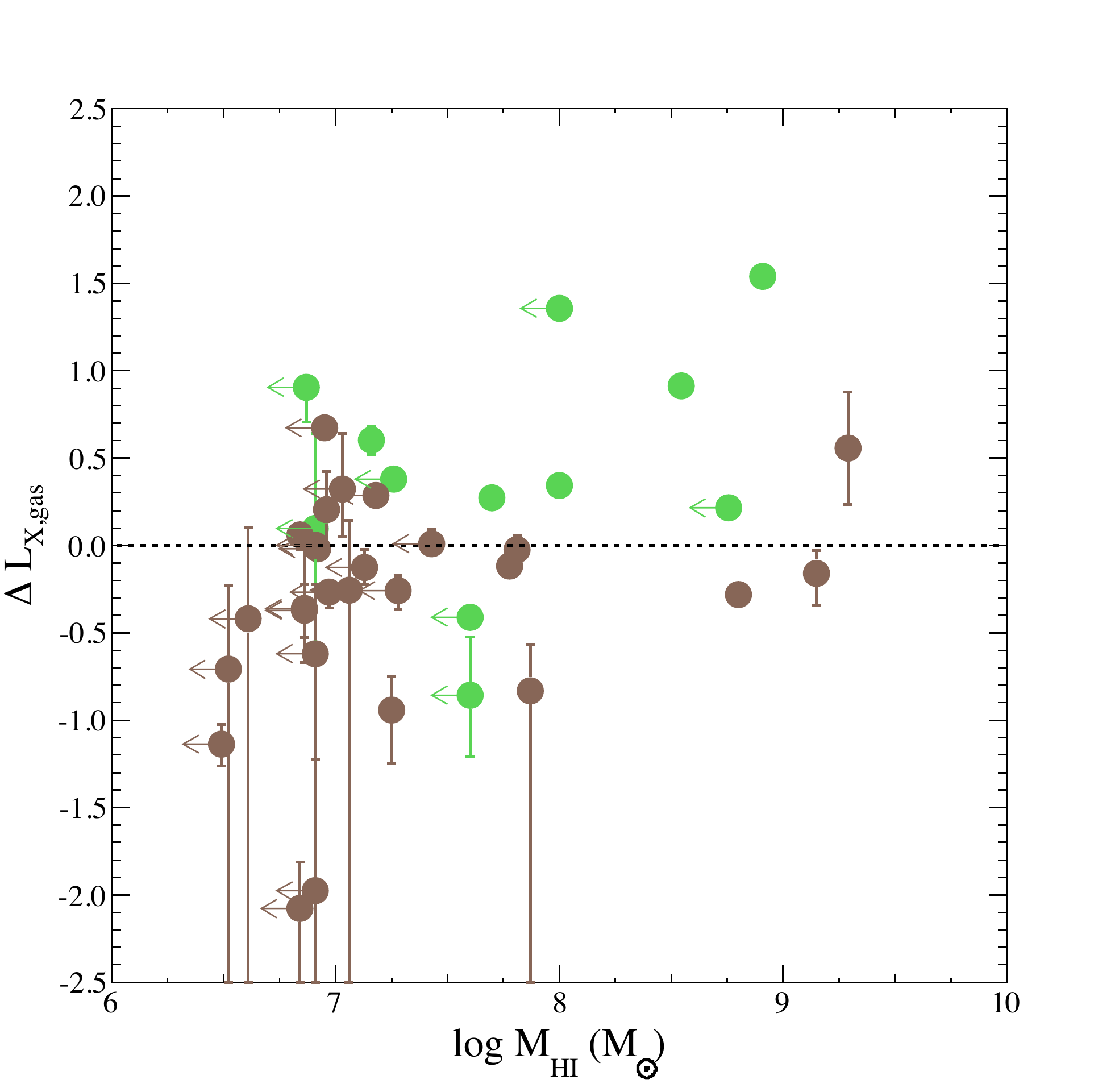}
\plottwo{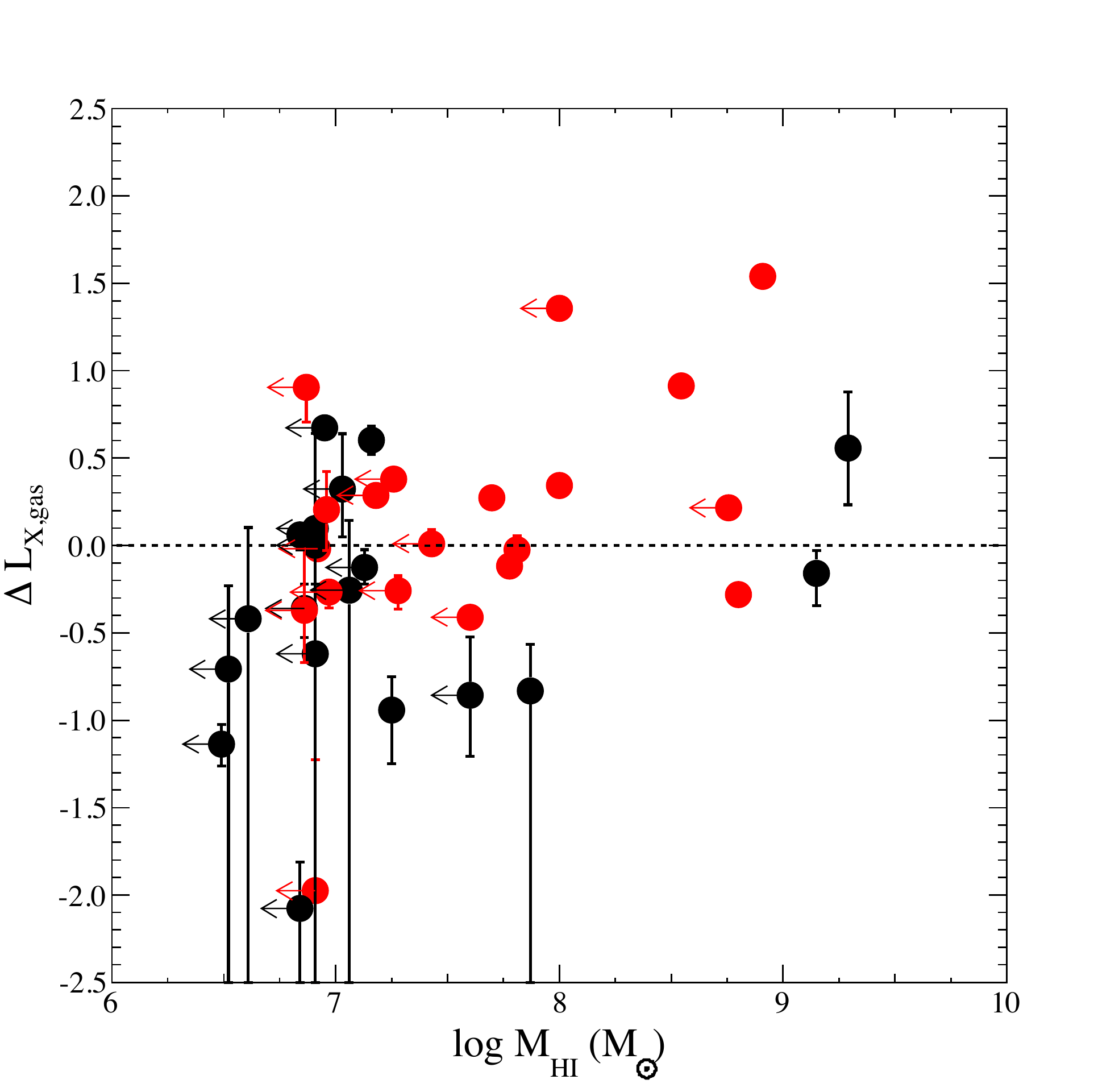}{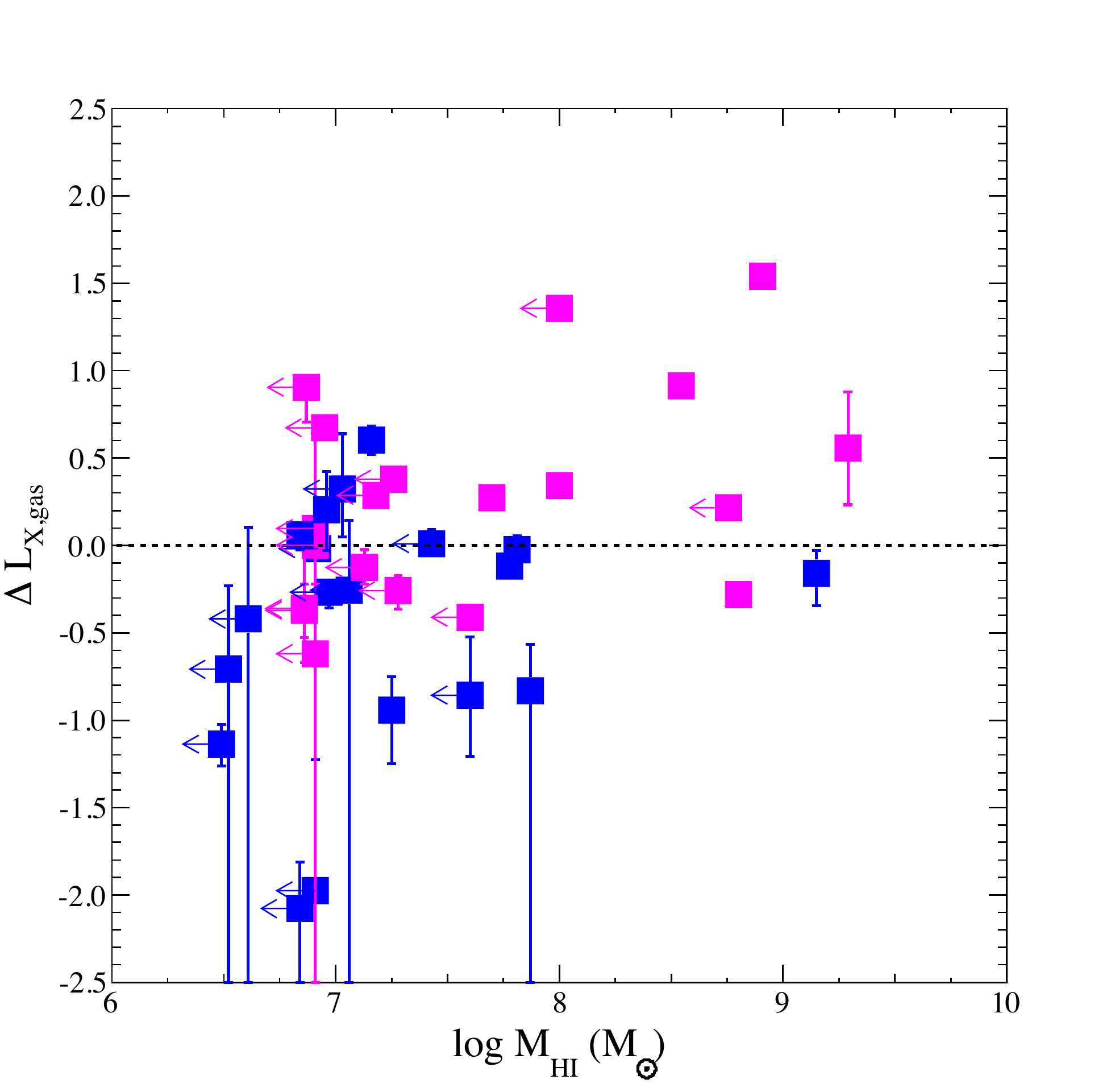}
\figcaption{\label{fig:hi} $\Delta L_{X_{\rm gas}}$ as a function of atomic gas mass.  {\it top-left}: color code is the same as in Figure~\ref{fig:lxk}. {\it top-right}: ``true lenticular" galaxies (brown) and ``true elliptical" galaxies (green). {\it bottom-left}: high mass galaxies (red) and low mass galaxies (black). {\it bottom-right}: field galaxies (blue) and galaxies in groups and clusters (magenta).}
\end{figure}

 \begin{figure} 
\epsscale{1.11}
\plottwo{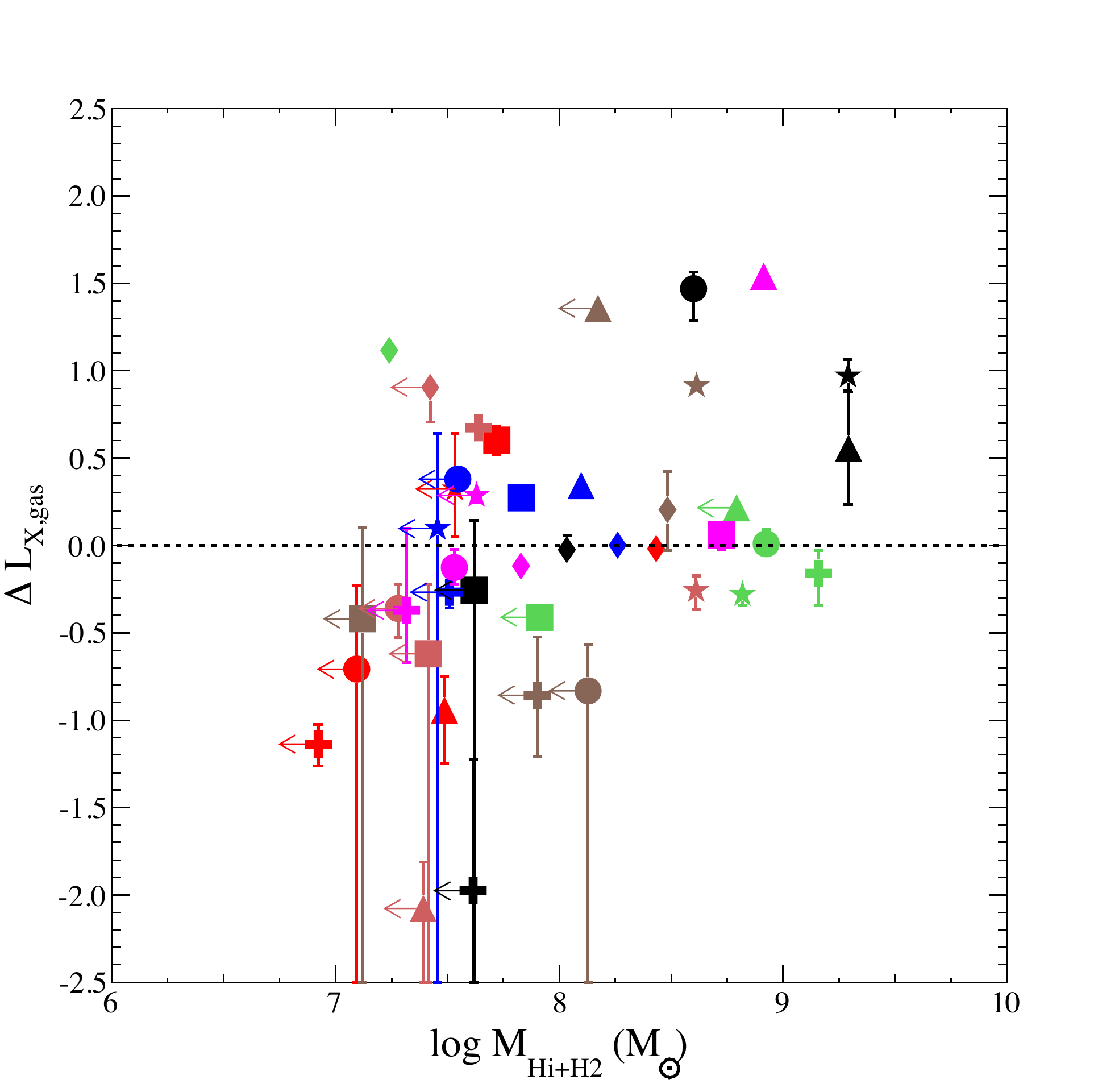}{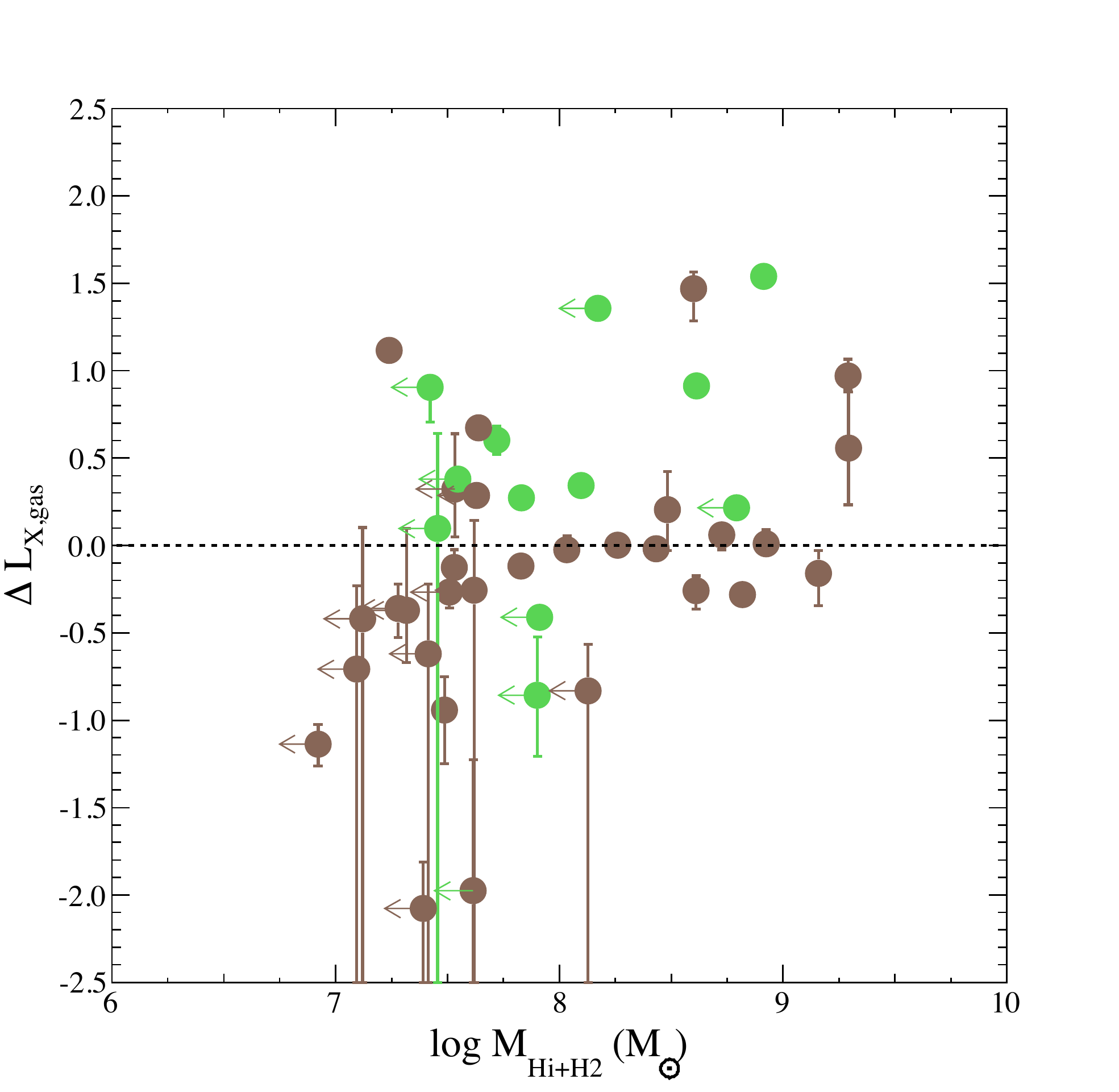}
\plottwo{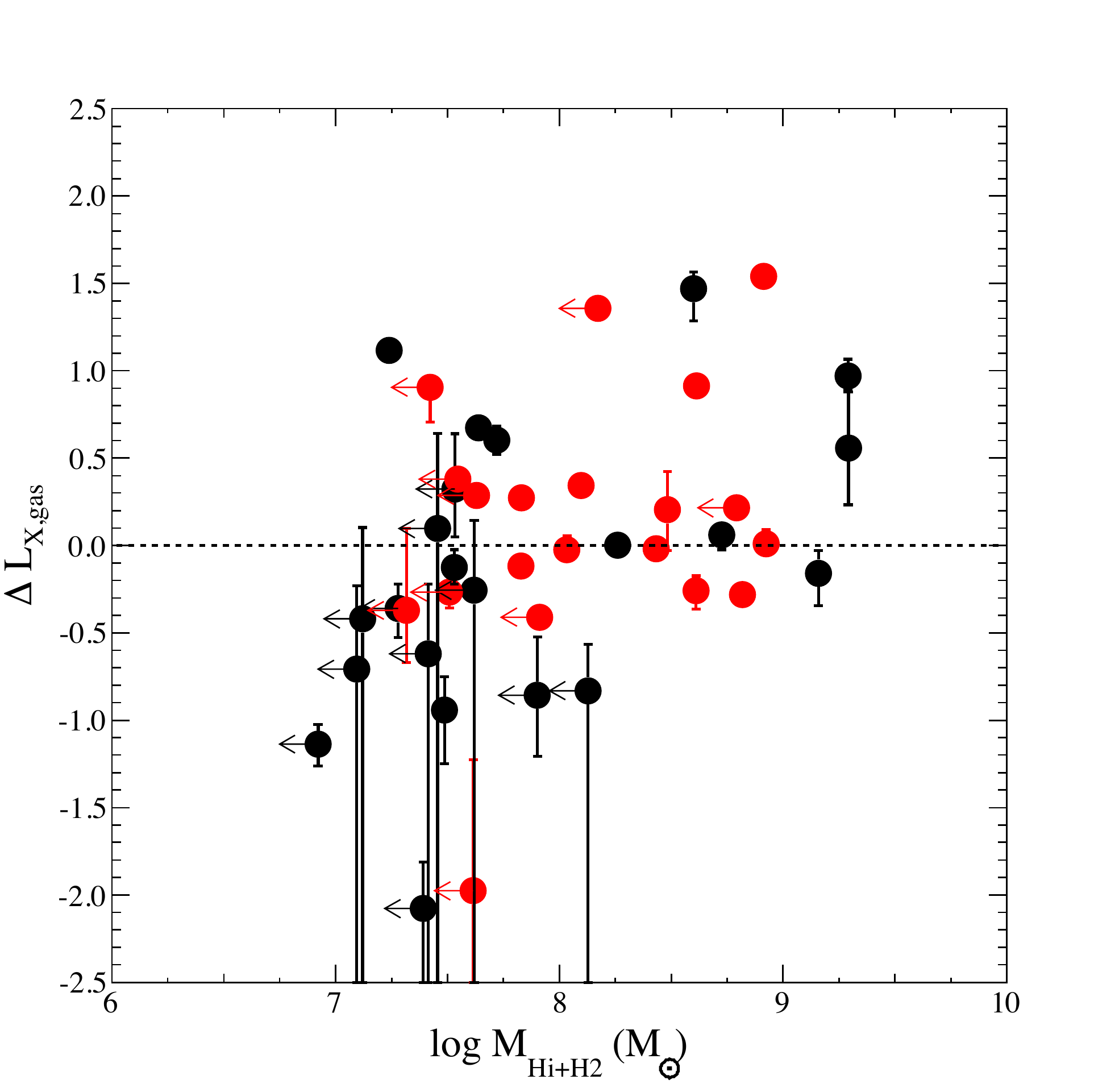}{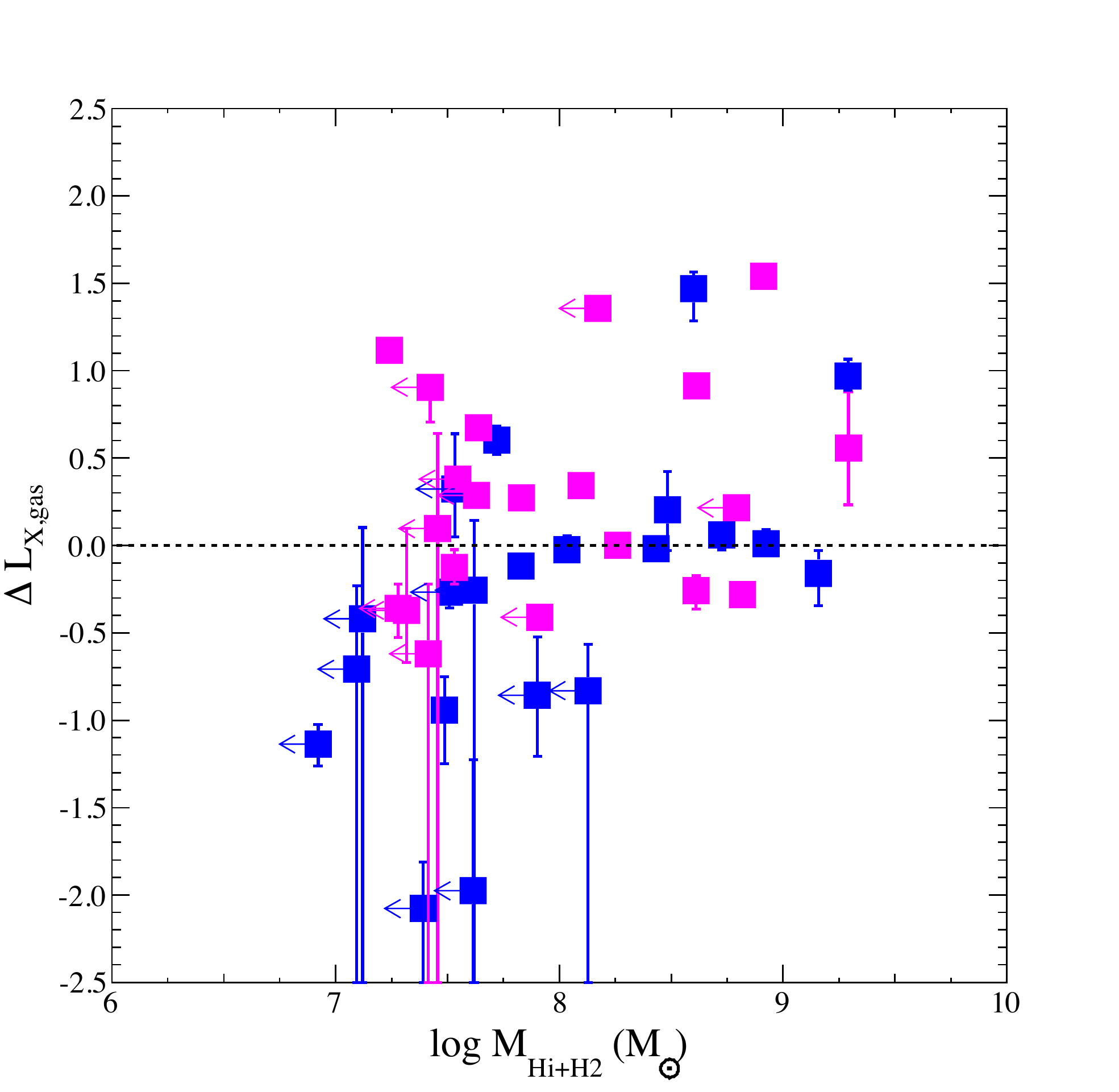}
\figcaption{\label{fig:cold} $\Delta L_{X_{\rm gas}}$ as a function of cold gas mass. {\it top-left}: color code is the same as in Figure~\ref{fig:lxk}. {\it top-right}: ``true lenticular" galaxies (brown) and ``true elliptical" galaxies (green). {\it bottom-left}: high mass galaxies (red) and low mass galaxies (black). {\it bottom-right}: field galaxies (blue) and galaxies in groups and clusters (magenta).}
\end{figure}

 \begin{figure} 
\epsscale{1.11}
\plottwo{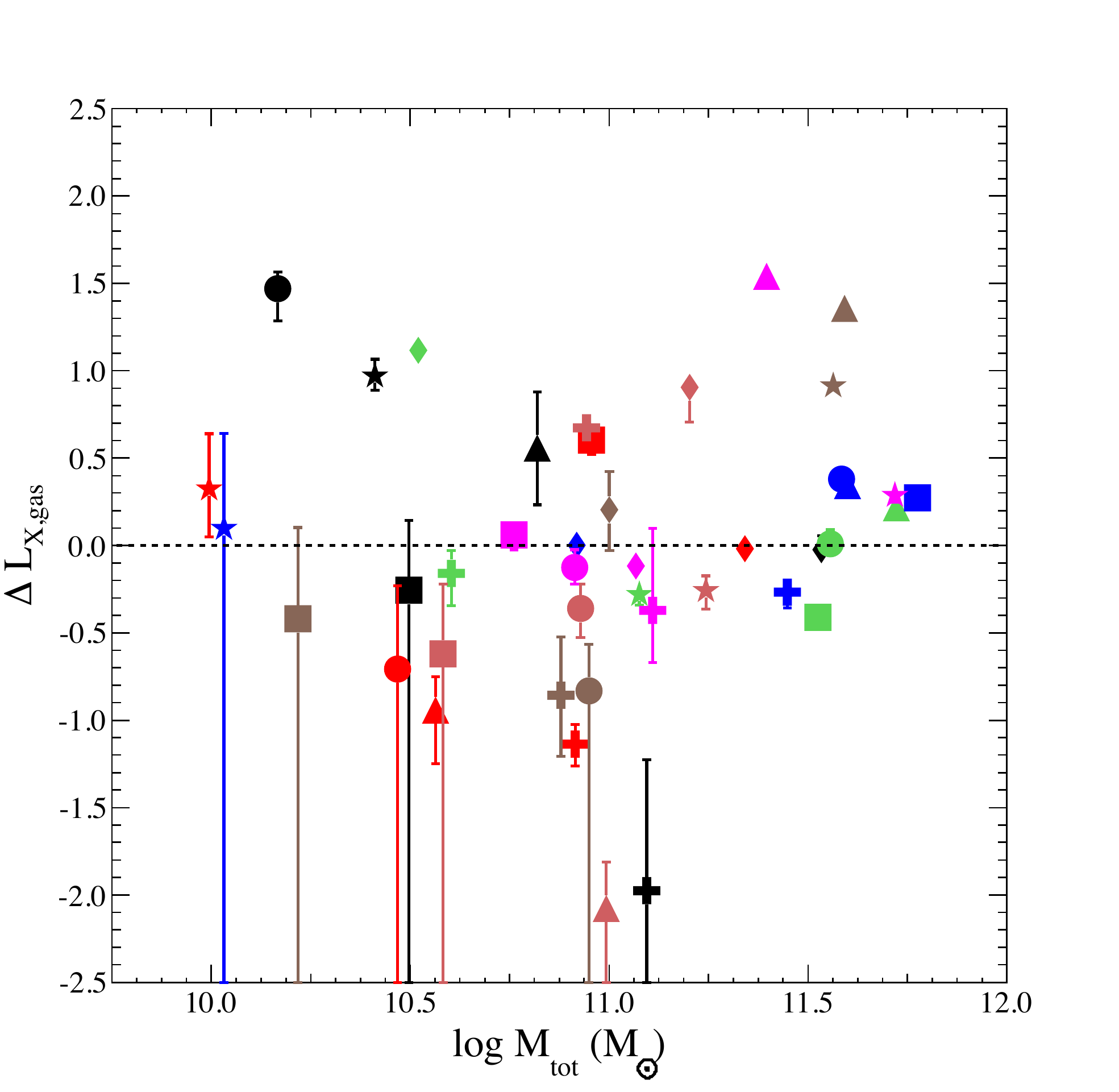}{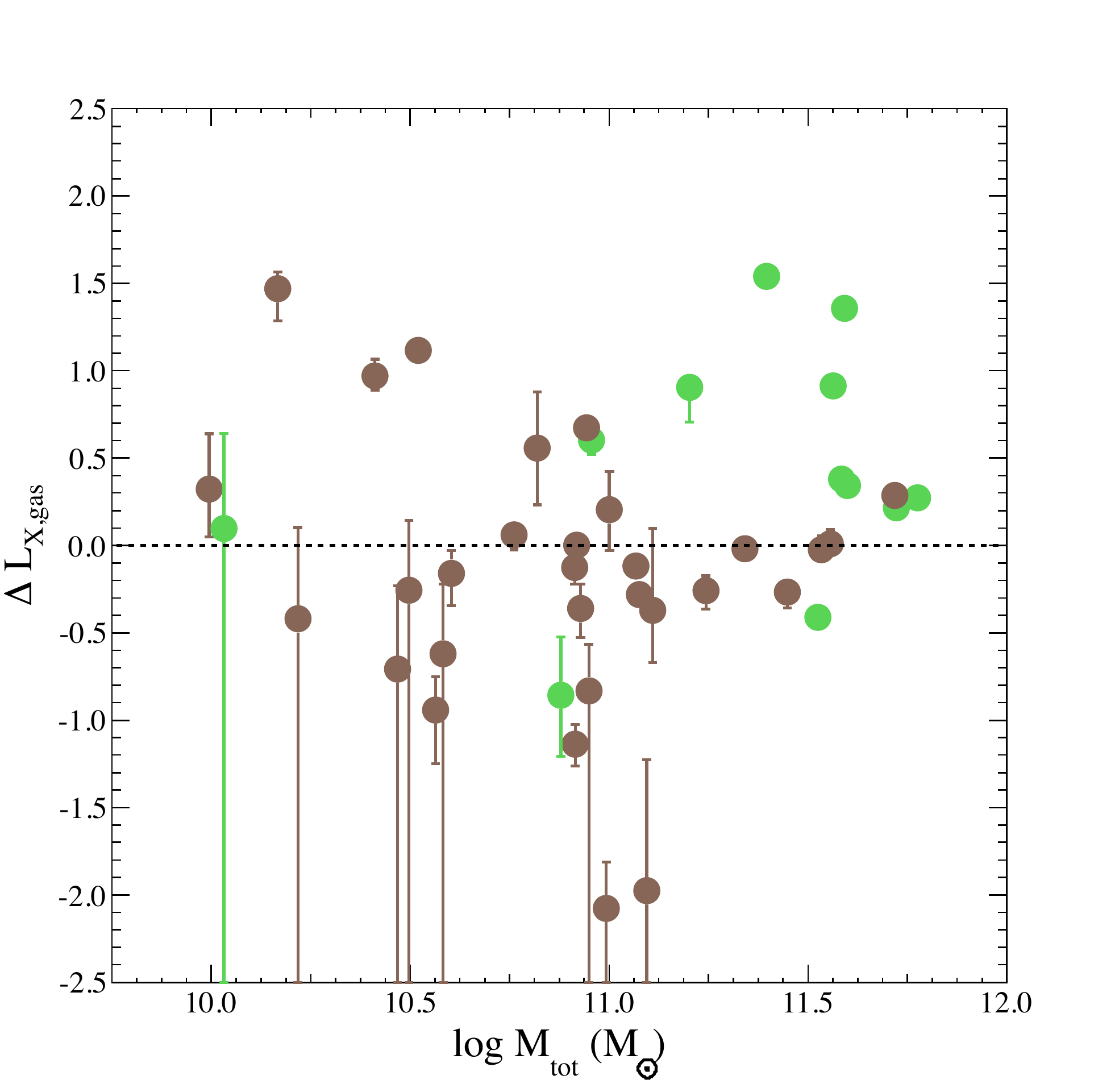}
\plottwo{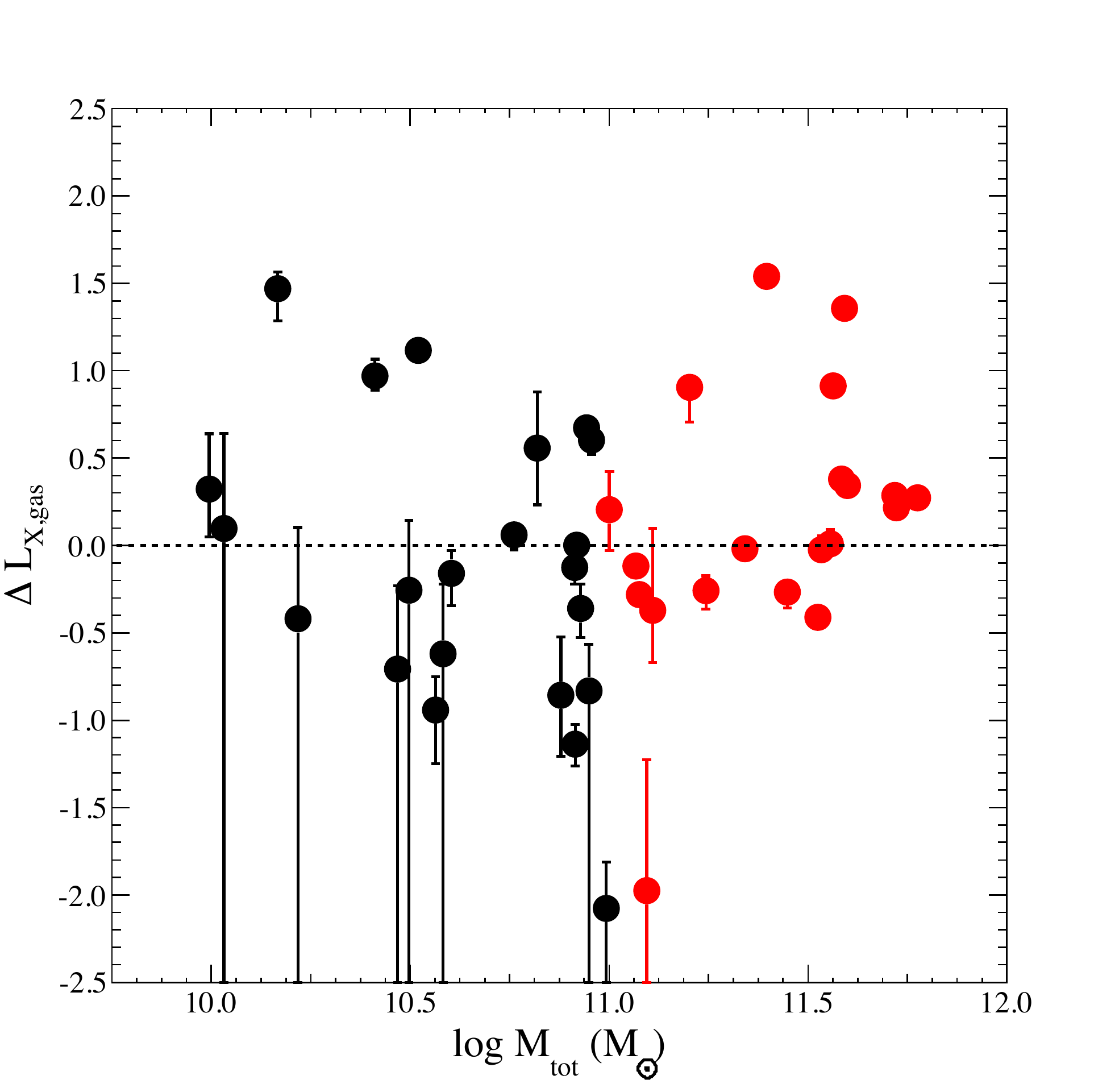}{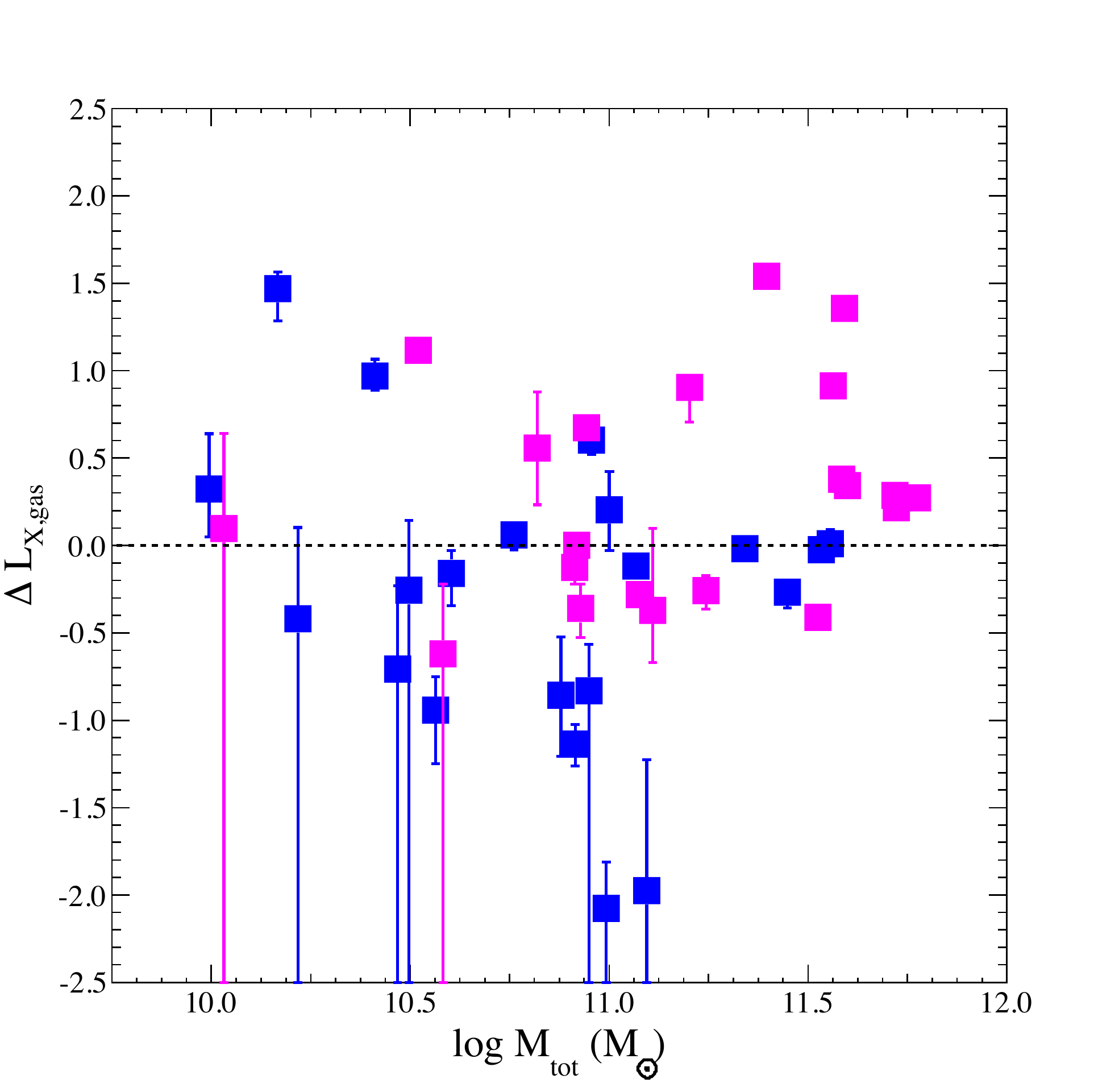}
\figcaption{\label{fig:mtot} $\Delta L_{X_{\rm gas}}$ as a function of total mass. {\it top-left}: color code is the same as in Figure~\ref{fig:lxk}. {\it top-right}: ``true lenticular" galaxies (brown) and ``true elliptical" galaxies (green). {\it bottom-left}: high mass galaxies (red) and low mass galaxies (black). {\it bottom-right}: field galaxies (blue) and galaxies in groups and clusters (magenta).}
\end{figure}

\clearpage

 \begin{figure} 
\epsscale{1.11}
\plottwo{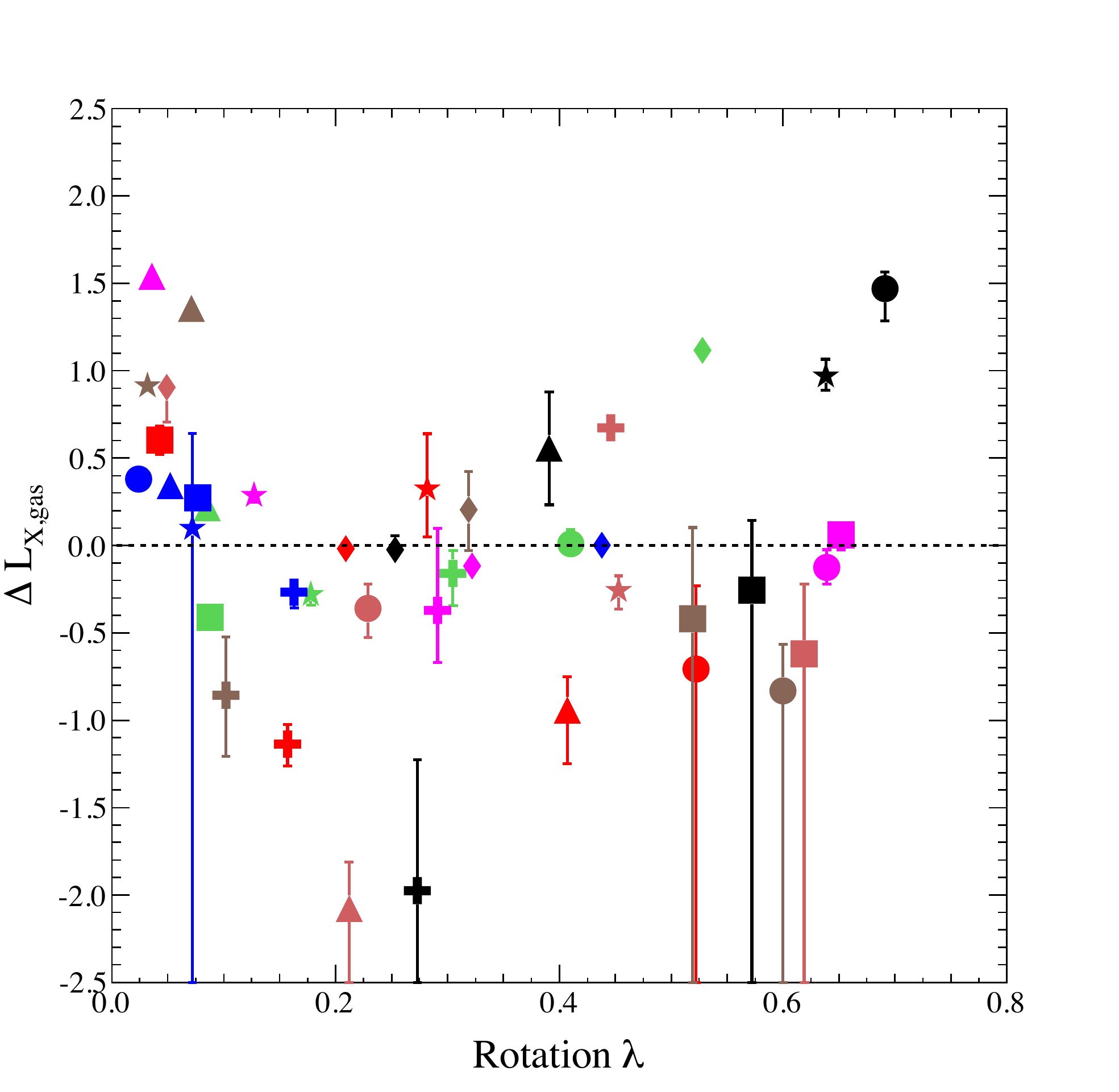}{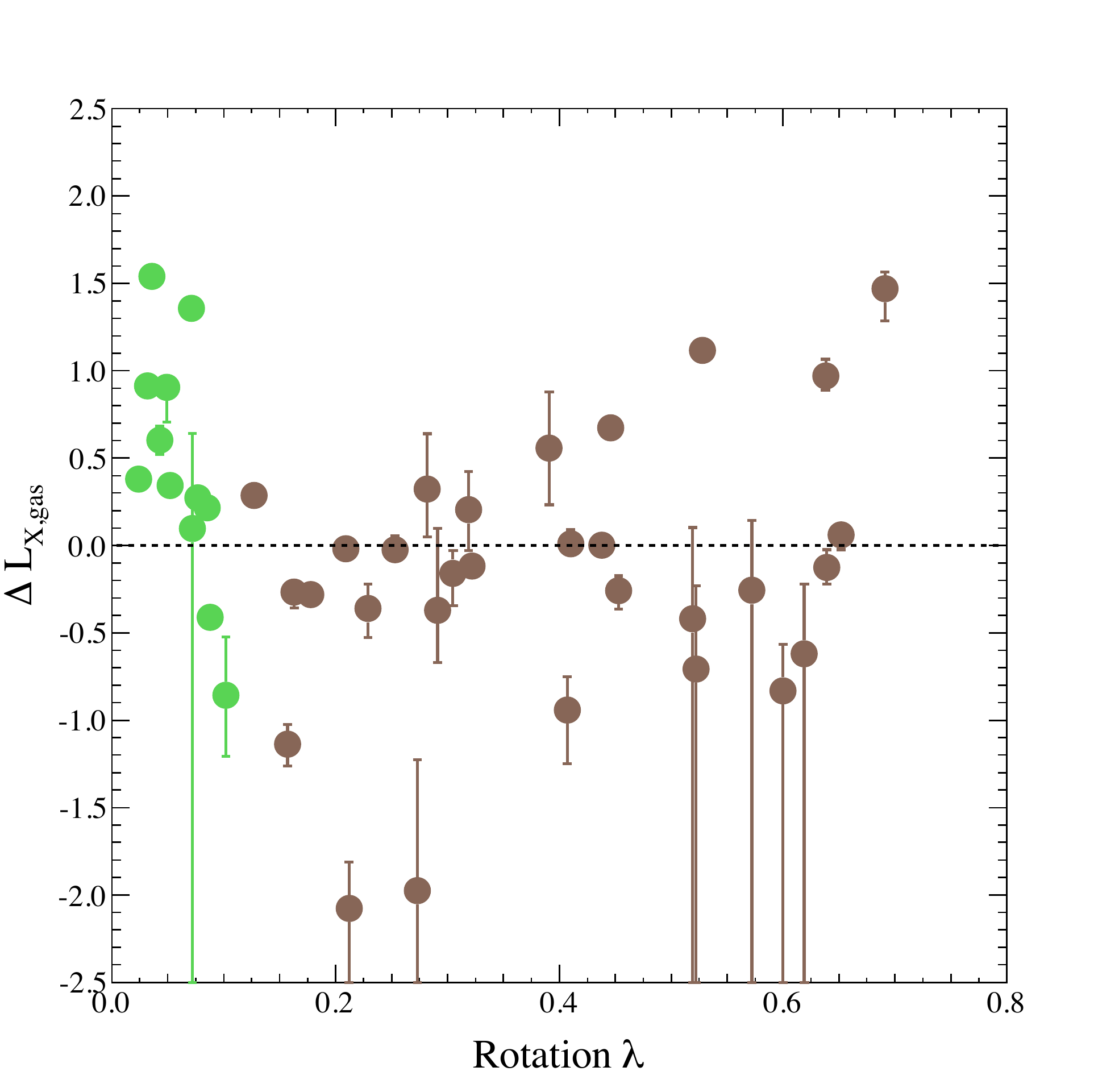}
\plottwo{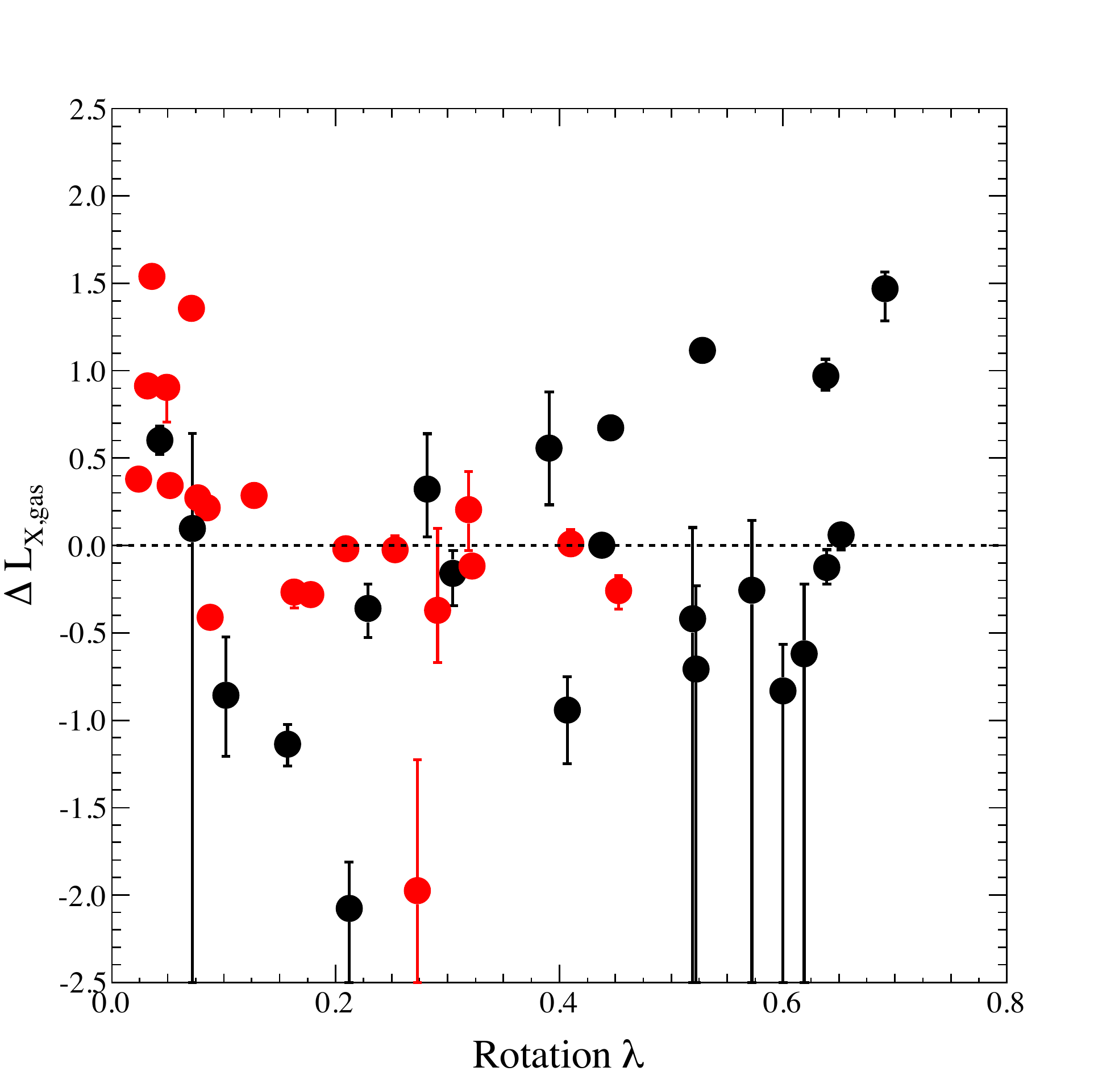}{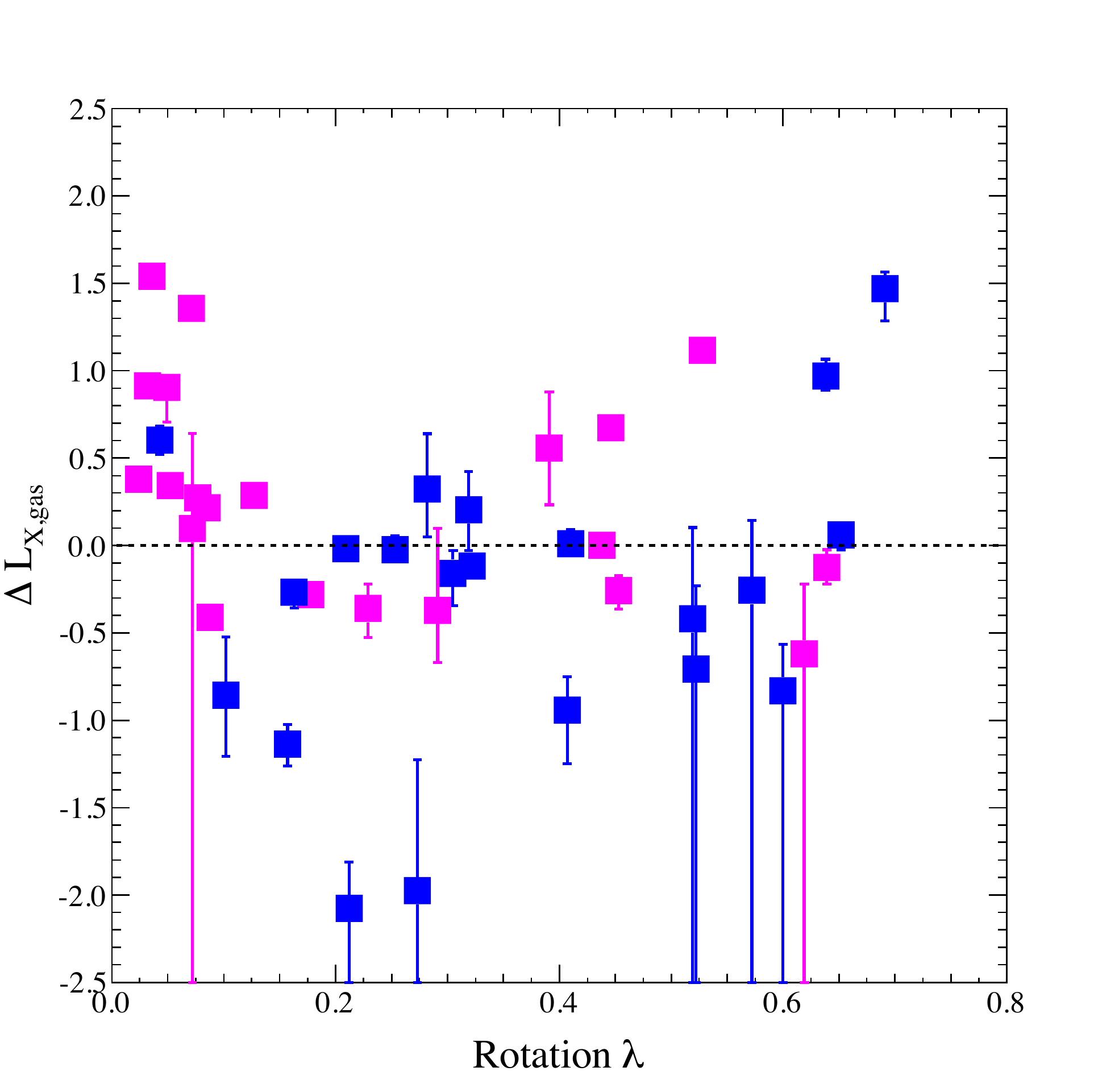}
\figcaption{\label{fig:rot} $\Delta L_{X_{\rm gas}}$ as a function of angular momentum $\lambda$.  {\it top-left}: color code is the same as in Figure~\ref{fig:lxk}. {\it top-right}: ``true lenticular" galaxies (brown) and ``true elliptical" galaxies (green). {\it bottom-left}: high mass galaxies (red) and low mass galaxies (black). {\it bottom-right}: field galaxies (blue) and galaxies in groups and clusters (magenta).}
\end{figure}

 \begin{figure} 
\epsscale{1.11}
\plottwo{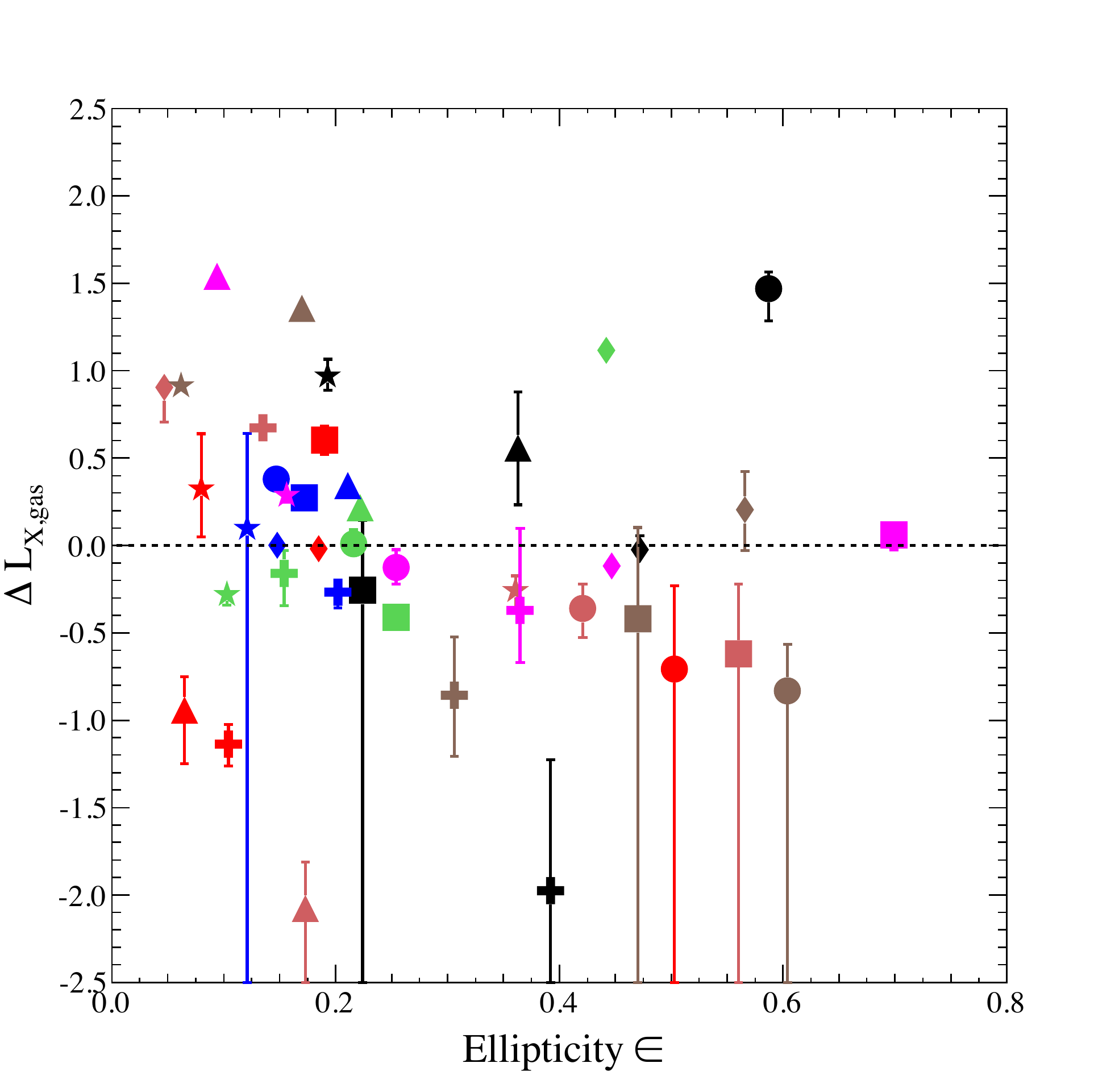}{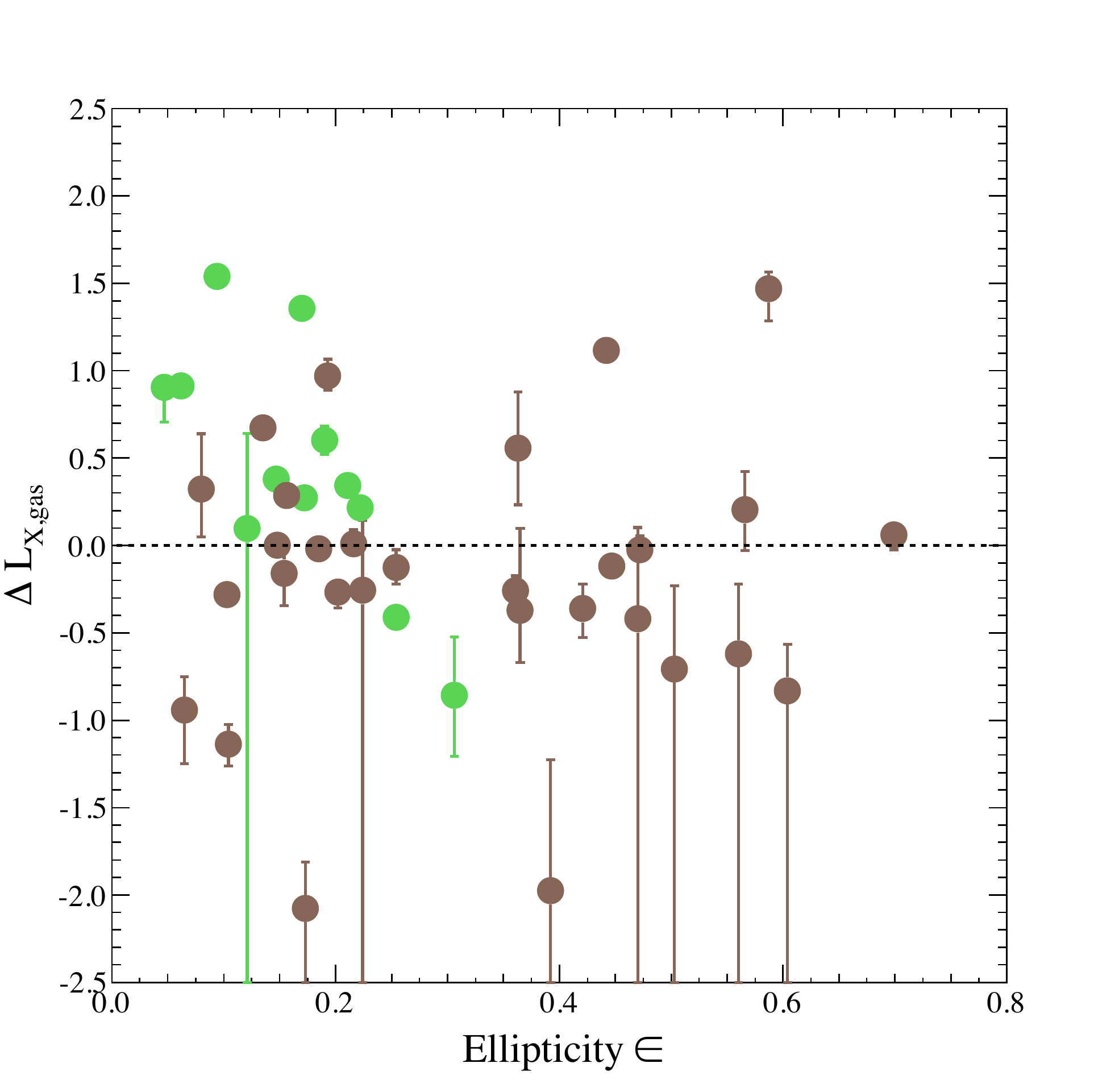}
\plottwo{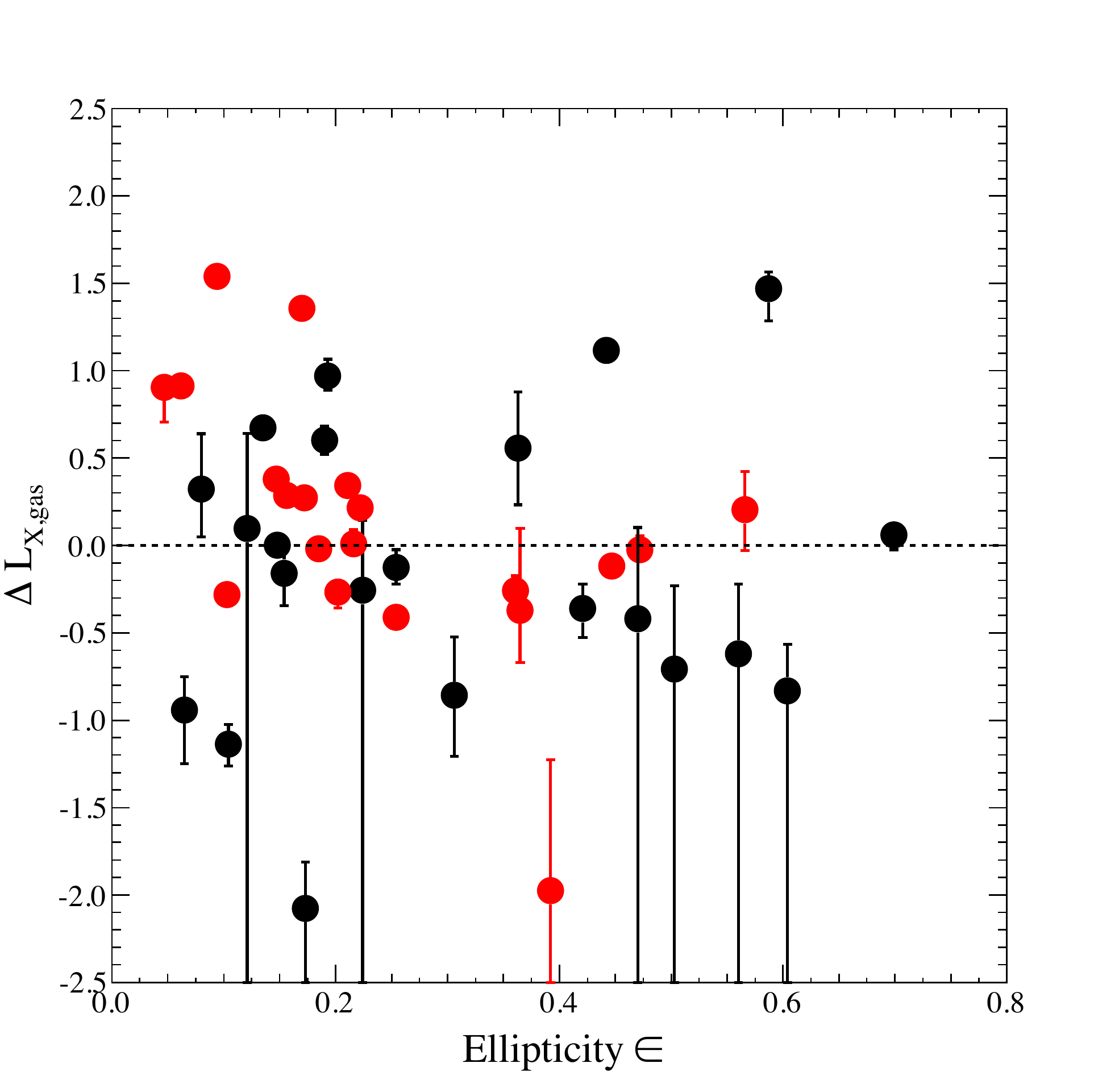}{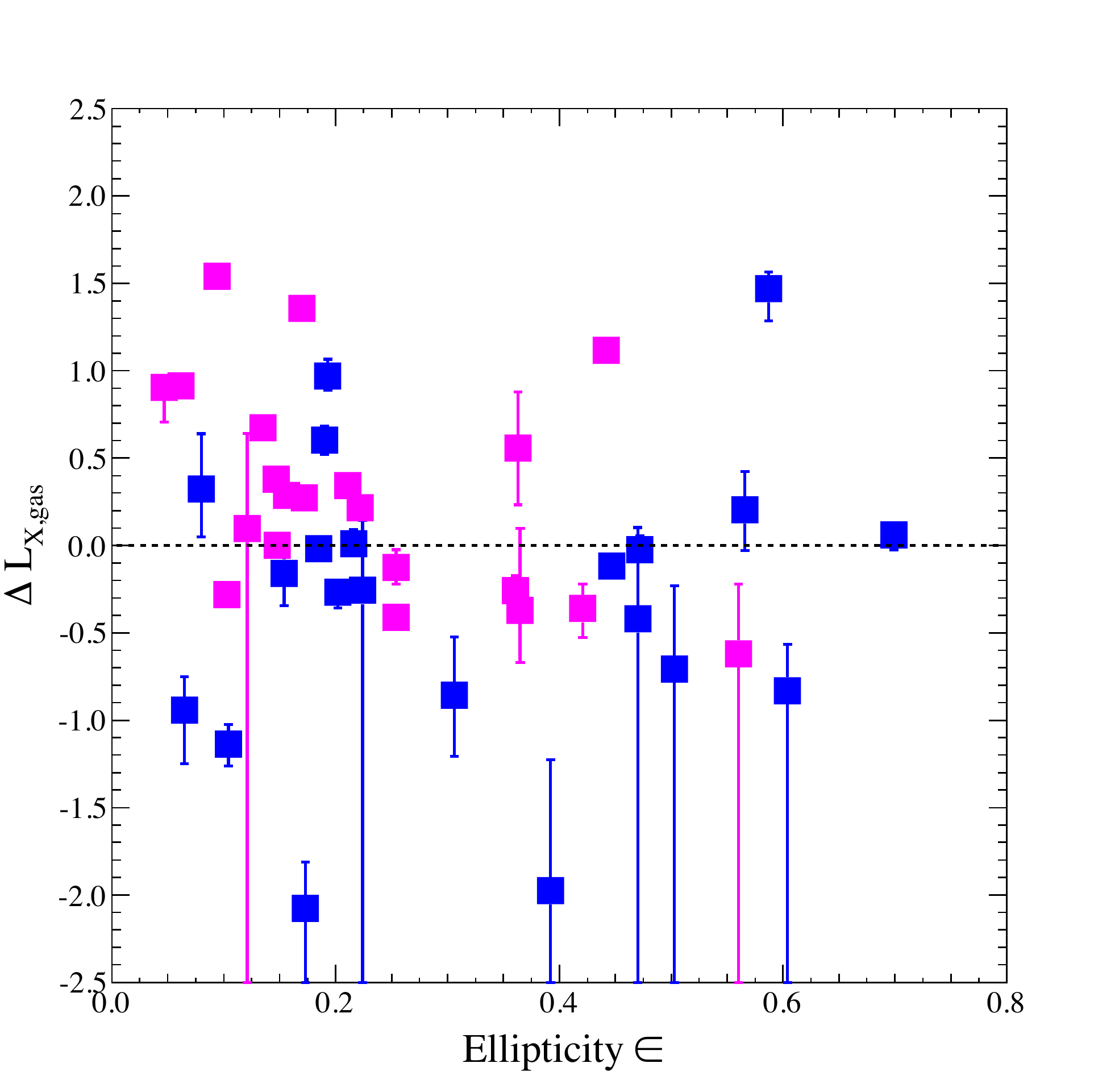}
\figcaption{\label{fig:flat} $\Delta L_{X_{\rm gas}}$ as a function of ellipticity $\epsilon$. {\it top-left}: color code is the same as in Figure~\ref{fig:lxk}. {\it top-right}: ``true lenticular" galaxies (brown) and ``true elliptical" galaxies (green). {\it bottom-left}: high mass galaxies (red) and low mass galaxies (black). {\it bottom-right}: field galaxies (blue) and galaxies in groups and clusters (magenta).}
\end{figure}

 \begin{figure} 
\epsscale{1.11}
\plottwo{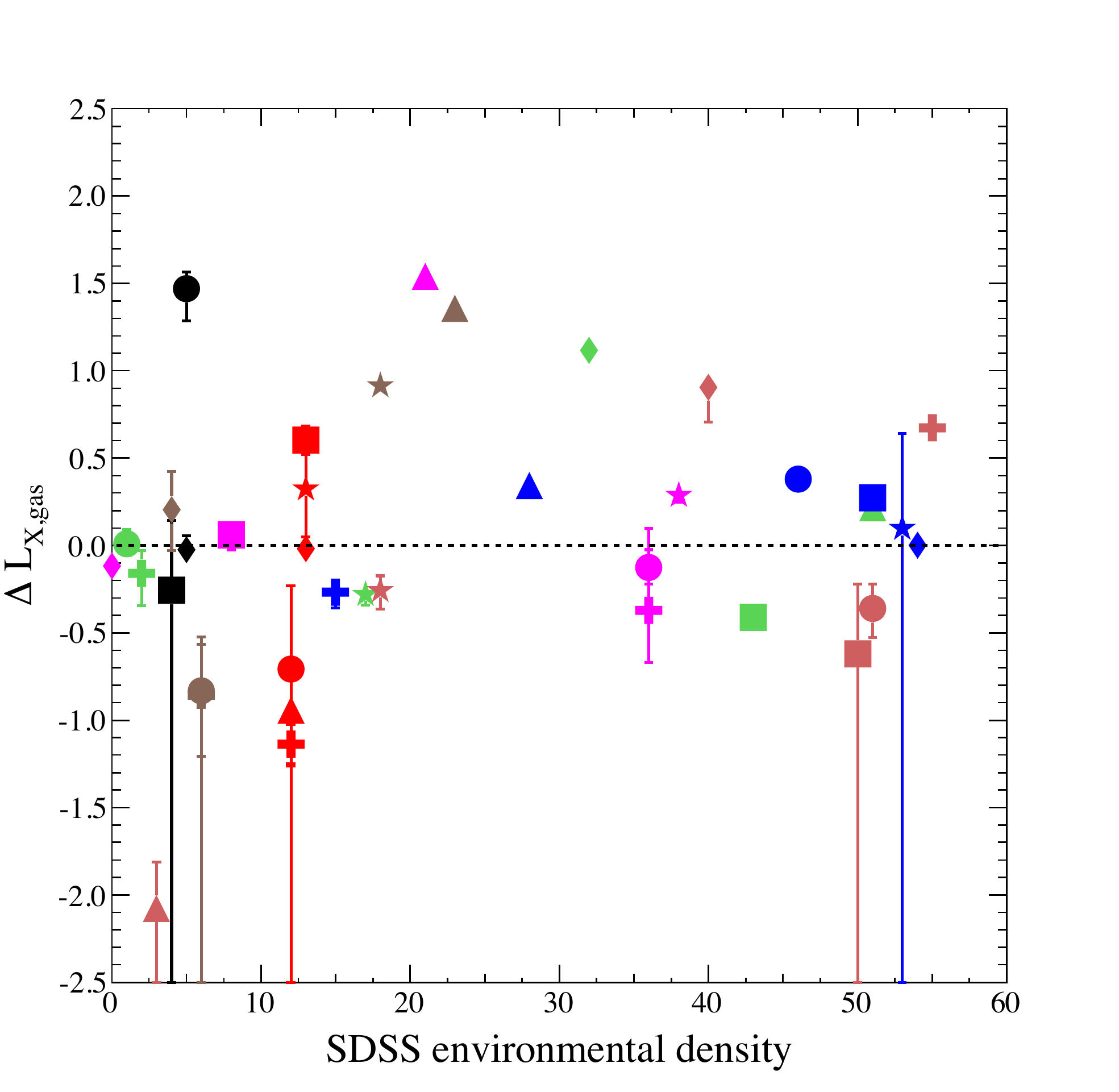}{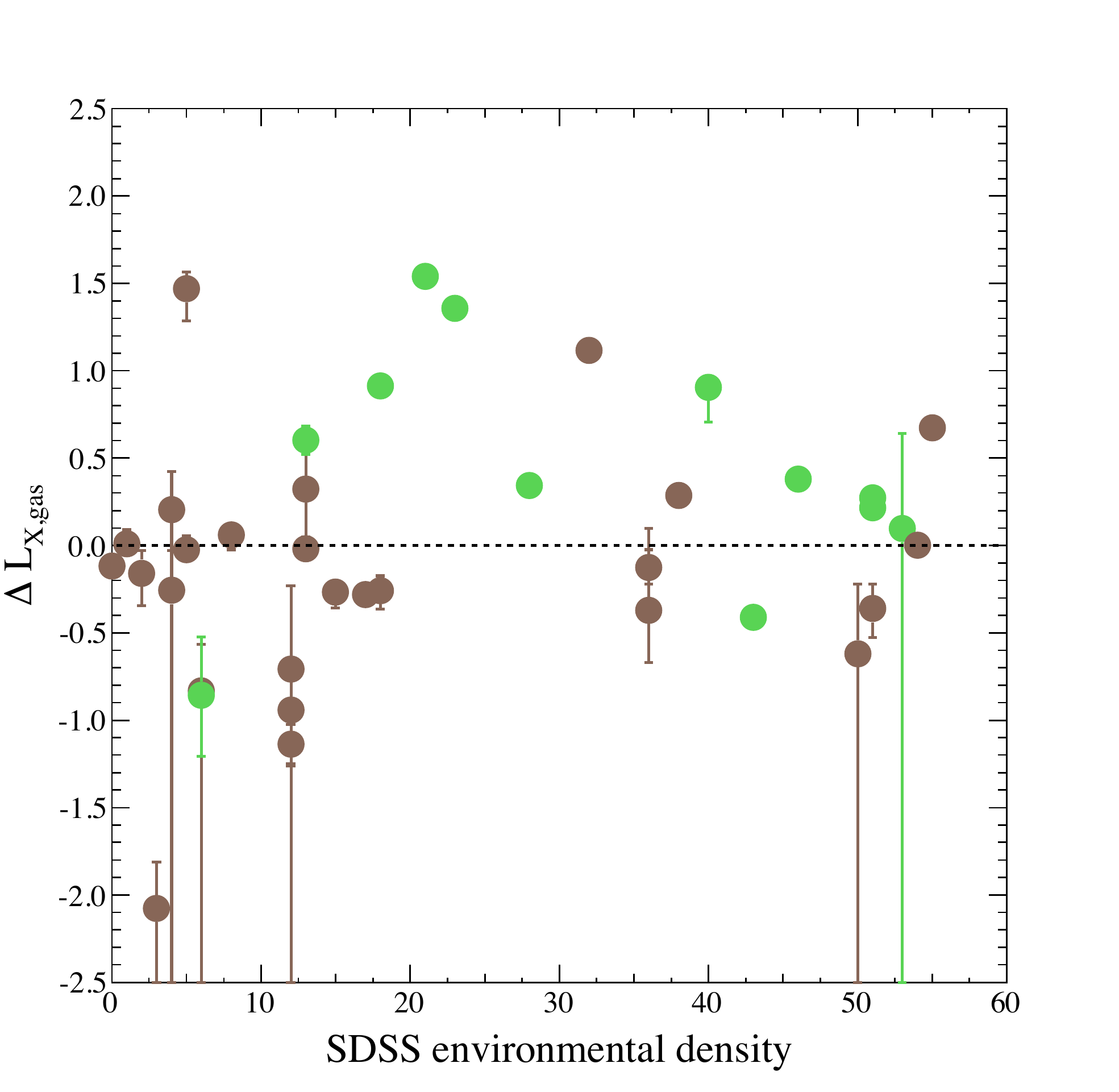}
\plottwo{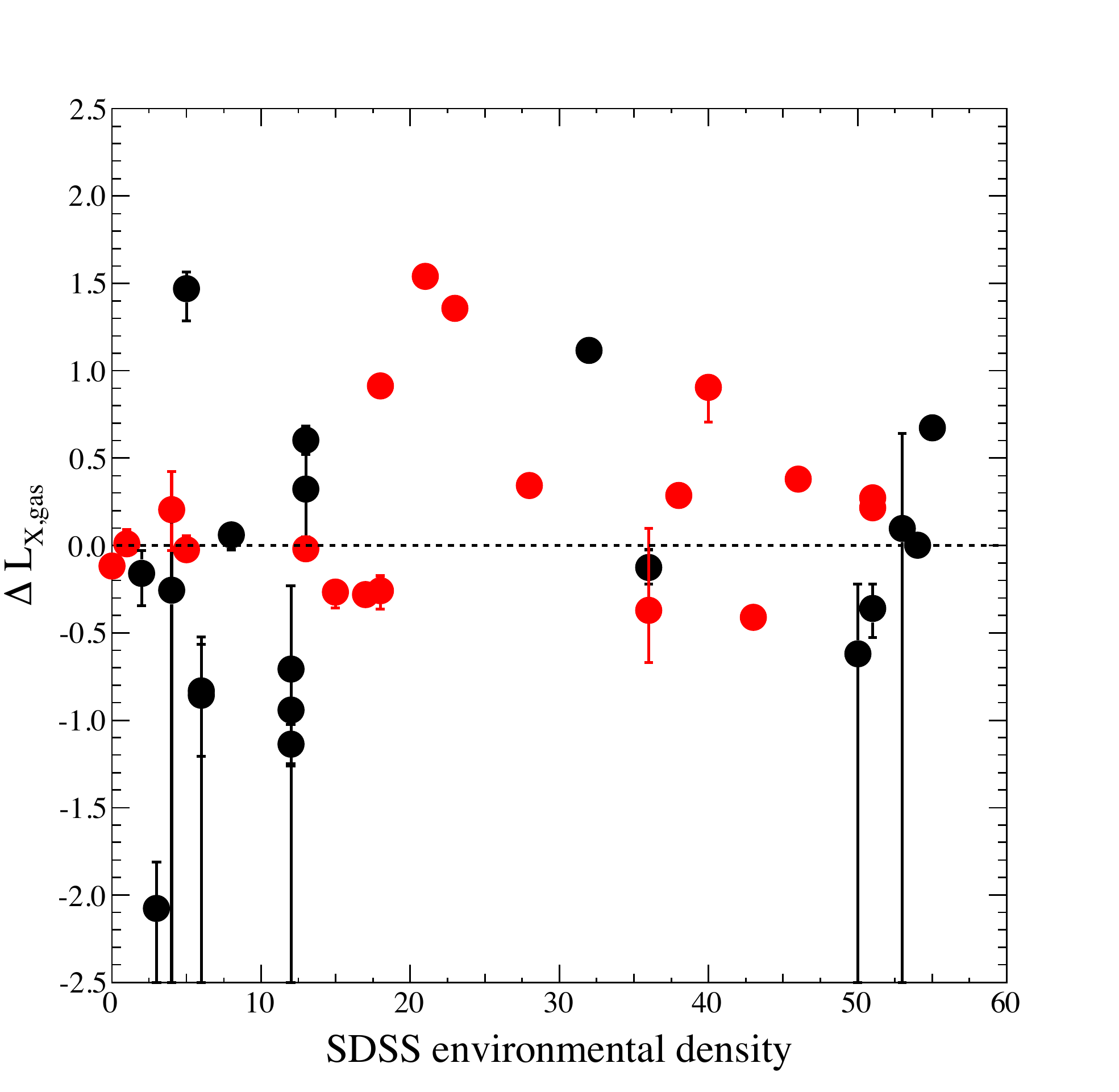}{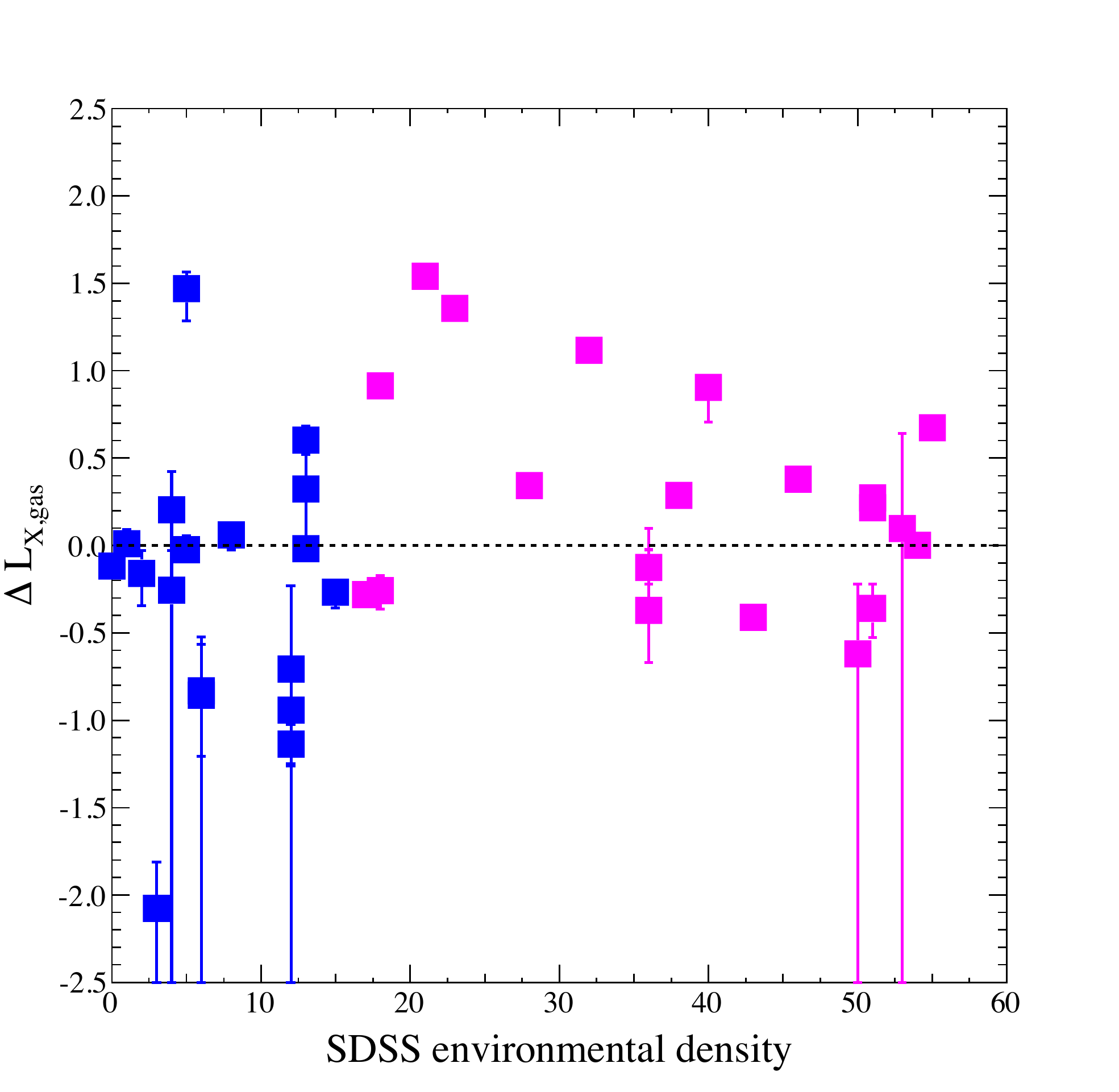}
\figcaption{\label{fig:sdss} $\Delta L_{X_{\rm gas}}$ as a function of the environmental  galaxy density. {\it top-left}: color code is the same as in Figure~\ref{fig:lxk}. {\it top-right}: ``true lenticular" galaxies (brown) and ``true elliptical" galaxies (green). {\it bottom-left}: high mass galaxies (red) and low mass galaxies (black). {\it bottom-right}: field galaxies (blue) and galaxies in groups and clusters (magenta).}
\end{figure}

\clearpage
%\newpage

\section{\bf B. Statistical Uncertainties and Systematic Tests}

The statistical uncertainties of these correlation coefficients were estimated through the Monte Carlo realizations of our measurements within their associated spectral fitting uncertainties (or uncertainties of data taken from literature, e.g.\  SFR). 
We listed such impact on each coefficient for the whole sample in Tables 4 and 5. 
The only significant impact is on relations with $T_X$ due to its relatively large uncertainties of spectral fitting. The $T_X$--$L_{X_{\rm gas}}/L_K$ relation has  $\rho=0.515\pm0.184$ and the $T_X$--$\Delta L_{X_{\rm gas}}$  relation has  $\rho=0.476\pm0.186$.

\begin{figure*}
   \begin{center}
     \leavevmode 
         \epsfxsize=10cm\epsfbox{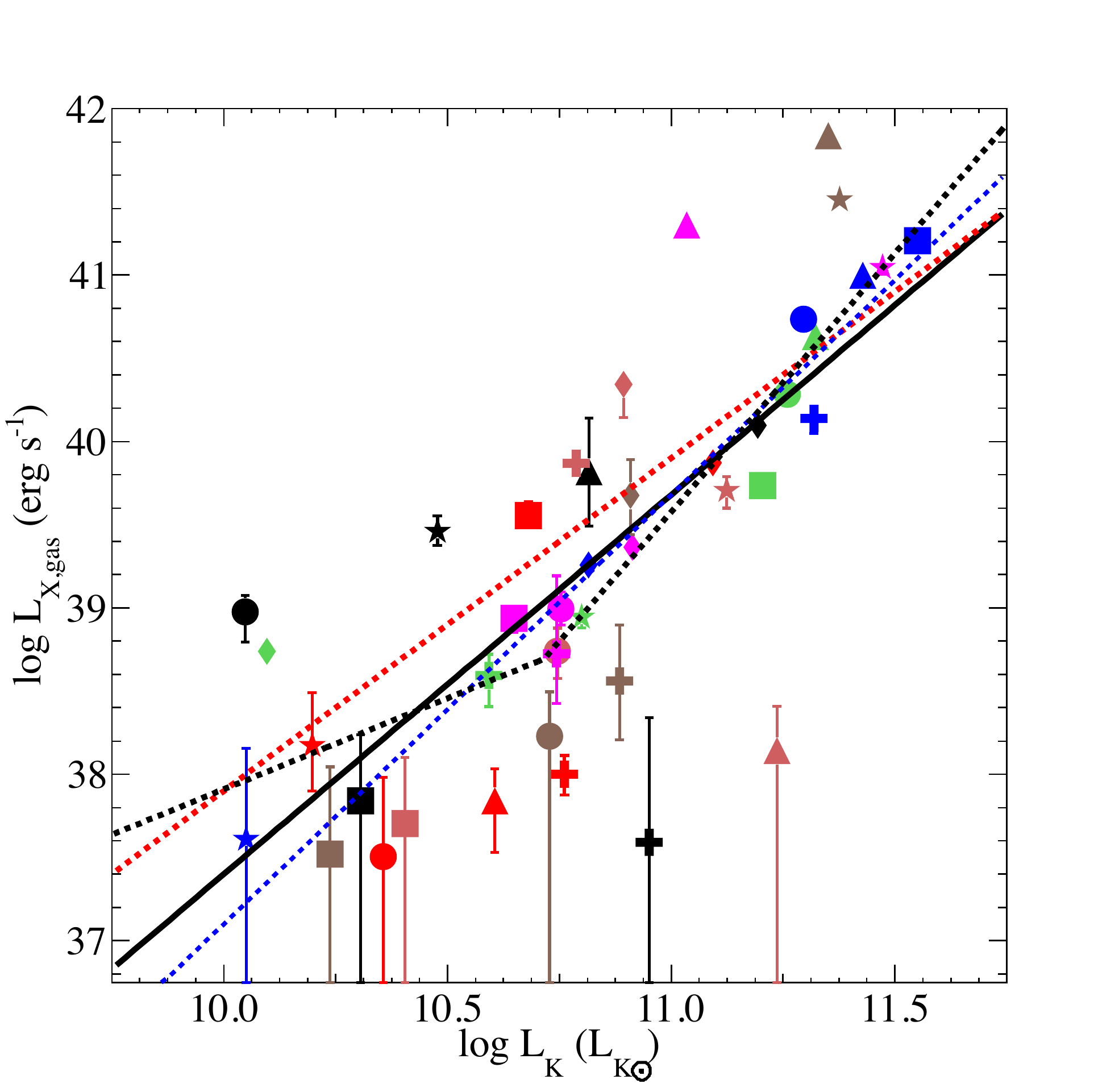}
\caption{Gaseous X-ray luminosities as a function of $K$-band luminosities for all galaxies in this work. 
The black solid line is the best-fit $L_{X_{\rm gas}}$--$L_K$ relation 
( ${\rm log}(L_{X_{\rm gas}})=2.3{\rm log}(L_K)+14.6$). Dashes black line is the best-fit $L_{X_{\rm gas}}$--$L_K$ relation to broken power law. Dashed red line indicates ${\rm log}(L_{X_{\rm gas}})=2.0{\rm log}(L_K)+17.9$ and dashed blue line indicates ${\rm log}(L_{X_{\rm gas}})=2.6{\rm log}(L_K)+11.3$. Color code is the same as in Figure~\ref{fig:lxk}.}
\label{fig:lxk_sys} 
\end{center}
\end{figure*}

The determination of the best-fit slope of the $L_{X_{\rm gas}}$--$L_K$ relation is very crucial to our study, since $\Delta L_{X_{\rm gas}}$ is very sensitive to this slope. 
In \S4, we fit the $L_{X_{\rm gas}}$--$L_K$ relation to a power law 
${\rm log}(L_{X_{\rm gas}})=A~{\rm log}(L_K)+B$ and obtain $A=2.3\pm 0.3$ and $B=14.6\pm 3.3$. The black solid line in Figure~\ref{fig:lxk} indicates the best-fit ${\rm log}(L_{X_{\rm gas}})=2.3~{\rm log}(L_K)+14.6$. We calculated $\Delta L_{X_{\rm gas}}$  again using 
$A=2.0$ and $A=2.6$ as indicated respectively by the dashed red line [${\rm log}(L_{X_{\rm gas}})=2.0~{\rm log}(L_K)+17.9$] and dashed blue line [${\rm log}(L_{X_{\rm gas}})=2.6~{\rm log}(L_K)+11.3$] in Figure~\ref{fig:lxk_sys}. Compared to the statistical uncertainties, the change of $A$ by $\pm$0.3 has little impact on our results as listed in Tables 4 and 5. Boroson et al.\ (2011) determined a best-fit of $A=2.6\pm0.4$ using a smaller sample of galaxies. We also performed the fit of the $L_{X_{\rm gas}}$--$L_K$ relation ignoring those poorly constrained datasets (8 extremely faint galaxies with their $L_{X_{\rm gas}}$ consistent with zero) and we obtained $A=2.1\pm0.3$. Both these cases are within our variations.  

We also performed the maximum likelihood procedure by Kelly (2007) to quantify the $L_{X_{\rm gas}}$--$L_K$ relation.
It is a Bayesian method that includes both individual measurement error and population intrinsic scatter within the linear regression. The best-fit determined by its {\tt linmix\_err} regression is ${\rm log}(L_{X_{\rm gas}})= (2.2\pm0.3)~{\rm log}(L_K)+(15.8\pm3.3)$, in fully consistent with the best-fit we obtained. It also falls into the range of variations we examined. We also listed the slope of the $L_{X_{\rm gas}}$--$L_K$ relation determined by {\tt linmix\_err} for each subgroup in Table~6, in agreement with the results we obtained.

The $L_X$--$L_{\rm opt}$ relation obtained by O'Sullivan et al.\ (2001) follows a broken power law (bknpower) (see Figure~\ref{fig:osullivan}). 
That may be a result of them including the LMXB components and biasing high the luminosities of faint galaxies due to the limited resolutions of {\it ROSAT}.  
We also fit the $L_{X_{\rm gas}}$--$L_K$ relation of galaxies in our sample obtained with high quality {\sl Chandra} observations to a bknpower and obtained a best-fit of
$$
{\rm log}(L_{X_{\rm gas}})=\left \{ 
\begin{array}{l}
1.1{\rm log}(L_K)+27.0  \quad ({\rm log}(L_K) < 10.7)\\
3.1{\rm log}(L_K)+5.4   \quad ({\rm log}(L_K) \ge 10.7)
\end{array} \right .\
$$

\begin{figure} 
\epsscale{1.11}
\plottwo{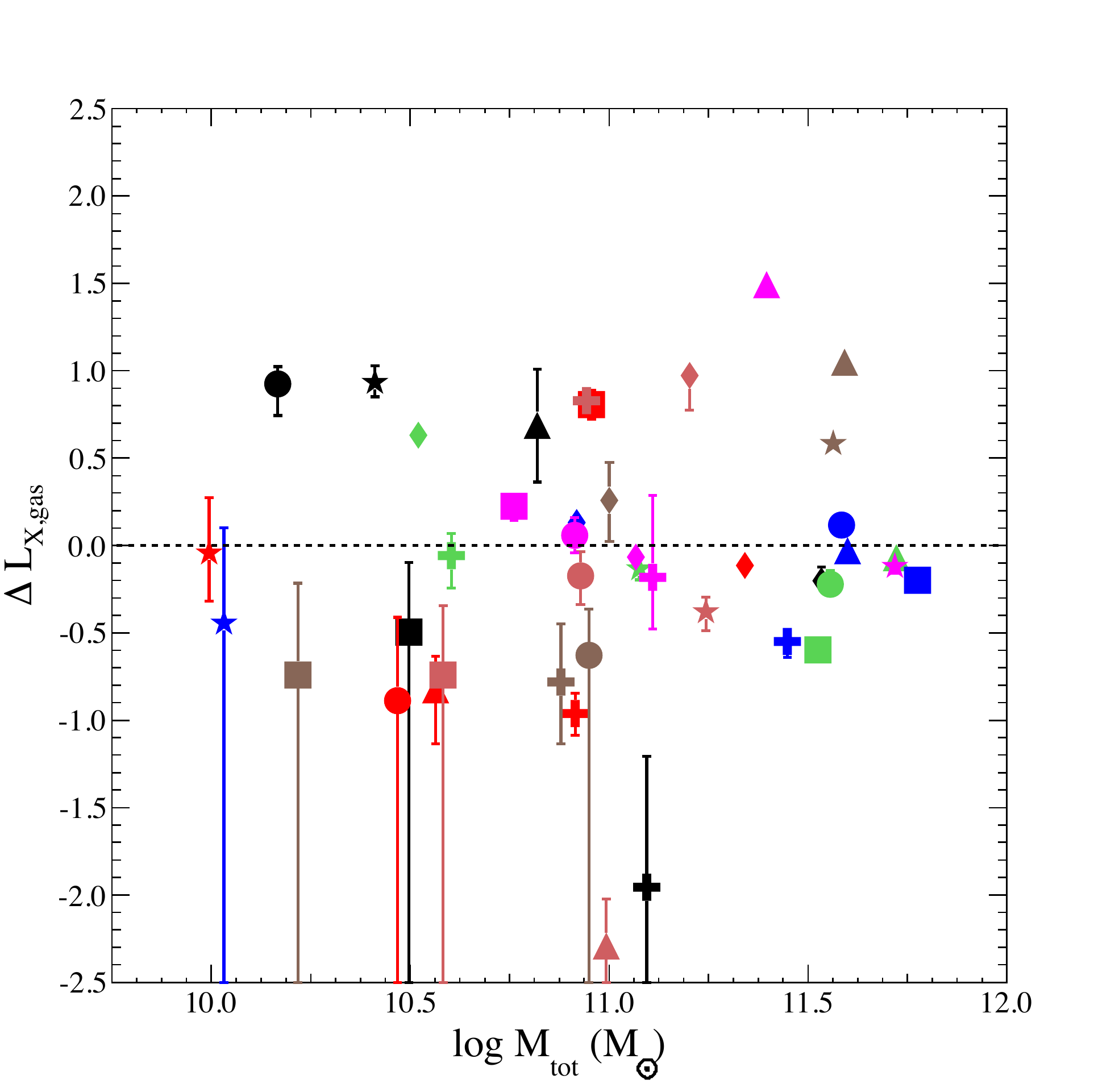}{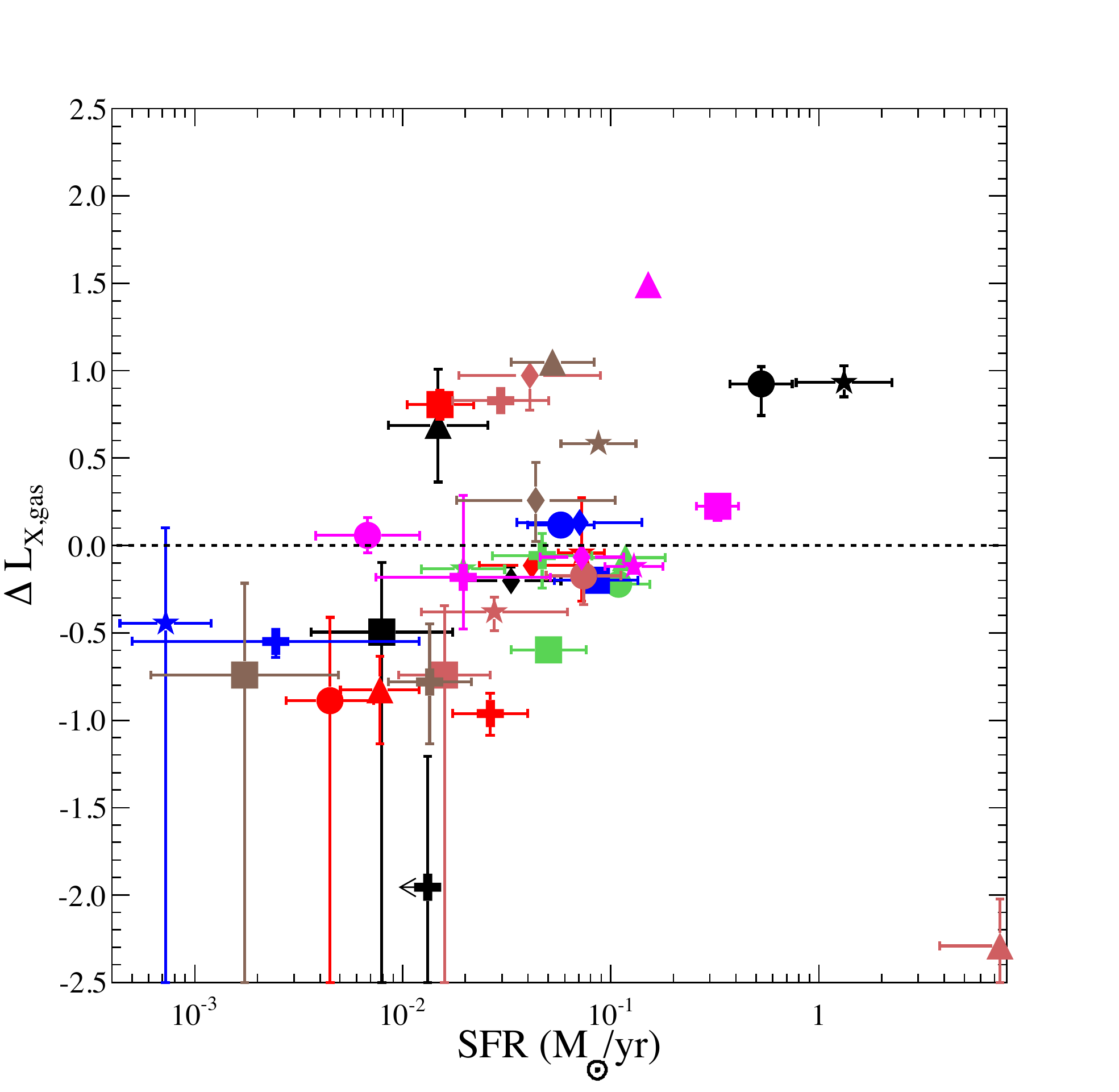}
\figcaption{\label{fig:bkn} $\Delta L_{X_{\rm gas}}$ as a function of total mass ({\it left}) and star formation rate ({\it right})
 when using a broken power law function as the best-fit of the $L_{X_{\rm gas}}$--$L_K$ relation. Color code is the same as in Figure~\ref{fig:lxk}.}
\end{figure}

The best-fits of the two slopes are $1.1\pm0.1$ and $3.1\pm0.5$. They are only different by 1.8\,$\sigma$.
Furthermore, we compared the standard deviations of the distributions of $\Delta L_{X_{\rm gas}}$ obtained by using bknpower and single power law respectively: $0.690\pm0.07$ and $0.737\pm0.109$. They are not significantly different from each other. 
We also performed {\tt ftest} and we do not find bknpower improve the fit significantly.
Therefore, we do not think it is statistically necessary to employ bknpower.
Still, we tested our results using a bknpower as the best-fit of the $L_{X_{\rm gas}}$--$L_K$ relation as listed in Table 5. Its impact does not change our results qualitatively.
The ``V-shaped" relation between $\Delta L_{X_{\rm gas}}$ and $M_{\rm tot}$ ($r_{\rm e}$, $M_{\rm tot}/r_{\rm e}$, and $\sigma$) diminished (Figure~\ref{fig:bkn}-{left}), while
$\Delta L_{X_{\rm gas}}$ is still most strongly correlated with star formation rate and hot gas temperature with even larger correlation coefficients (see Figure~\ref{fig:bkn}-{right}).

In addition to the Spearman correlation, we also use Kendall's tau correlation\footnote{{\url http://cran.r-project.org/web/packages/NADA/NADA.pdf}} to quantify the relations between various factors. None of these results are qualitatively different from using the Spearman correlation. The tau correlation coefficient of each relation is listed in Tables 4 and 5. 

\newpage
\bigskip

\begin{references}
\reference{} Abazajian, K.\ N., Adelman-McCarthy, J.\ K., AgŸeros, M.\ A. et al.\ 2009, ApJS, 182, 543 
\reference{} Alatalo, K., Lacy, M., Lanz, L.\ et al.\ 2014, arXiv1410.4556
\reference{} Arnold, J., Romanowsky, A., Brodie, J., et al.\ 2014, ApJ, 791, 80
%\reference{} Balogh, M., McCarthy, I., Bower, R., et al.\ 2008, MNRAS, 385, 1003
\reference{} Bentz, M., Walsh, J., Barth, A., et al.\ 2008, ApJ, 689, L21
\reference{} Binney, J.\ \& Tremaine, S.\ 1987, Galactic dynamics, Princeton, NJ, Princeton University Press, 1987, 747
%\reference{} Birzan, L., Rafferty, D., McNamara, B.\ et al.\ 2004, ApJ, 607, 800
\reference{} Boroson, B., Kim, D.\ \& Fabbiano, G.\ 2011, ApJ, 729, 12
\reference{} Brown, B.\ \& Bregman, J.\ 2000, ApJ, 539, 592
\reference{} Bruggen, M.\ \& De Lucia, G.\ 2008, MNRAS, 383, 1336
\reference{} Canizares, C., Fabbiano, G., \& Trinchieri, G.\ 1987, ApJ, 312, 503
\reference{} Cappellari, M., Emsellem, E., Krajnovic, D., et al.\ 2011, MNRAS, 413, 813
\reference{} Cappellari, M., Scott, N., Alatalo, K.\ et al.\ 2013, MNRAS, 432, 1709 
\reference{} Cen, R., Pop, A., Bahcall, N.\ 2014, arXiv1405.0537
\reference{} Chernin, A., Dolgachev, V., Domozhilova, L.\ 2010, ARep, 54, 902
\reference{} Civano, F., Fabbiano, G., Pellegrini, S., et al.\ 2014, ApJ, 790, 16
\reference{} Close, J., Pittard, J., Hartquist, T.\ et al.\ 2013, MNRAS, 436, 302
\reference{} Crocker, A., Krips, M., Bureau, M., et al.\ 2012, MNRAS, 421, 1298
\reference{} David, L., Nulsen, P., McNamara, B.\ et al.\ 2001, ApJ, 557, 546
\reference{} David, L., Jones, C., Forman, W., et al.\ 2006, ApJ, 653, 207
\reference{} David, L., Lim, J., Forman, W., et al.\ 2014, 792, 94
\reference{} Dickey, J.\ \& Lockman, F.\ 1990, ARA\&A, 28, 215 
\reference{} Deason, A., Belokurov, V., Evans, N.\  et al.\ 2012, ApJ, 784, 2
%\reference{} de Vaucouleurs et al.\ 1991
%\reference{} Dressler, A.\ 1980, ApJ, 236, 351
\reference{} Eskridge, P., Fabbiano, G., Kim, D 1995, ApJ, 442, 523
\reference{} Emsellem, E., Cappellari, M., Davor, K., Katherine, A.\ et al.\ 2011, MNRAS, 414, 888
\reference{} Fabbiano, G., Kim, D.-W., Trinchieri, G., 1992, ApJ, 80, 531
%\reference{} Fabian, A., Stewart, G., Nulsen, P., et al.\ 1984, Natur, 307, 343
\reference{} Forman, W., Nulsen, P., Heinz, S., et al.\ 2005, ApJ, 635, 894
\reference{} Finoguenov, A.\ \& Jones, C.\ 2002, ApJ, 574, 754
\reference{} Gallagher, J., Garnavich, P.\ M., \& Caldwell, N.\ et al.\ 2008, ApJ, 685, 752
%\reference{} Gomez et al.\ 2003
%\reference{} Goto et al.\ 2003
\reference{} Gunn, J.\ E.\ \& Gott, J.\ R.\ III 1972, ApJ, 176, 1
%\reference{} Helmboldt, J.\ F.\ 2007, MNRAS, 379, 1227
\reference{} Howell, J.\ 2005, AJ, 130, 2065
\reference{} Humphrey, P.\ J.\ \& Buote, D.\ A.\ 2006, ApJ, 639, 136
\reference{} Humphrey, P.\, Buote, D.\ A., \& Canizares, C.\ R.\ et al.\ 2011, ApJ, 729, 53
\reference{} Idiart, T.\ P., Silk, J., de Freitas Pacheco, J.\ 2007, MNRAS, 381, 1711
\reference{} Jesseit, R., Cappellari, M., Naab, T.\ et al.\ 2009, MNRAS, 397, 1202
\reference{} Irwin, J.\ A.\ \& Sarazin, C.\ L.\ 1996, ApJ, 471, 683
\reference{} Irwin, J.\ A., Athey, A.\ E., \& Bregman, J.\ N.\  2003, ApJ, 587, 356 
%\reference{} Kapferer, W., Ferrari, C., Domainko, W.\ et al.\ 2006, A\&A, 447, 827
\reference{} Kelly, B.\ C. 2007, ApJ, 665, 1489
\reference{} Kim, D.\ \& Fabbiano, G.\ 2013, ApJ, 776, 116
\reference{} Knapp, G.\ R., Gunn, J.\ E., Wynn-Williams, C.\ G.\ 1992, ApJ, 399, 76
\reference{} Kuntschner, H., Emsellem, E., Bacon, R., et al.\ 2010
\reference{} Lagos, C.\ P., Davis, T.\ A., Lacey, C.\ G.\ et al.\ 2014, MNRAS, 443, 1002
\reference{} Lavalley, M., Isobe, T., \& Feigelson, E.\ 1992, ASPC, 25, 245
\reference{} Lees, J.\ F., Knapp, G.\ R., Rupen, M.\ P., Phillips, T.\ G.1991, ApJ, 379, 177
%\reference{} Li, J., Wang, Q.\ D., Li, Z., \& Chen, Y.\ 2011, ApJ, 737, 41 
\reference{} Lucero, D.\ M.\ \& Young, L.\ M.\ 2013, AJ, 145, 56
\reference{} McCarthy, I.\ G., Frenk, C.\ S., \& Font, A.\ S., et al.\ 2008, MNRAS, 383, 593
\reference{} McDermid, R.\ M., Emsellem, E., Shapiro, K., L.\ et al.\ 2006, MNRAS, 373, 906
\reference{} Machacek, M., Dosaj, A., \& Forman, W.\ R., et al.\ 2005, ApJ, 621, 663
\reference{} Mathews, W.\ G.\& Brighenti, F.\ 1998, ApJ, 503, L15
\reference{} Mathews, W.\ G.\& Brighenti, F.\ 2003, ARA\&A, 41, 191
\reference{} Mathews, W.\ G., Brighenti, F., Faltenbacher, A.\ et al.\ 2005, ApJ, 652, L17
%\reference{} Mathews et al.\ 2006
\reference{} Mannucci, E., Della Valle, M., Panagia, N.\ et al.\ 2005 A\&A 433, 807
\reference{} Mulchaey, J.\ S.\ \& Jeltema, T.\ E.\ 2010, ApJ, 715, L1
\reference{} Negri, A., Posacki, S., Pellegrini, S., Ciotti, L.\ 2014, MNRAS, 445, 135
\reference{} Oosterloo, T., Morganti, R., Crocker, A.\ et al.\ 2010, MNRAS, 409, 500
\reference{} O'Sullivan, E., Forbes, D.\ A., \& Ponman, T.\ J.\ 2001, MNRAS, 328, 461
%\reference{} Peterson, J.\ R., Kahn, S.\ M., Paerels, F.\ B.\ S.\ et al.\ 2003 ApJ, 590, 207
\reference{} Pinto, C., Fabian, A.\ C., Werner, N et al.\ 2014, A\&A, 572, L8
%\reference{} Pipino, A., Kawata, D., Gibson, B.\ K., \& Matteucci, F.\ 2005, A\&A, 434, 553
\reference{} Posacki, S., Pellegrini, S., Ciotti, L.\ 2013, MNRAS, 433, 2259
\reference{} Randall, S., Nulsen, P., \& Forman, W.\ R., et al.\ 2008, ApJ, 688, 208
\reference{} Revnivtsev, M., Churazov, E., Sazonov, S.\ et al.\ 2007, A\&A, 473, 857 
\reference{} Revnivtsev, M., Churazov, E., Sazonov, S.\ et al.\ 2008, A\&A, 490, 37
\reference{} Revnivtsev, M., Churazov, E., Sazonov, S.\ et al.\ 2009, Nature, 458, 1142
\reference{} Sarzi, M., Alatalo, K., Blitz, L., et al.\ 2013, MNRAS, 432, 1845
\reference{} S‡nchez-Bl‡zquez, P., Gorgas, J.,  Cardiel, N., Gonz‡lez, J.\ J.\ 2006, A\&A, 457, 809
\reference{} Serra, P.\ \& Oosterloo, T.\ A.\ 2010, MNRAS, 401, L29
\reference{} Serra, P., Oosterloo, T., Morganti, R.\ et al.\ 2012, MNRAS, 422, 1835
\reference{} Sil'Chenko, O.\ K.\ 2006, ApJ, 641, 229
\reference{} Sil'Chenko, O.\ K.\ \& Chilingarian, I.\ V.\ 2011, AstL, 37, 1
%\reference{} Solanes et al.\ 2001
\reference{} Skrutskie, M.\ F.\ , Cutri, R.\ M.\, Stiening, R., et al.\ 2006, AJ, 131, 1163
\reference{} Su, Y.\ \& Irwin, J.\ 2013, ApJ, 766, 61
\reference{} Su, Y., Gu, L., White, R.\ E.\ III, Irwin, A.\ J.\ 2014, ApJ, 786, 152
\reference{} Sun, M., Jones, C., \& Forman, W.\ et al.\ 2007 ApJ, 657, 197
\reference{} Terlevich, A.\ I.\ \& Forbes, D.\ A.\ 2002, MNRAS, 330, 547
\reference{} Trager, S.\ C., Faber, S.\ M., Worthey, Guy, \& Gonzalez, J.\ J.\ 2000, AJ, 119, 1645
%\reference{} Trager 1998
\reference{} Urban, O., Werner, N., Simionescu, A., et al.\ 2011, MNRAS, 414, 210
\reference{} White, R.\ E.\ III \& Sarazin, C.\ L.\ 1991, ApJ, 367, 476
\reference{} Young, L.\ M., Bureau, M., Davis, T.\ A.\ et al.\ 2011, MNRAS, 414, 940
\reference{} York, D.\ G., Adelman, J., Anderson, J.\ E., Jr. et al.\ 2000, AJ, 120, 1579 
\reference{} Zhang, Y., Gu, Q.-S., \& Ho, L.\ C.\ 2008, A\&A, 487, 177
\end{references}
\end{document}